\documentclass{article}

\usepackage{arxiv}

\usepackage[utf8]{inputenc} % allow utf-8 input
\usepackage[T1]{fontenc}    % use 8-bit T1 fonts
\usepackage{hyperref}       % hyperlinks
\usepackage{url}            % simple URL typesetting
\usepackage{booktabs}       % professional-quality tables
\usepackage{amsfonts}       % blackboard math symbols
\usepackage{nicefrac}       % compact symbols for 1/2, etc.
\usepackage{microtype}      % microtypography
\usepackage{lipsum}
\usepackage{graphicx}
\graphicspath{ {./figs/} }

\usepackage{float}
\usepackage{nccmath}
\usepackage[title]{appendix}
\usepackage{nameref}
\usepackage{chngcntr}

\usepackage{amssymb}

\usepackage[square,sort,comma,numbers]{natbib}
\usepackage{algorithm}
\usepackage{algpseudocode}

\usepackage{array, makecell, rotating}
\usepackage{mdframed}
\setcellgapes{2pt}

\usepackage[table,xcdraw]{xcolor}
\usepackage{caption}
\title{Measuring unequal distribution of pandemic severity across census years, variants of concern and interventions}

\author{
   Quang Dang Nguyen \textsuperscript{*}\\
   Centre for Complex Systems\\
   Faculty of Engineering\\
   The University of Sydney\\
   New South Wales, Australia \\
   \texttt{dang.q.nguyen@sydney.edu.au} \\
    \And
   Sheryl L. Chang \textsuperscript{*} \\
   Centre for Complex Systems\\
   Faculty of Engineering\\
   The University of Sydney\\
   New South Wales, Australia \\[3pt]
   Sydney Institute for Infectious Diseases \\
   The University of Sydney\\
   New South Wales, Australia \\
   \texttt{sheryl.chang@sydney.edu.au} \\
    \And
    Christina M. Jamerlan\\
   Centre for Complex Systems\\
   Faculty of Engineering\\
   The University of Sydney\\
   New South Wales, Australia \\
   \texttt{ma.jamerlan@sydney.edu.au} \\
   \And
  Mikhail Prokopenko  \\
   Centre for Complex Systems\\
   Faculty of Engineering\\
   The University of Sydney\\
   New South Wales, Australia \\ [3pt]
   Sydney Institute for Infectious Diseases \\
   The University of Sydney\\
   New South Wales, Australia \\
 \texttt{mikhail.prokopenko@sydney.edu.au}
 }

\begin{document}
\maketitle
\vspace*{0.5cm}
\begin{abstract}
Diverse and complex intervention policies deployed over the last years have shown varied effectiveness in controlling the COVID-19 pandemic. However, a systematic analysis and modelling of the combined effects of different viral lineages and complex intervention policies remains a challenge. Using large-scale agent-based modelling and a high-resolution computational simulation matching census-based demographics of Australia, we carried out a systematic comparative analysis of several COVID-19 pandemic scenarios. The scenarios covered two most recent Australian census years (2016 and 2021), three variants of concern (ancestral, Delta and Omicron), and five representative intervention policies. In addition, we introduced pandemic Lorenz curves measuring an unequal distribution of the pandemic severity across local areas. We quantified nonlinear effects of population heterogeneity on the pandemic severity, highlighting that (i) the population growth amplifies pandemic peaks, (ii) the changes in population size amplify the peak incidence more than the changes in density, and (iii) the pandemic severity is distributed unequally across local areas. We also examined and delineated the effects of urbanisation on the incidence bimodality, distinguishing between urban and regional pandemic waves. Finally, we quantified and examined the impact of school closures, complemented by partial interventions, and identified the conditions when inclusion of school closures may decisively control the transmission. Our results suggest that (a) public health response to long-lasting pandemics must be frequently reviewed and adapted to demographic changes, (b) in order to control recurrent waves, mass-vaccination rollouts need to be complemented by partial NPIs, and (c) healthcare and vaccination resources need to be prioritised towards the localities and regions with high population growth and/or high density.
\end{abstract}

% % keywords can be removed

\textsuperscript{*}: \textit{these authors contributed equally}\\
%\textsuperscript{\#}: \textit{corresponding author}

\section{Introduction}
\label{sec:main}
On 30 January 2020 the COVID-19 was recognised by the World Health Organisation (WHO) as a public health emergency of international concern: the WHO's highest level of alert. On 11 March 2020 this was followed by the WHO declaring the outbreak a pandemic~\cite{WHO11032020}. On 5 May 2023, that is, 170 weeks since announcing the global health emergency, the WHO declared an end to the emergency, while continuing to refer to the COVID-19 as a pandemic~\cite{WHO05052023}. Over this time, the COVID-19 pandemic has had a profound impact, causing significant loss of life, reducing life expectancy~\cite{andrasfay2021reductions,aburto2022quantifying}, seriously challenging healthcare systems~\cite{miller2020disease}, and adversely affecting socio-economic activity worldwide~\cite{del_rio-chanona_supply_2020}. By mid-June 2023, the pandemic had caused almost 800 million confirmed cases and 7 million confirmed deaths~\cite{owidcoronavirus}. 

Over the course of pandemic, the severe acute respiratory syndrome coronavirus 2 (SARS-CoV-2) which causes COVID-19 has mutated from its ancestral strain into a number of lineages and sub-lineages, varying in terms of infectivity and virulence. Several of these variants have been designated by the WHO as variants of concern, including the highly transmissible lineages B.1.617.2 (Delta)~\cite{campbell2021increased,milne2022mitigating} and B.1.1.529 (Omicron)~\cite{karim2021omicron,duong2022sars}. In Australia, these variants have triggered significant new waves of the pandemic~\cite{covid19data,chang_simulating_2022,porter2022new,chang_persistence_2023}. 
%The third pandemic wave, created by the Delta variant, peaked in Australia at approximately 100 daily cases per million (mid-October 2021)~\cite{covid19data,chang_simulating_2022}. The fourth stage triggered by the Omicron variant~\cite{porter2022new} included several recurrent waves. The first and largest peak was observed across the nation in mid-January 2022 (around 4,250 daily cases per million), while subsequent peaks also proved to be significant: 2,200 daily cases per million in early April 2022; and 2,000 daily cases per million in mid-May 2022~\cite{chang_persistence_2023}. 

In response, the public health systems worldwide employed a diverse range of intervention policies. The initial reactions involved non-pharmaceutical interventions (NPIs)~\cite{walker2020impact,flaxman2020estimating,chang_modelling_2020}. Typically, the NPI interventions combined various components, such as border closures and travel restrictions, case isolation, home quarantine, school closures, and comprehensive ``stay-at-home'' orders comprising social distancing.  Once safe and effective vaccines became available, this was followed by mass vaccination campaigns around the globe~\cite{moore2021vaccination,harris2021impact,zachreson_how_2021}. Vaccination rollouts differed with respect to (a) population coverage, ranging from partial to nearly complete; (b) rollout schemes, e.g., preemptive, progressive, or boosting; as well as (c) different vaccine combinations, e.g., priority and general vaccines~\cite{zachreson_how_2021}. Each vaccine had efficacy variations with respect to (i) the susceptibility-reducing efficacy, (ii) the disease-preventing efficacy, and (iii) the transmission-limiting efficacy~\cite{moore2021vaccination,zachreson_how_2021}. In addition, vaccine effectiveness was non-linearly diminishing over time~\cite{szanyi2022log}. %Often the intervention policies were adapted in response to a rapidly unfolding crisis, under different socio-economic constraints which needed to reconcile the health effects and economic losses~\cite{nguyen_general_2022}. 
The practice of dealing with multiple variables, objectives and constraints confounded many potential effects of various intervention policies: for example, school closures have been found to  contribute differently under different circumstances~\cite{chang_modelling_2020,viner2020school}.  

The combined effects of the evolving viral lineages and complex intervention policies have been difficult to systematically analyse, model and predict. For example, the persistence of Omicron variant in Australia and the resulting recurrent waves were explained by a nuanced combination of the new transmissible sub-variants, the fluctuating adoption of NPIs, and the waning immunity from prior infections and vaccinations~\cite{chang_persistence_2023}. Importantly, such complex effects become sensitive to demographic variations in heterogeneous populations spanning different age groups, household sizes, socio-economic profiles and jurisdictions. In general, to study a pandemic which has lasted more than three years, one needs to account for demographic changes which play an increasingly salient role. This influence often remains concealed due to the lack of high-resolution data and presence of jurisdictional barriers, socio-political biases,  and other factors~\cite{kontis2020magnitude,volk2021influence}. 

Here, we aim to examine some of these public health challenges and carry out a systematic simulation-based analysis of several COVID-19 pandemic scenarios, using Australia as a case study. We use two most recent census years (2016 and 2021) as the alternative demographic settings within which each pandemic scenario is simulated. Our comparative analysis contrasts three variants of concern which made an impact in Australia: the ancestral strain, the Delta and the Omicron lineages.  For every scenario, five representative intervention policies are compared, ranging from (1) baseline (i.e., no interventions), to (2) partial NPIs without vaccination, (3) partial preemptive vaccination without NPIs, (4) mixed intervention with both partial NPIs and partial vaccination, and (5) partial lockdown including school closures. 

In order to compare 30 possible scenarios ($2 \times 3 \times 5$) we apply an agent-based model (ABM) which simulates an artificial population generated using the high-resolution Australian census data. %thus comprising 23.4 million agents for 2016 and 25.4 million agents for 2021. 
The ABM has been previously calibrated and validated for several pandemic stages in Australia during the last four years~\cite{chang_modelling_2020,zachreson_how_2021,chang_simulating_2022,nguyen_general_2022,chang_persistence_2023}.

This study identifies and explains, in context of different variants and policies, several coupled nonlinear effects of the population growth and heterogeneity on the pandemic severity. In particular, we study how pandemic peaks may be amplified by distributed changes in the population size or the changes in population density.

Importantly, the study introduces a novel measure of pandemic inequality --- \textit{pandemic Lorenz curves} --- and demonstrates that the pandemic severity may be distributed unequally across local areas. For example, while the pandemic inequality may reduce when the population or the disease transmissibility grow, it may increase with more stringent interventions or in non-urban areas. We also measure \textit{pandemic biomodality}, which characterises formation of distinct urban and regional pandemic waves. In addition, we measure \textit{bifurcations in the effectiveness} of interventions, e.g., school closures, relative to variants of concern.

\section{Methods}
\label{sec:methods}

\subsection{Agent-based modelling COVID-19 pandemic with Australian census}  

We simulated several scenarios of the COVID-19 pandemic in Australia using a well-established agent-based model previously validated for several pandemic waves and variants of concern \cite{chang_modelling_2020,zachreson_how_2021,chang_simulating_2022,chang_persistence_2023}. Our model has two fundamental components: (i) a simulated Australian population generated to represent key demographic features of the Australian census data, and (ii) a stochastic agent-based model for disease transmission and control,  detailed in Appendix \ref{sec_supp:pop_gene} and Appendix \ref{sec_supp:model} respectively.

Our model comprises stochastically-generated anonymous agents that represent the population of Australia: about 23.4 million using 2016 census, and 25.4 million using 2021 census. The population is partitioned into Statistical Areas (SAs) at different resolutions, e.g., SA2 level represents suburbs.  
%(between 3,000 and 25,000 individuals) in which people interact socially and economically.

Disease transmission follows a discrete-time simulation, updating states of each agent over time. Following initial infections ``seeded'' around international airports, the transmission is probabilistically simulated by considering agent interactions across multiple  social layers (mixing contexts), given different contact and transmission rates within both residential and work/study contexts. Epidemiological characteristics and variant-specific natural history of the disease are described in Appendix \ref{sec_supp:model}, and in Section~\ref{sec:variant}.
%
%In addition to social interactions, disease transmission in our model is also dependent on:
%\begin{enumerate}[(i)]
%    \item Epidemiological characteristics 
%    defined by parameters such as the fractions of symptomatic/asymptomatic cases for adults and children, the reduced probabilities of transmission for asymptomatic agents, case detection rates for symptomatic/asymptomatic cases, and other parameters, described in Section 3 in Supplementary Material.
%    \item A variant-specific natural disease history model for an infected agent (symptomatic or asymptomatic). An infected agent transitions through different states, starting from exposure, peaking at maximum infectivity, and proceeding to recovery. The infectivity profile  varies across the variants of concern, based on available epidemiological evidence, as described in Section \ref{sec:variant} and section 3 in Supplementary Material.
%\end{enumerate}
%
The transmission process is also affected by intervention policies, including NPIs and  vaccinations. The transmission probabilities are modified across  social contexts of the agents in response to their compliance with specific intervention policies. 
%In this study, we explored several intervention scenarios with different combinations of NPIs and vaccination coverage, across three variants of concern, as detailed in Section \ref{sec:scenario}.

\subsection{Simulated pandemic scenarios} 
\label{sec:scenario}
\subsubsection{Variants of concern}
\label{sec:variant}
\begin{figure}[ht]
    \centering
    \includegraphics[width=0.75\textwidth]{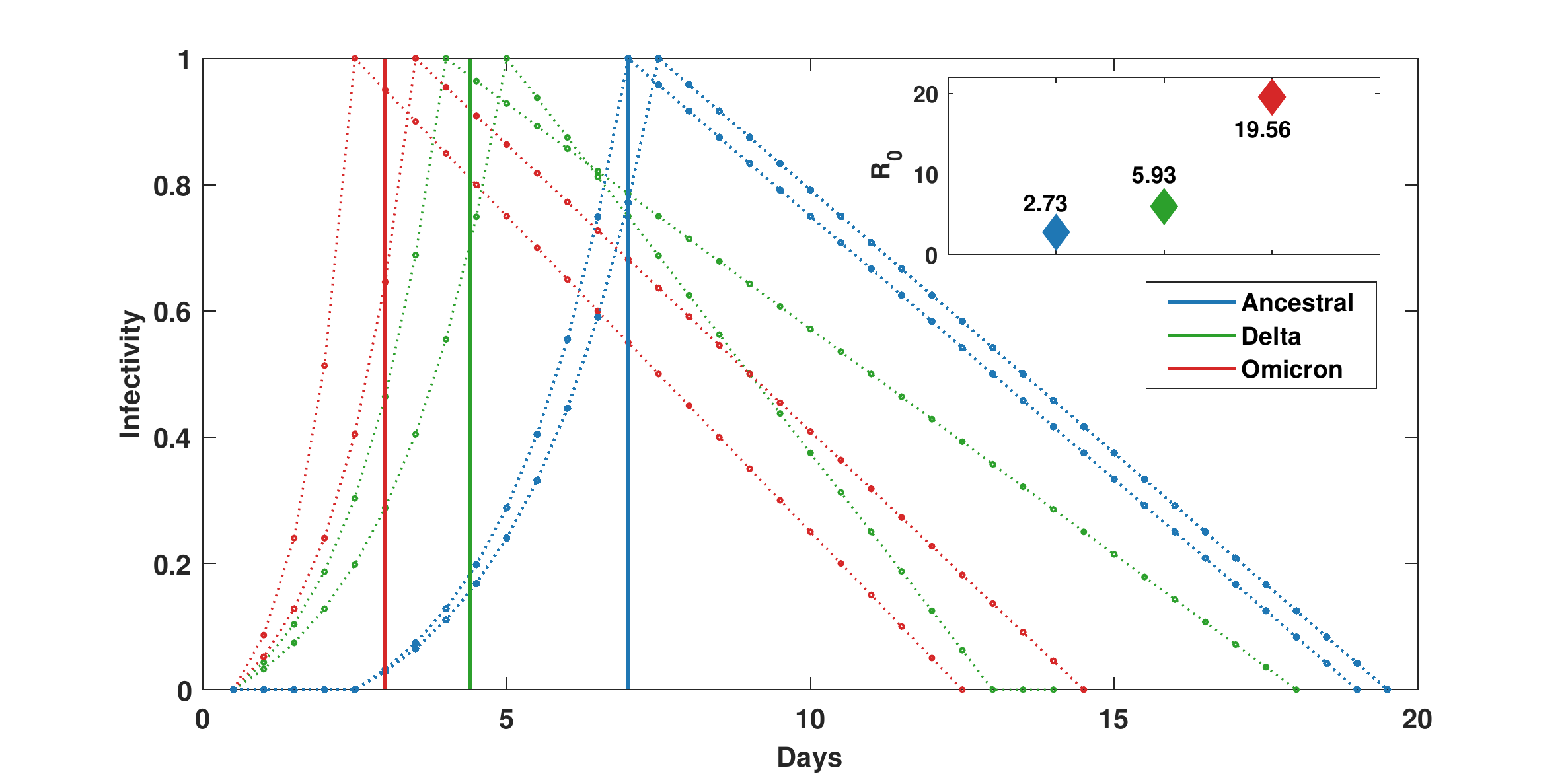}
    \caption{Model of the natural history of three COVID-19 variants: ancestral (blue), Delta (green), and Omicron (red). The illustrated profiles are sampled from 2 random agents. Each profile rises exponentially until reaching the infectivity peak, followed by a linear decrease until full recovery. Vertical lines mark the mean incubation period for the three considered variants (ancestral: blue, Delta: green, and Omicron: red), with the means following a log normal distribution. The mean incubation period and recovery period for each of the variants are reported in Appendix Table \ref{tab_supp:epi_para}. The inset shows $R_0$ of the three considered variants. }
    \label{fig:selected_variants_of_concern}
\end{figure}

Our model has been calibrated to match key COVID-19 characteristics across three variants of concern: ancestral (i.e., the strain initially detected in Wuhan, which was prevalent in Australia in 2020)~\cite{chang_modelling_2020}, Delta (i.e.,  B.1.617.2 variant, prevalent in Australia in 2021)~\cite{chang_simulating_2022}, and Omicron (i.e.,  B.1.1.529 variant, prevalent in Australia during 2022)~\cite{chang_persistence_2023}. Over the last four years, these variants have not only  evolved towards higher infectivity (i.e.,  higher basic reproductive number, $R_0$), but have also exhibited distinct characteristics in the disease natural history. In this study, we performed a lateral comparison of these three variants, each following a different natural disease history (as shown in Figure \ref{fig:selected_variants_of_concern}), combined with the intervention policies described in Section \ref{sec:policy}, and using the data from two most recent census years (2016 and 2021). See Appendix \ref{sec_supp:model} for a detailed parameterisation of the relevant epidemiological characteristics.

\subsubsection{Intervention policies}
\label{sec:policy}

NPIs considered in our model include case isolation (CI), home quarantine (HQ), social distancing (SD), and school closures (SC), each affecting different agents based on their health states (infected or susceptible), age groups (school-aged or not), and household compositions (if there is an infected household member).  
%While CI and HQ are activated from the beginning of the simulation and last throughout the entire simulation period, SD and SC are triggered only if the cumulative incidence exceeds certain threshold. 
%The selection of agents following SD is determined by Bernoulli sampling. 

In simulating pandemic scenarios, we assumed a moderate level of preemptive vaccination coverage  of the population (50\%), accounting for the combined effects of a relatively high vaccination coverage in Australia~\cite{vac_coverage} and low diminishing vaccine efficacy \cite{vac_eff_diminishing}. In line with prior studies~\cite{chang_simulating_2022,chang_persistence_2023,zachreson_how_2021}, the vaccination scheme distributes two types of vaccines (priority and general), each with the vaccine efficacy defined in terms of reducing susceptibility, preventing symptoms of the disease, and limiting further transmission, as described in Appendix Section \ref{sec_supp:vac}.

\begin{figure}[ht]
    \centering
    \includegraphics[width=\textwidth]{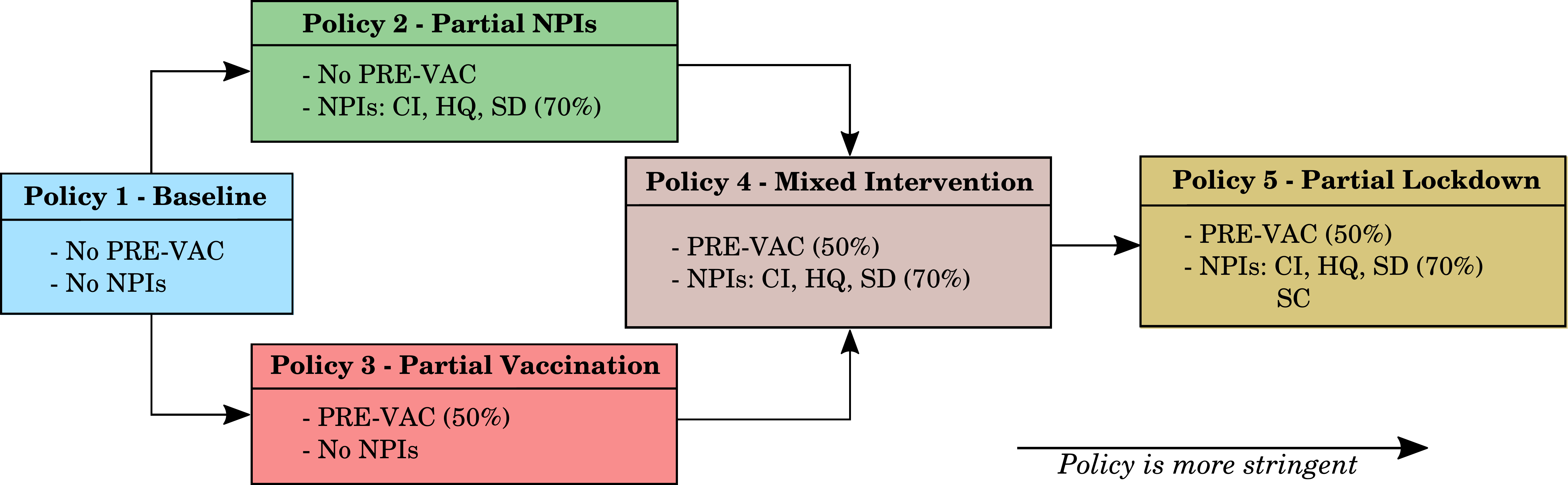}
    \caption{Five simulated intervention policy scenarios. PRE-VAC: preemptive vaccination  prior to the pandemic. NPIs: non-pharmaceutical interventions. Policies are considered to be more stringent moving from left to right. The macro- and micro-parameters for NPI-related policies are summarised in Appendix Table \ref{tab:NPI}.  Parameters relating to the vaccination coverage and vaccine efficacy are summarised in Appendix Table \ref{tab:vac_parameters}. }
    \label{fig:policies_diamond_flow_chart}
\end{figure}

We simulated five specific policies, each with a different combination of preemptive vaccination coverage and NPIs. These policies cover a wide range of scenarios, starting from the ``live-as-usual'' intervention-free scenario to the lockdown-like scenario with strong restrictions limiting population mobility and social interactions, as shown in Figure \ref{fig:policies_diamond_flow_chart}. Specifically, these policies can be summarised as follows:
\begin{itemize}
    \item Policy 1 --- Baseline: no NPIs and no preemptive vaccination coverage, representing a scenario without any interventions.
    \item Policy 2 --- Partial NPIs: some NPIs implemented (CI, HQ and SD at 70\% compliance level), with no preemptive vaccination coverage. This represents a pandemic intervention scenario feasible without vaccines.
    \item Policy 3 --- Partial vaccination: 50\% preemptive vaccination of the population before a pandemic wave. This represents a scenario developing in a population with partially acquired immunity,  but without any restrictions on social interactions during the pandemic.
    \item Policy 4 --- Mixed intervention: some NPIs implemented during a pandemic wave (CI, HQ and SD at 70\% compliance level) assuming that 50\% of the population has been preemptively vaccinated prior to the pandemic wave. This represents a scenario with partial acquired immunity in the population, followed by further restrictions on social interactions during the wave.
    \item Policy 5 --- Partial lockdown: all NPIs implemented (CI, HQ, SD at 70\% compliance level, and SC in addition) and 50\% preemptive vaccination of the population. This scenario represents a lockdown of the partially immunised population using strong restrictions on social interactions. 
\end{itemize}

\subsection{Lorenz curves: measuring unequal distribution of pandemic severity}
\label{sec:Lorenz}
Different communities may experience impacts of interventions in significantly different ways, and these complex effects may not uniformly aggregate into the national pandemic dynamics (e.g., nationwide incidence and cumulative incidence). To examine distribution of the overall pandemic severity across different local areas, we introduce a novel technique based on Lorenz curves.  

Lorenz curves, initially proposed to evaluate the degree of inequality in wealth distribution in society~\cite{lorenz_methods_1905}, have since been applied in many other domains, such as economics~\cite{chand_applications_nodate, ABS_Lorenz_curve_2022}, underpinning the well-known Gini index which measures income/wealth distribution across a population~\cite{ceriani2012origins},  and biology~\cite{wittebolle_initial_2009, damgaard_describing_2000}. 
In this study, we proposed and constructed the pandemic Lorenz curves that capture inequality in the distribution of cumulative incidence at the SA2 level, and compared their shapes across the considered scenarios. 
The pandemic Lorenz curve  dissects the nationwide cumulative incidence at the SA2 level, tracing it across all SA2 areas and assessing their relative contribution to the pandemic severity.

\begin{figure}[ht]
    \centering
    \includegraphics[width=0.45\textwidth]{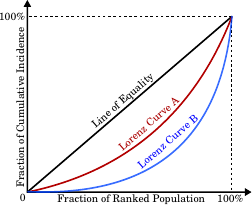}
    \caption{Pandemic Lorenz curves measuring inequality in distribution of the pandemic severity. The black line represents the line of equality where each SA2 contributes equally to the cumulative incidence. A curve closer to the line of equality (i.e., Lorenz Curve A, shown in red) indicates that the contributions of SA2 residential areas towards the aggregate cumulative incidence in response to a specific intervention policy A are more equally distributed than the contributions of these areas under policy B which are traced by the curve shaped further away from the line of equality (i.e., Lorenz Curve B, shown in blue).  }
    \label{fig:lorenz_curve_for_attack_rate}
\end{figure}

Figure \ref{fig:lorenz_curve_for_attack_rate} shows a simplified example to demonstrate possible shapes of pandemic Lorenz curve where the x-axis represents the cumulative fraction of SA2 residential population, ranked by their local attack rate (cumulative incidence over the SA2 residential population), and the y-axis represents the fraction of cumulative incidence at the `global' national level, contributed by the corresponding fraction of SA2 residential population.  Appendix \ref{sec_supp:lorenz} provides a detailed explanation of the Lorenz curves introduced in our study to measure the unequal distribution of pandemic severity.

\section{Results}
\label{sec:results}
We present our results in three parts matching our key objectives, firstly examining the effects of population heterogeneity on pandemic severity across two census years  (Section \ref{sec:pop_hetero}), then exploring pandemic spread, under the considered policies, in terms of urbanisation (Section \ref{sec:urbanisation}), and finally examining  varying effects of a specific intervention policy --- school closures --- across the three considered variants of concern (Section \ref{sec:school}). 

\subsection{Effects of population heterogeneity on pandemic severity}
\label{sec:pop_hetero}
We related the population heterogeneity with the pandemic severity observed in simulated scenarios across different intervention policies. Here, we measured the pandemic severity as the normalised incidence per million (unless specified otherwise), computed as the ratio between the detected incidence cases to the total population (in millions) for the considered census year. 
We assessed the population heterogeneity in terms of population increase at the `global' national level (Section \ref{sec:hetero_policy}) and at the `local' SA2 level (Sections \ref{sec:hetero_SLA} and \ref{sec:hetero_lorenz}).  
Appendix \ref{sec_supp:pop} provides more information about the population structure captured by the Australian Bureau of Statistics (ABS).

\subsubsection{Population growth amplifies pandemic peaks}
\label{sec:hetero_policy}
\begin{figure}[ht]
    \centering
    \includegraphics[width=\textwidth]{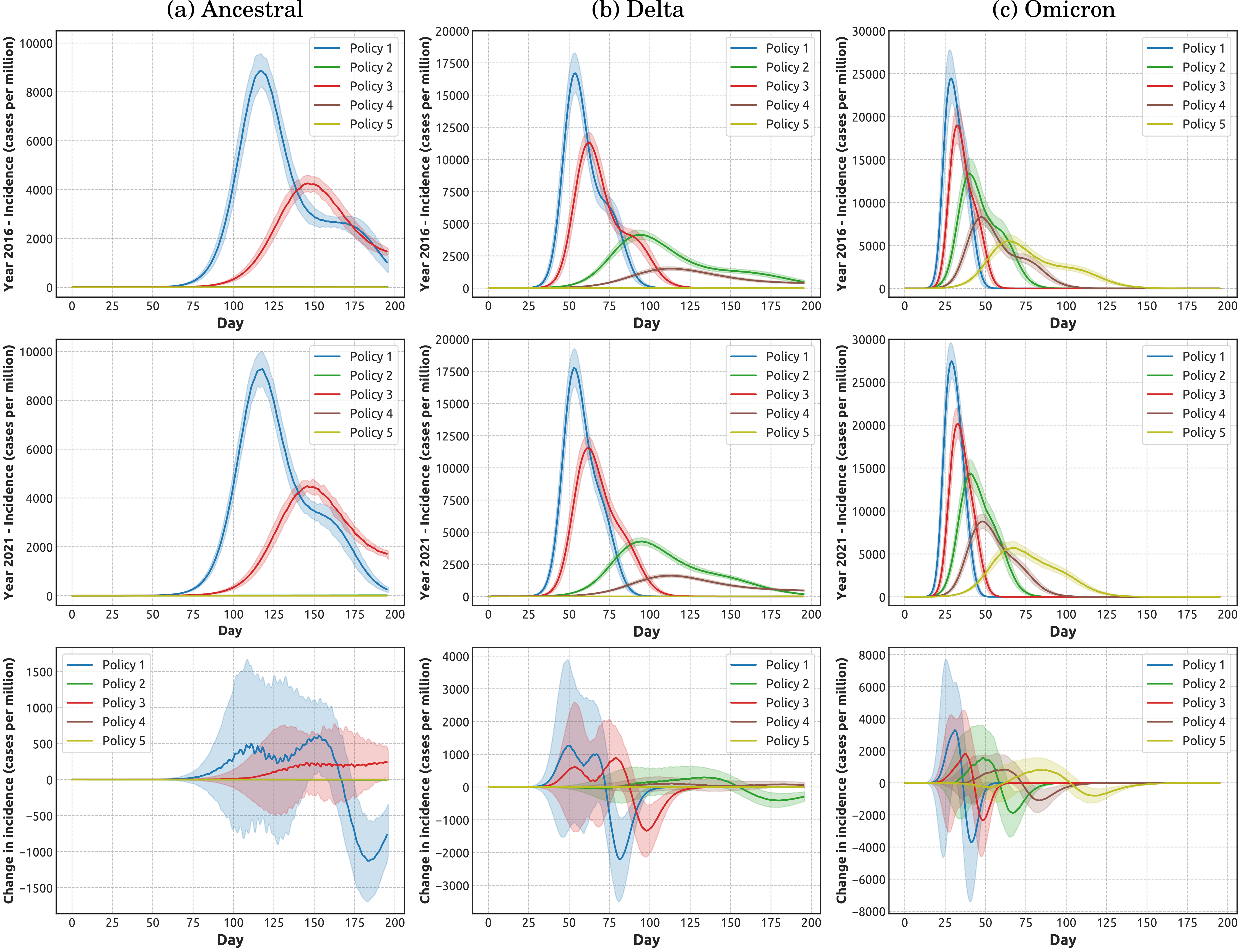}
    \caption{Impact of different intervention policies on pandemic severity for three considered variants simulated for two census years (top row: 2016; middle row: 2021; bottom row: relative change between years). Each column compares the impact of five intervention policies for one variant of concern: (a) ancestral; (b) Delta; (c) Omicron. See Figure \ref{fig:policies_diamond_flow_chart} for a detailed description of the considered intervention policies. Coloured shaded areas around  solid lines show standard deviation. Each profile corresponds to one intervention policy and is computed as the average over 100 runs.}
    \label{fig:sim1}
\end{figure}

\begin{figure}[ht]
    \centering
    \includegraphics[width=\textwidth]{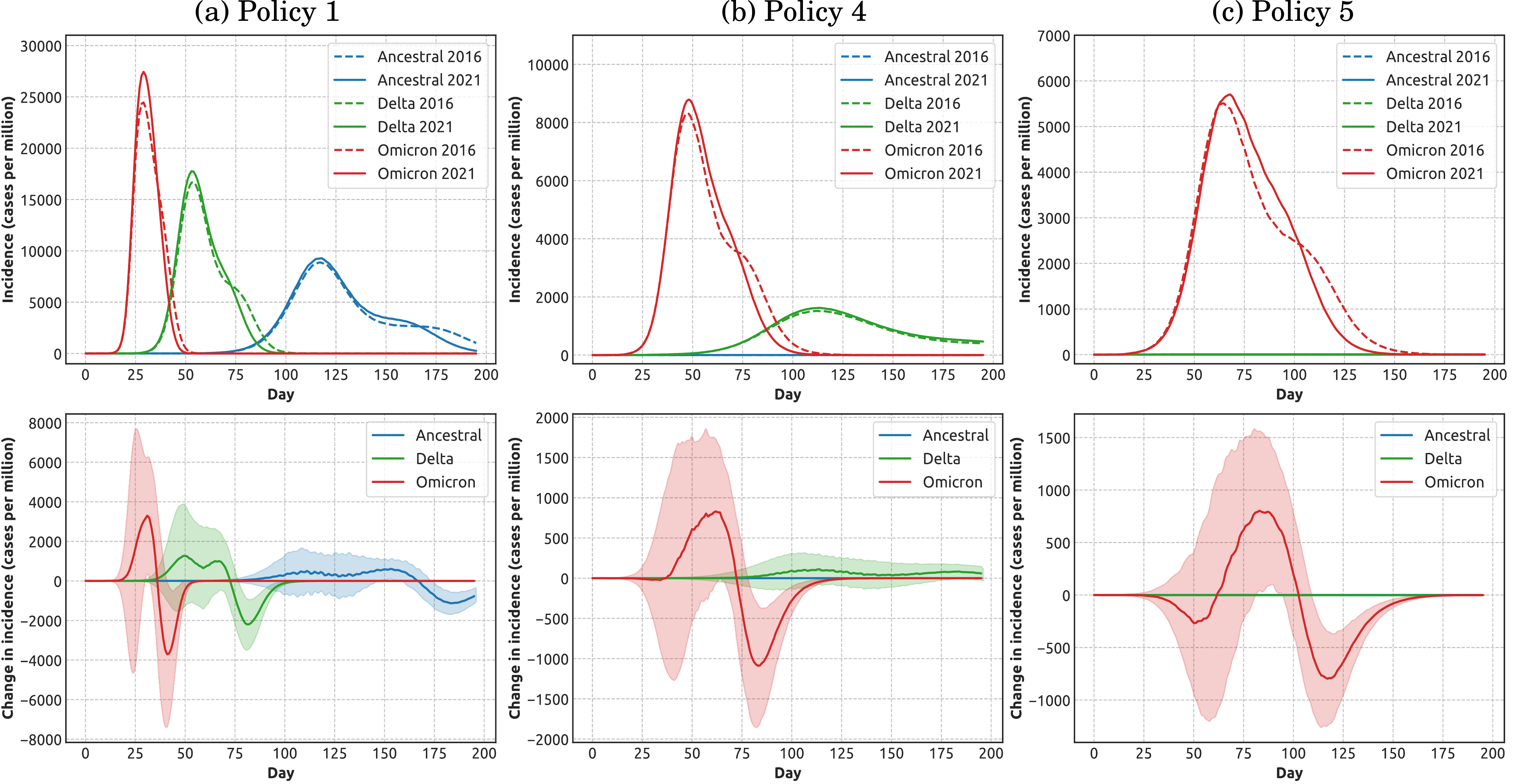}
    \caption{A comparison of pandemic severity for different policies across three considered variants (ancestral: blue; Delta: green; Omicron: red) and two census years (solid line: 2021; dashed line: 2016). The severity of each variant is measured by cases per million (top row). The change in incidence (bottom row) is calculated as the difference of incidence cases per million between two census years.  Each column compares the impact of three variants for one intervention policy: (a) Policy 1; (b) Policy 4; and (c) Policy 5.  See Figure \ref{fig:policies_diamond_flow_chart} for a detailed description of the considered intervention policies. Coloured shaded areas around  solid lines (in bottom row) show standard deviation. Each profile (solid and dashed lines) corresponds to one intervention policy and is computed as the average over 100 runs. }
    \label{fig:sim2}
\end{figure}

Our results show that the population growth, i.e., the 8.6\% population increase between 2016 and 2021, produced a non-linear response effect on the pandemic severity. If the  pandemic severity was proportional to the population growth, the relative change in incidence between the two years could be expected to be zero (i.e., flat line in Figure  \ref{fig:sim1}, bottom row). However, comparison of the simulated pandemic scenarios between 2016 and 2021 produces a relative change in normalised incidence  which non-trivially diverges from a flat profile (Figure \ref{fig:sim1}, bottom row), especially in scenarios with less stringent policies (e.g., policies 1, 2 and 3). It is clear that the divergence is positive (i.e., a higher normalised incidence in 2021, indicating an amplified non-linear response to population growth) around the incidence peak, followed by a negative oscillation (i.e., a lower normalised incidence in 2021, suggesting a negative  response to population growth) after the peak. Figure \ref{fig:sim2} (top row) directly contrasts the pandemic profiles across  two census years.  

This non-linear response pattern (an early amplification compensated by a late negative oscillation) is manifested for the baseline scenarios across all variants of concern, as well as the scenarios with less stringent policies for the Omicron variant, as shown in Figure \ref{fig:sim1} (bottom row). 
In general, the amplification effect is more notable in the scenarios associated with more transmissible  variants (Figure \ref{fig:sim2}). Specifically, for the Omicron variant, the incidence peak for Policy 1 (baseline) simulated for 2021 census data (Figure \ref{fig:sim2} (a), red profiles, solid and dashed lines), is 12.21\% higher than its counterpart produced for 2016 census. This non-linearity arises due to the population distribution which non-uniformly affects the pandemic severity, amplifying the peak incidence (see the following subsection). 

\subsubsection{Changes in population size amplify incidence peak more than changes in density}
\label{sec:hetero_SLA}
To investigate possible causes that have contributed to the non-linear response effects observed at national level, we considered the demographic changes which occurred between the two census years at SA2 level. The higher resolution offered at this level enabled us to trace how the local areas contribute to the disproportionate response in relative incidence between the two census years, 2016 and 2021 (see Appendix \ref{sec_supp:pop} for the demographic statistics and structure captured by census data). Specifically, we examined changes in the incidence peaks across 2,147 SA2 areas which were registered in both 2016 and 2021 census years. In doing so, we used the baseline scenario implementing Policy 1 (no interventions) and correlate the peak incidence with specific demographic changes: the residential population size and the population density at SA2 level. 

\begin{figure}[ht]
    \centering
    \includegraphics[width=\textwidth]{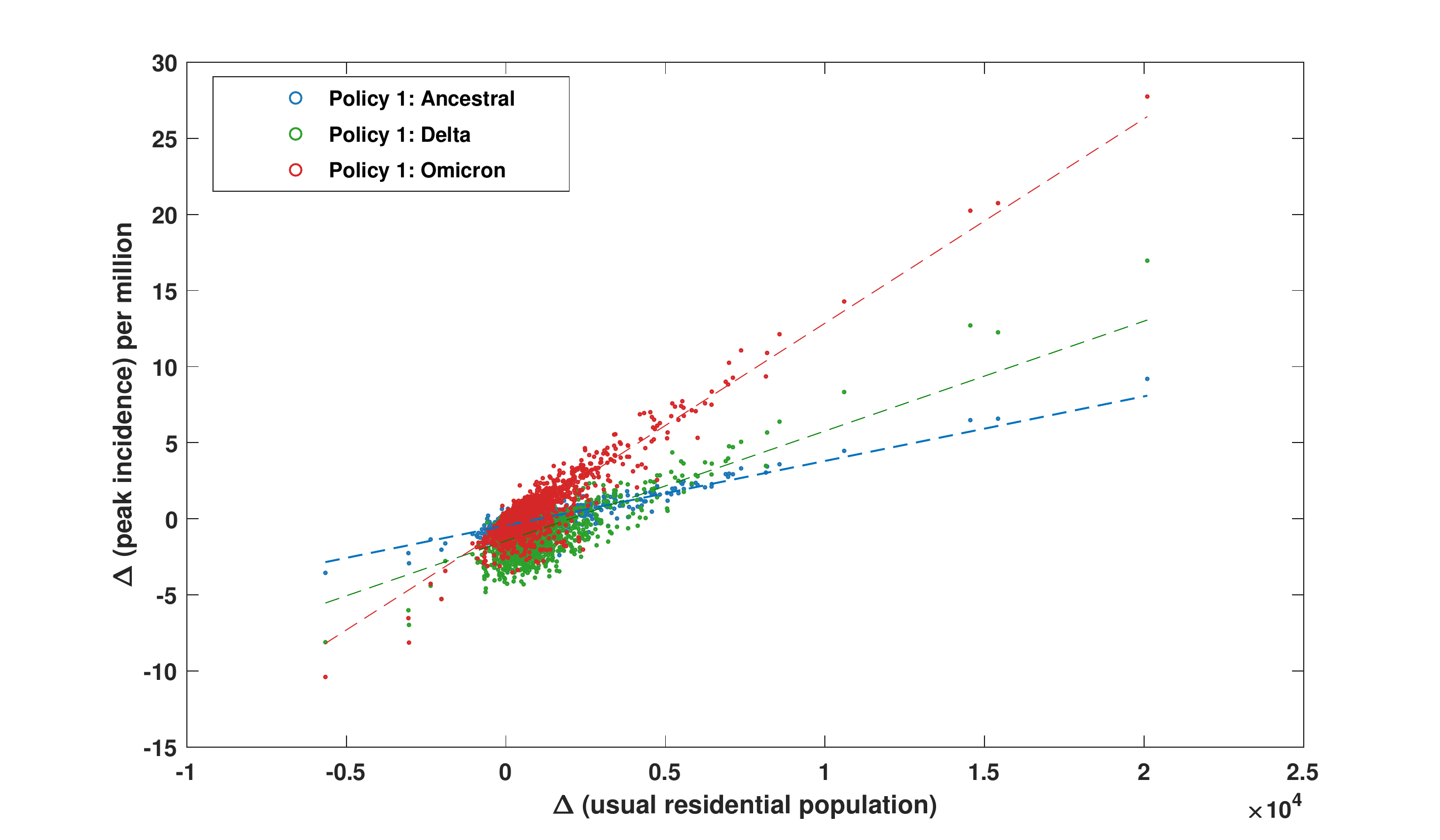}
    \caption{Positive correlation between the usual residential population difference and the peak incidence  difference between 2016 and 2021 at SA2 resolution for three considered variants: Ancestral (blue), Delta (green), and Omicron (red). Dashed lines represent linear fitting for each of the profiles (see Appendix Table \ref{tab_sup:t10} for statistical analysis). Data points corresponding to each SA2 are computed as the average over 100 runs. Total number of overlapping SA2 between 2016 and 2021 census years: 2147. Pearson correlation coefficients: $r_{Ancestral}=0.7717$, $r_{Delta}=0.6447$, $r_{Omicron}=0.9002$.}
    \label{fig:inc_pop_delta}
\end{figure}

We found that there is a strong positive correlation between the changes of peak incidence and the changes in the ``usual residential'' population. This finding is supported by high correlation coefficients across all variants ($0.64 \leq r \leq 0.91$, as shown in Figure \ref{fig:inc_pop_delta}). That is, an SA2 with a greater net population influx in the five-year period between 2016 and 2021 is highly likely to have a higher spike in the incidence peak. The impact of population increase is further amplified for highly transmissible variants, resulting in a greater peak incidence  difference shown by the steeper slope observed for the Omicron variant  (Figure \ref{fig:inc_pop_delta}, in red). In addition to the correlation with the usual residential population,  we also found a weaker positive correlation ($0.34 \leq r \leq 0.45$) between the changes in peak incidence and the changes in population density, suggesting that SA2 areas which develop a higher population density may also have a higher incidence peak across all variants (Appendix Figure \ref{fig_sup:pop_density}).

\subsubsection{Pandemic severity is distributed unequally across local (SA2) areas}
\label{sec:hetero_lorenz}
\begin{figure}[ht]
    \centering
    \includegraphics[width=\textwidth]{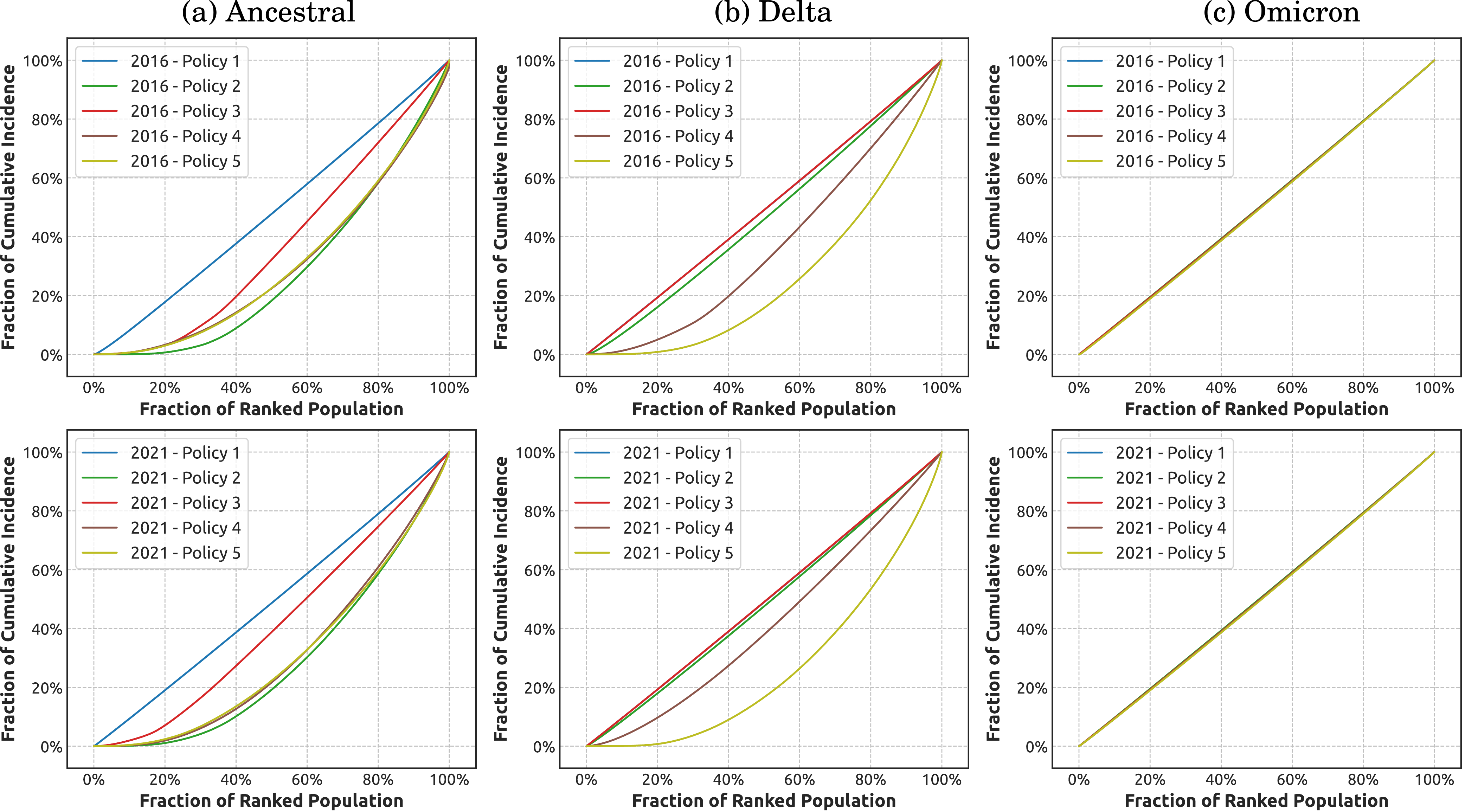}
    \caption{Pandemic Lorenz curves measuring distribution of pandemic effects across SA2 areas for considered variants, years and policies. Each column compares the impact of five intervention policies for one variant: (a) ancestral; (b) Delta; (c) Omicron. Top raw: 2016; bottom raw: 2021. Refer to Figure \ref{fig:policies_diamond_flow_chart} for a detailed description of the considered intervention policies. Each profile corresponds to one intervention policy and is computed as the average over 100 runs. }
    \label{fig:lorenz}
\end{figure}

Using Lorenz curves, we examined how the local (SA2) areas are impacted by different intervention policies across two demographic settings representing two census years. Figure \ref{fig:lorenz} (a) and (b) show that for the baseline scenario without any interventions (Policy 1) the pandemic effects are distributed equally across the areas, for all considered variants simulated for both census years (Appendix Figure \ref{fig_sup:Lorenz_policy} provides a different layout of the same results). Comparing across two census years, we note that Lorenz curves produced for 2021 census are closer to the line of equality (i.e., the diagonal line)  for the less transmissible variants (i.e., ancestral and Delta). This suggests that for these two variants, the SA2 areas contribute to the aggregate attack rate at the national level more equally in 2021 scenarios than in their 2016 counterparts.

For the ancestral and Delta variants, more stringent intervention policies led to a more diverging contribution pattern (i.e., Lorenz curve shaped further away from the line of equality). For example, for the Delta variant, using 2016 census data, we found that the increase in the fraction of cumulative incidence from 20\% to 80\% under Policy 1 (Figure \ref{fig:lorenz} (a) top row, blue profile) is attained by an equal increase (i.e., 20\% to 80\%) in the population ranked by their local SA2 attack rate. However, the same increase in the incidence fraction (i.e., 20\% to 80\%) under Policy 4 (Figure \ref{fig:lorenz} (a) top row, purple profile), was attained by a smaller set of SA2s comprising only approximately 40\% to 85\% of the population ranked by their local attack rate. This finding indicates that the pandemic severity is distributed more unequally under more stringent interventions (including NPIs and vaccinations). Specifically, SA2 areas with a higher local attack rate 
%(i.e., towards the right-hand side of the ranked population on x-axis) 
(i.e., ranked higher on x-axis) 
account for a higher fraction of the cumulative incidence.

We also note that the unequal distribution of  pandemic severity diminishes for the highly transmissible Omicron variant (shown in Figure \ref{fig:lorenz} (c)) where all simulated intervention policies failed to adequately slow down the transmission. This resulted in all SA2 areas contributing equally to the national-level cumulative incidence, with all Lorenz curves overlapping with the line of equality.

\subsection{Effects of urbanisation on pandemic spread}
\label{sec:urbanisation}
\begin{figure}[ht]
    \centering
    \includegraphics[width=\textwidth]{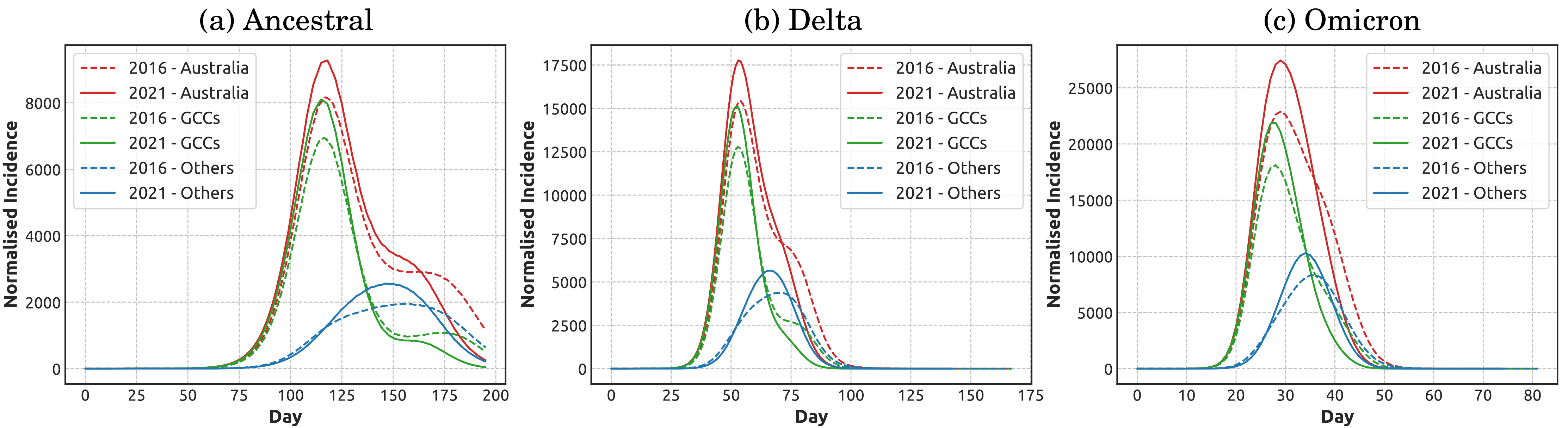}
    \caption{Comparison of pandemic waves in  Greater Capital Cities (GCCs) and all other areas. 
    % (ii) bottom row: SA2 areas identified as GCCs together with SA2 areas around the international airports not listed as GCCs versus other areas. 
    Each column compares the baseline scenario (Policy 1) for 2016 and 2021 census data, for a variant of concern: (a) ancestral; (b) Delta; (c) Omicron. Each profile is computed as the average over 100 runs.}
    \label{fig:urbanisation}
\end{figure}

The bimodal pandemic dynamics is characterised by the incidence peak (attributable to  predominantly urban spread), followed by an inflexion point which shapes around a smaller secondary wave (attributable to mostly regional and rural areas)~\cite{zachreson_urbanization_2018}.
Typically, the first wave triggered by the international air traffic is rapidly shaped in  urban populations concentrated near international airports. 
In contrast, the pandemic progression into non-urban regions (i.e., areas further away from international airports) is significantly slower. The confluence of these factors resulting in  bimodal dynamics in Australia was detected and described in context of simulating the pandemic influenza using 2011 and 2016 census data \cite{zachreson_urbanization_2018}. Our simulations of the COVID-19 pandemic scenarios in this study also produced  bimodal dynamics, especially for the ancestral and Delta variants using 2016 census data (Figure \ref{fig:urbanisation}). However, the bimodality is less prominent in scenarios using 2021 census data, indicating a shift of the pandemic dynamics  between urban and regional regions. 

This observation is also supported by the pandemic dynamics examined at the SA2 level. Appendix Figure \ref{fig_sup:SA2_peak} shows that in comparison to the simulated results for 2016 census, the time gap between the first and second pandemic waves (defined in terms of the number of SA2 areas that peaked on a given day) for 2021 census has become shorter at the SA2 level. These findings indicate that there are more intricate structural demographic changes in the population which occurred between 2016 and 2021, beyond a uniform population growth.

To explain this phenomenon, we assessed the pandemic progression during the baseline scenario (Policy 1), by tracing pandemic waves in urban and non-urban SA2 areas (see Appendix Table \ref{tab_sup:GCC_others} for the population statistics). Figure \ref{fig:urbanisation} compares the Greater Capital Cities (GCCs) against other areas (i.e., all remaining SA2s). We note that the two distinct pandemic progression profiles (urban and non-urban) are mostly separated by the initial conditions: the national incidence peak is largely attributed to urban SA2 areas while the inflexion point observable at a later stage of the pandemic is caused by a secondary wave emerging in non-urban SA2 areas. 

%\subsubsection{Greater Capital Cities vs other areas}
%\label{sec:GCCs}
For the baseline scenarios traced for two census years, our results showed that the bimodal dynamics is diminishing from 2016 to 2021 data. This can be explained by two factors: (i) a higher incidence peak occurring in both GCCs and other urban areas, and (ii) an accelerated pandemic progression into non-urban areas (or at least, into areas outside of GCCs). This effect was observed for scenarios across all variants of concern, with more transmissible variants showing a greater increase in the incidence peak. This observation suggests a faster pandemic spread for 2021 scenarios, indicating that the urbanisation increased over the five-year period, captured by the census statistics (see Appendix Section \ref{sec_supp:pop}), reducing the bimodal dynamics. 

We also observed that the bimodality is weakened for the Omicron variant. This is explained by a reduced  difference in the peak incidence timing between GCCs and non-GCCs. In other words, the two waves, urban (primary) and non-urban (secondary), tend to merge into a primary significant wave with a single incidence peak. This can be attributed to the high transmissibility of the Omicron variant  which suppresses the impact of population heterogeneity. When the  transmission was adequately slowed down  (due to interventions, for example, implementing Policy 4 and Policy 5), the two pandemic waves became more separable, leading to notable bimodality (Appendix Figure \ref{fig_sup:GCC_inc}).

It is also worth noting that we do not consider re-infections in these scenarios. In other words, a higher incidence peak occurring earlier in the simulation corresponds to a reduction of the susceptible population, thus exhausting the susceptible population sooner and consequently weakening bimodality. 

%\subsubsection{Greater Capital Cities and cities with international airports vs other areas}
%There are several non-GCC cities in Australia with international airports (see Figure 4 in Supplementary Material, shown in green). The inclusion of the SA2 areas surrounding these cities into analysis of the urban areas, does not affect the observations reported in Section \ref{sec:GCCs}. Specifically, scenarios for 2021 census data produce higher incidence peaks in both urban and rural areas, with the latter areas showing an accelerated pandemic progression profile, relative to the 2016 profile. This finding confirms that 

\subsection{Effects of school closures across variants of concern}
\label{sec:school}
\begin{figure}[ht]
    \centering
    \includegraphics[width=\textwidth,trim={0.5cm 0.3cm 0.5cm 0.3cm},clip]{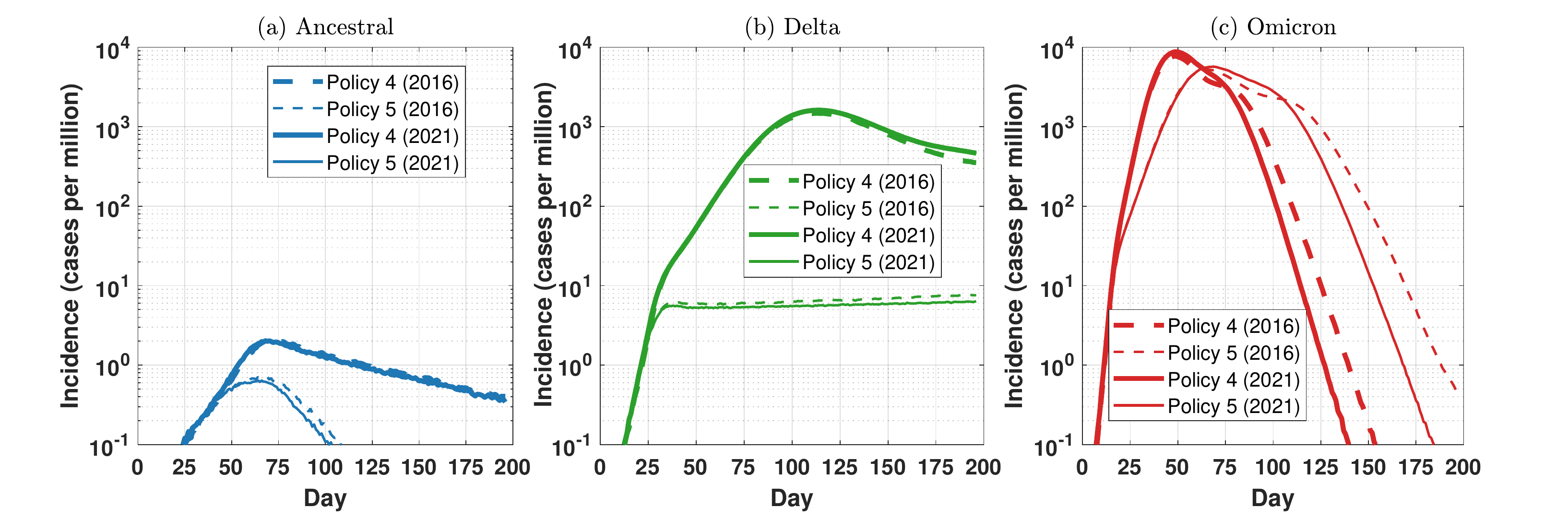}
    \caption{Effects of school closures combined with NPIs for three considered variants (log scale):  (a) ancestral; (b) Delta; (c) Omicron. School closures effectively control the spread of Delta variant, producing a sharp difference in incidence. Such a bifurcation is not observed for the ancestral and Omicron variants. Each profile corresponds to one intervention policy and is computed as the average over 100 runs. Appendix Figure \ref{fig_sup:school_linear} shows these plots on linear scale.}
    \label{fig:school_closure}
\end{figure}

Finally, we examined the effects of school closures across three variants of concern by comparing pandemic scenarios between Policy 4 (Mixed intervention) and Policy 5 (Partial lockdown).
Our results suggested that the effectiveness of school closures varies significantly for different variants. The effects of school closures were most prominent for the Delta variant with a two-order reduction in peak incidence  (from over 1,000 cases per million to under 10 cases per million), resulting in a sharp difference shown in Figure \ref{fig:school_closure} (b). Such a bifurcation was observed only in scenarios for the Delta variant and was not detected for variants with either lower $R_0$ (i.e., ancestral variant, Figure \ref{fig:school_closure} (a)) or a higher $R_0$ (i.e., the Omicron variant, Figure \ref{fig:school_closure} (c)). 

For the ancestral variant, Policy 4 was sufficiently effective in containing the spread with new cases kept at a very low level. Although school closures could further reduce the peak incidence, the reduction (from around 2 cases per million to around 0.6 cases per million) would be marginal. This could be observed for both 2016 and 2021 census years (Figure \ref{fig:school_closure} (a)). 

For the Omicron variant, school closures delay the incidence peak by approximately 25 days with a sizable reduction from nearly 9,000 cases per million to under 6,000 cases per million (Figure \ref{fig:school_closure} (c)). This reduction, however, would still be insufficient to curb the spread. This finding suggests that school closures could slow down the spread of the Omicron variant to some extent but would be inadequate for suppressing incidence, due to the extremely high $R_0$ of the Omicron variant. 

Overall, these observations suggest that when coupled with NPIs and partial vaccination, school closures can be a highly effective policy for pandemic suppression of variants comparable in transmissibility with the Delta variant.
These benefits, however, may not eventuate for either significantly less or significantly more transmissible variants of concern.  

\section{Discussion}
\label{sec:discussion}
In this study, we carried out a systematic comparison of pandemic scenarios across census years, variants of concern and intervention policies. The considered scenarios combined five different intervention policies with three dominant variants impacting Australia between 2020 and 2023 (ancestral, Delta, and Omicron), using a surrogate population generated based on 2016 and 2021 Australia census data. Our simulation results revealed that the population growth and heterogeneity non-linearly affect pandemic dynamics, amplifying the peak incidence for all variants of concern. These non-linear response effects highlight a complex interplay between the demographic and pandemic characteristics, which may amplify pandemic impacts on continuously growing  populations worldwide. 

Firstly, we focused on the Australian population which grew by 8.6\% between 2016 and 2021 (net gain of nearly 2 million people). However, the population growth has been distributed non-uniformly across the country, affecting both local residential population size and population density. Such unevenly distributed demographic changes directly affect pandemic progression, and our results demonstrated that the estimation of pandemic severity cannot be accomplished by a simple scaling of the outcomes obtained for previous datasets, such as 2016 census data. This strongly suggests that pandemic models and public health policies need to be frequently reviewed and adapted to changes in the population heterogeneity, especially when dealing with long-lasting pandemics similar to COVID-19.

We showed that the effectiveness of intervention policies differed across variants of concern (Figures~\ref{fig:sim1} and~\ref{fig:sim2}). Coupled with the observed amplification effects of demographic changes (Figure~\ref{fig:inc_pop_delta}), these findings call for specific interventions aimed to reduce the amplification effect, e.g., prioritising (1) the SA2 areas with a higher population growth, and (2) the SA2 areas with a higher density increase. 

At the same time, some policy-related findings were independent of the considered variants or demographics. In particular, a partial preemptive vaccination rollout with 50\% coverage and limited vaccine efficacy (Policy 3) can be effective only in a combination with partial NPIs, comprising a mixed intervention (Policy 4). Applied on its own, Policy 3 was unable to control the pandemic spread across all considered scenarios, as shown in Figure~\ref{fig:sim1}. Thus, the ongoing and future immunity boosting vaccination rollouts need to aim at high population coverage comparable with  original mass-vaccination campaigns, while still being complemented by  partial NPIs. Similarly, partial NPIs (Policy 2) cannot provide a principled solution on its own. While managing to control pandemic scenarios for the less transmissible ancestral variant, partial NPIs did not succeed in preventing sizable incidence peaks in scenarios for the Delta and Omicron variants, but only delayed these peaks (Figure~\ref{fig:sim1}). In general, such  delays may be useful in rolling out booster vaccinations, reinforcing the point that a mixed intervention which combines partial NPIs and partial vaccination (such as Policy 4) may provide an adequate intervention.  

Our study has also highlighted the ``pandemic inequality'', with certain SA2 areas contributing to the nationwide cumulative incidence stronger than others.
We found that the pandemic inequality reduces when the population or the disease transmissibility grows (Figure~\ref{fig:lorenz}). However, the opposite tendency --- increasing pandemic inequality --- was observed with more stringent interventions (Appendix Figure \ref{fig_sup:Lorenz_policy}) or in non-urban areas (Appendix Figure \ref{fig_sup:Lorenz_urban}). 
While this inequality was lesser in scenarios for the  highly transmissible  Omicron variant, the findings still suggest that a resource prioritisation scheme is needed, with the interventions targeting the localities and regions which have been experiencing a high population growth and/or developing a high density. 

Secondly, we quantified the pandemic effects of urbanisation, distinguishing between the first wave affecting  Greater Capital Cities and a delayed second wave developing in regional and rural areas (Figure~\ref{fig:urbanisation}). This bimodality indicates that, to be effective, the intervention efforts need to adapt during a pandemic progression, with the initial focus on metropolitan centres followed by a shift to non-urban areas, in anticipation of the corresponding peaks. Such geo-spatial redistribution of healthcare and vaccination resources needs to account for the transmissibility of  dominant variants of concern, as more transmissible variants accelerate the pandemic, bring the urban and non-urban waves closer in time, and shorten the time between their peaks. 

Finally, we evaluated the role of school closures in suppressing the pandemic transmission caused by different variants of concern. Crucially, we demonstrated that the effects of school closures are highly dependent on the dominant variant, with the more decisive effect observed only for the Delta variant (given the range of other policy-defining parameters, i.e., the vaccination coverage of 50\% and the NPI compliance with 70\% SD), as illustrated in Figure~\ref{fig:school_closure}. This also reinforces the suggestions that policy makers should not assume that interventions will have the same effect across different variants. In addition, this highlights the possibility that some interventions can compensate others in specific circumstances: for example, when the preemptive vaccination or booster uptake is lower or slower than anticipated, school closures and stricter NPIs may be required to compensate for the lack of immunity. 

In summary, the study highlighted the need for geo-spatially and demographically tailored, proactive and agile interventions, in contrast to general-purpose, reactive and rigid policies.

\section{Limitations and future work}

This study of pandemic severity did not include considerations of
(a) socio-economic factors, and (b) disease burden in terms of hospitalisations, ICU occupancy and mortality. As demonstrated in our previous studies~\cite{nguyen_general_2022,chang_simulating_2022,chang_persistence_2023}, these components can be included within an ABM study but substantially increase its scope.  

Our simulations ran over a period of 196 days, without considering re-infections. 
%In other words, once recovered, the agents are excluded from susceptible population. 
Given the considered simulation horizon, this limitation has a minor effect discussed in our study of recurrent waves~\cite{chang_persistence_2023}. 

Our focus on the three dominant variants of concern rather than on their numerous sub-lineages (which may co-circulate) allowed us to distill some of the salient public health lessons. We believe these lessons would remain relevant across other sub-variants, including co-circulating ones.

Finally, we did not model the differences in vaccine efficacy across variants of concern (including ancestral, Delta, and Omicron variants). This should not impact the outcomes over the considered simulation horizon. Nevertheless, our model will be extended in near future, addressing these limitations (reinfections and multiple co-circulating variants with different immunity profiles).

\section{Conclusion}
In pursuing our objectives, we solved several methodological challenges, extending the range of applicability for agent-based pandemic modelling. Firstly, we incorporated the ABS census data for 2021, thus accounting for the most recent demographic information for Australia, in terms of the population structure, age distribution, household composition, and commuting flow patterns. This ``upgrade'' is important because previous similar studies used the data from the Australian census of 2016, scaling the modelling outcomes by approximately 10\% to account for the larger population. Our results showed that the demographic changes  over the five-year period contribute to the pandemic outcomes in more subtle non-linear ways that often cannot be captured by a uniform scaling. To study these nuanced contributions we employed Lorenz curves characterising  an unequal distribution of pandemic effects. 

Secondly, we addressed a well-known inconsistency between low-resolution and aggregated high-resolution census data brought about in 2016 by the ABS anonymity policy compliance system~\cite{ABS_data_confidentiality_guide,fair_creating_2019}. 
In order to reduce this mismatch, 
%which is still present in the 2021 census data, we improved the re-sampling technique proposed earlier~\cite{fair_creating_2019}, and 
we reconstructed a surrogate high-resolution 2021 commuter topology (Section \ref{sec_supp:TTW} of Supplementary Material). 
This allowed us to examine nuanced effects of the pandemic scenarios on urban and regional areas, and measure pandemic bimodality.

Overall, the extended ABM, coupled with the reconstruction techniques, offered a versatile approach to model comparative scenarios with multiple variants of concern, simulated across different demographic settings (census years) and for distinct intervention policies. A combination of census-based ABM and pandemic Lorenz curves provided a unique high resolution method to not only simulate different pandemic scenarios across varying demographics and variants, but also evaluate the unequally distributed effects of feasible intervention policies. This, in particular, allowed us to emphasise the divergent role of school closures as a complementary NPI -- with respect to the disease transmissibility, exemplifying a bifurcation in the effectiveness of school closures.

In summary, the presented results illustrate how comparative analysis measuring distribution of the pandemic severity across different dimensions can help in improving public health preparedness and response to future pandemics. In particular, the study highlights that rigorous pandemic modelling can provide insights into the impact of complex demographic factors on the spread of infectious diseases over medium- to long-term.

\section*{Acknowledgements}
 This work was supported by the Australian Research Council grant DP220101688 (MP, SLC and CMJ) and the University of Sydney's Digital Science Initiative (DSI) Research Pilot Project funding scheme (SLC, MP and QDN). Additionally, SLC is supported by the University of Sydney Infectious Diseases Institute Seed Grant 220182. CMJ is supported by the University of Sydney Faculty of Engineering Research Stipend Scholarship (ERSS). We are grateful to Oliver Cliff, Cameron Zachreson and Nathan Harding for contributing to earlier versions of the open source code of AMTraC-19. We also thank Vitali Sintchenko and Tania Sorrell for their insights on epidemiological and public health aspects of this study, as well as Tim C. Germann and Sara Del Valle for helpful discussions regarding pandemic agent-based modelling. We thankfully acknowledge the use of high-performance computing cluster, Artemis, provided by the Sydney Informatics Hub at the University of Sydney.

\section*{Author contributions}
MP, QDN, and SLC designed the computational experiments. QDN implemented and tested the software code and supporting algorithms. QDN and CMJ performed data curation and validation. QDN and SLC carried out the computational experiments, processed simulation data, and prepared figures. MP supervised the study.  All authors drafted the original article, contributed to editing of the article, and approved the final manuscript.

\section*{Data sharing}
We used anonymised data from the 2016 and 2021 Australian Census obtained from the Australian Bureau of Statistics (ABS), the Australian Curriculum and Assessment and Reporting Authority (ACARA), and the Bureau of Infrastructure and Transport Research Economics (BITRE). These datasets are accessible publicly, except the travel-to-work data and household-composition data which can be obtained from the ABS on request. 
Simulation  and post-processing data are available at Zenodo \cite{AMTrac_dataset}. The source code of AMTraC-19 is also available at Zenodo \cite{chang_amtract_user_guide_2022_7325675}.

% ----------------------------------------------------------
% ----------------  SUPPLEMENTARY MATERIAL  ----------------
% ----------------------------------------------------------

\begin{appendices}
\counterwithin{figure}{section}
\counterwithin{table}{section}

\section{Population data}
\label{sec_supp:pop}
The population data used in our model is drawn from 2016 and 2021 Australian Census data published by the Australian Bureau of Statistics (ABS). Australian Census comprises a large number of hierarchical data fields categorised by geographical, demographic, and socio-economic parameters~\cite{abs_general_webpage}. In our model, we use the demographic fields in Australian census (extracted utilising ABS TableBuilder \cite{ABS_census}) to generate an artificial population as a ``digital twin'' of the Australian population, aiming to capture its main characteristics with high fidelity. Here, we  describe the demographic census data structures and examine salient features of the Australian population mobility and the international air traffic.    

\subsection{Census data structure}
\begin{figure}[ht]
    \centering
    \includegraphics[width=\textwidth]{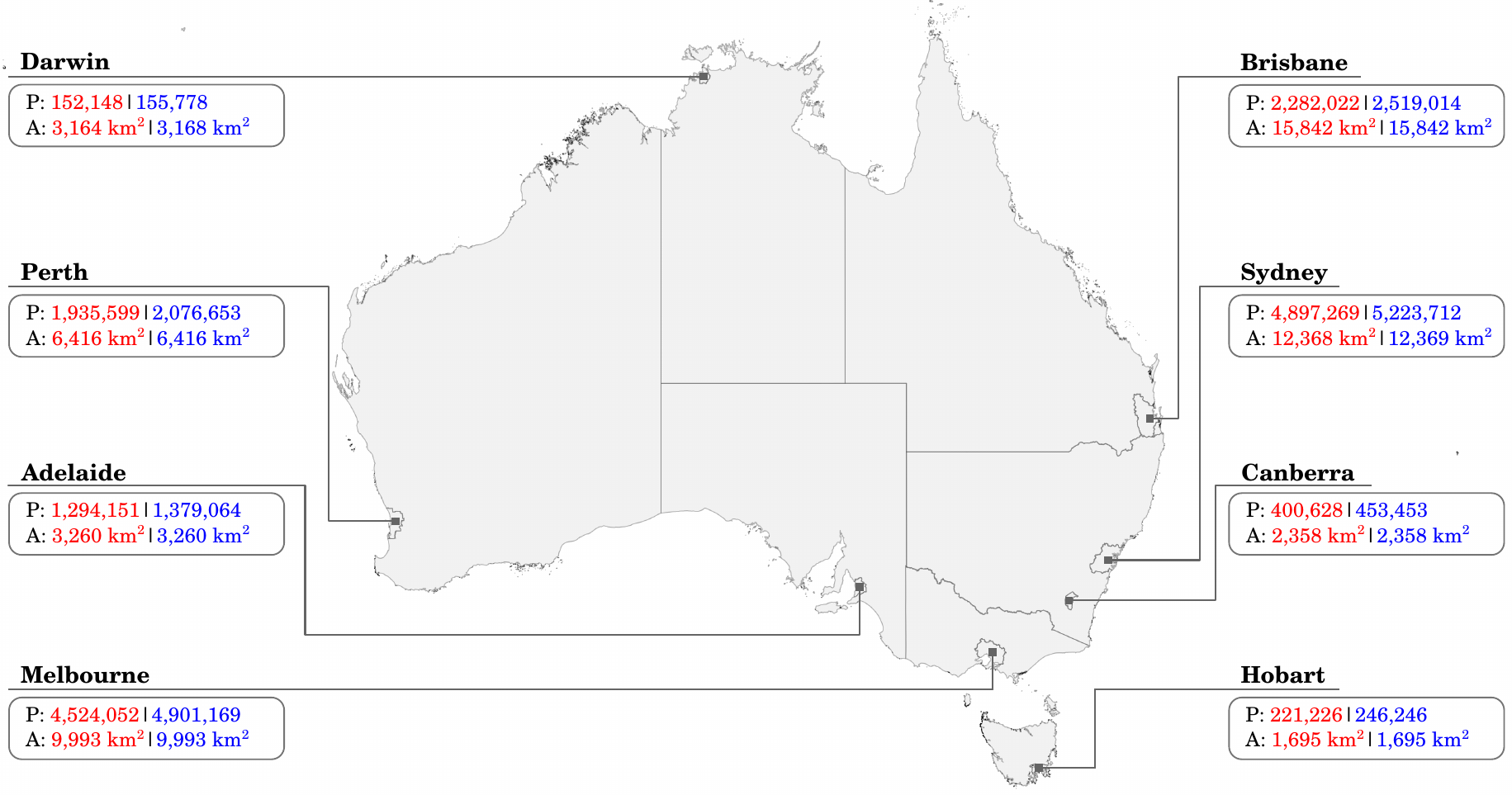}
    \caption{Snapshot of the Australian population: locations and geographical areas. Capital city of each state is annotated with their usual residential population (P) and geographical area (A) in 2016 (in red) and in 2021 (in blue) respectively. The state names are abbreviated as follows: Western Australia (WA), Northern Territory (NT), South Australia (SA), Queensland (QLD), New South Wales (NSW), Victoria (VIC) and Tasmania (TAS). Note that changes in the size of geographical areas are due to boundary changes.}
    \label{fig_sup:capital_cities_pop_area}
\end{figure}

The 2021 census reports that the Australian population reached 25.4 million, following an 8.6\% increase  (2,020,896 million people) since 2016 census counts. There are eight states and territories in Australia, each with a capital city. Figure \ref{fig_sup:capital_cities_pop_area} shows the geographical representation of the states and their corresponding capital cities which comprise about two thirds of Australian population. 

%Specifically, Australian census documents demographic characteristics including the residential population composition (e.g., household size, residential population size, etc) and the commuting patterns between the places of residence and places of work or study. These characteristics are reported by partitioning the population into: (i) place of residence (i.e., usual-residence, UR) at multiple Statistical Area Levels (SAs), each representing a different resolution (i.e., SA1 to SA4 with ascending resolution); and (ii) by destination zone where the individuals commute to for work- or study-related activities depending on their age group (i.e., place-of-work, POW). See Section 1 in Supplementary Material for more details.

Each state is  further divided into different levels of statistical areas (without gaps or overlaps) following a nested hierarchical framework of geographic levels \cite{abs_general_webpage}. Each succeeding level is a breakdown of the previous level into smaller areas defined as follows: States and Territories (ST/T), Statistical Area Level 4 (SA4), Statistical Area Level 3 (SA3), Statistical Area Level 2 (SA2), Statistical Area Level 1 (SA1), and Mesh Blocks (MB). Table \ref{tab_sup:SA} summarises the number of statistical areas at each level and the corresponding usual residential population range. Note that 2021 census reported more statistical areas across all levels due to changes in the partitioning approach. Due to merging and splitting of areas, the new partitioning approach  resulted in a net increase of 7.5\% for the number of SA1 areas and 7\% increase for the number of SA2 areas.

\begin{table}[ht]
\centering
    \begin{tabular}{c|c|c|c}
    \hline
         &  2016 census & 2021 census & population range (approx.) \\
         \hline \hline
        MB & 358,122 & 368,286 & N.A  \\
        SA1 & 57,523 & 61,845 &200 - 800 \\
        SA2 & 2,310 & 2,473 & 3,000 - 25,000 \\
        SA3 & 358 & 359 & 30,000 - 130,000 \\
        SA4 & 107 & 108 & 100,000 - 500,000 \\
        \hline
    \end{tabular}
    \caption{The hierarchical levels in Australian census structure. The population for each level refers to the usual residential population, excluding visitors.}
    \label{tab_sup:SA}
\end{table}

In addition to the residential population, a wide range of other demographic features are also reported for each level. The following features are incorporated into the artificial population generation: age, gender, household composition, and commuting patterns. See Section \ref{sec_supp:pop_gene} for more details regarding how these features are incorporated in our model.

In order to reduce the risk of individual- or household-level identification, all data reported in the census are anonymised via several perturbation methods including, but not limited to, the removal of aggregated microdata falling under a set threshold. It is important to note that the ABS data perturbation policies may differ across datasets and may change between census years.

\subsection{Mobility: travel to work}

The short-distance commuting patterns are reported via the travel-to-work (TTW) dataset, detailing the accumulated number of employed individuals commuting between their place of usual residence (UR) and a place of work (POW). We use this dataset to construct a commuter mobility network (detailed in Section \ref{sec_supp:TTW}). Commuter counts are accumulated on several statistical levels:
\begin{itemize}
    \item UR: reported at SA1, SA2, SA3, SA4 level
    \item POW: reported at DZN, SA2, SA3, SA4 level
    \item number of employed commuters (either full-time or part-time)
\end{itemize}

Another level (DZN, destination zone) is used to describe the POW population at a resolution comparable to SA1 but following a different set of partition rules. Both SA1 and DZN accumulate to attain exact partitions at SA2 level.

The perturbation protocol that the ABS adopted for privacy protection removes some edges between UR and POW (that is, UR $\leftrightarrow$ POW edges). This removal is done if the commuter counts are below a certain threshold. In our previous work~\cite{fair_creating_2019}, we showed that the edge removal causes large discrepancies in the edge counts between different statistical area levels. When generating the artificial population, we address this problem by minimising the discrepancies between different SA levels, improving consistency of the population representation across the levels (see Section \ref{sec_supp:TTW_improve}). 

\subsection{Urban structure and international air traffic}
Australia is a highly urbanised society with heterogeneously distributed population as annotated, see Figure \ref{fig_sup:capital_cities_pop_area}. The eight capital cities, despite their rather small geographical area size, account for more than two thirds of the Australian residential population, while the remaining land (mostly regional areas) comprises only one third of the population.  

The capital cities and their surrounding areas are defined by the ABS as the Greater Capital Cities (GCC) including the urban areas (city centres) and the areas in some proximity, to account for the residents who work, shop, and socialise within the city. We examine the urban structure, particularly GCCs, for two reasons.

\begin{figure}[ht]
    \centering
    \includegraphics[width=0.9\textwidth]{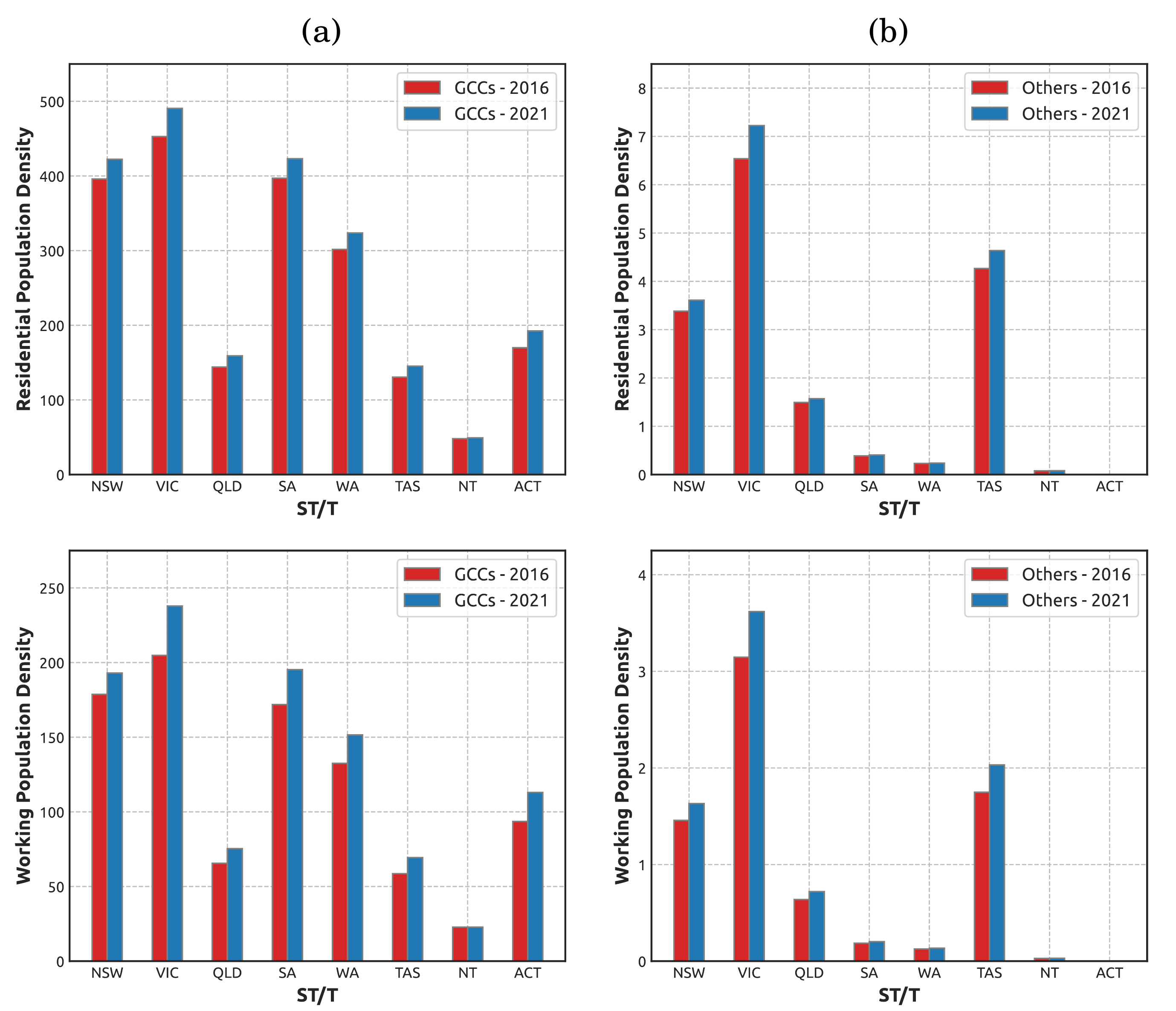}
    \caption{Comparison of residential population density and working population density in different states and territories (ST/T) of Australia partitioned by GCCs (a), and other areas (b). Note that the population density scale in (b) is more than 50 times lower than that in (a). The Australian Capital Territory (ACT)  predominantly comprises urban areas and thus  does not show a comparable density for non-GCC areas.}
    \label{fig_sup:urban_density_graphs}
\end{figure}

\begin{figure}[ht]
    \centering
    \includegraphics[width=\textwidth]{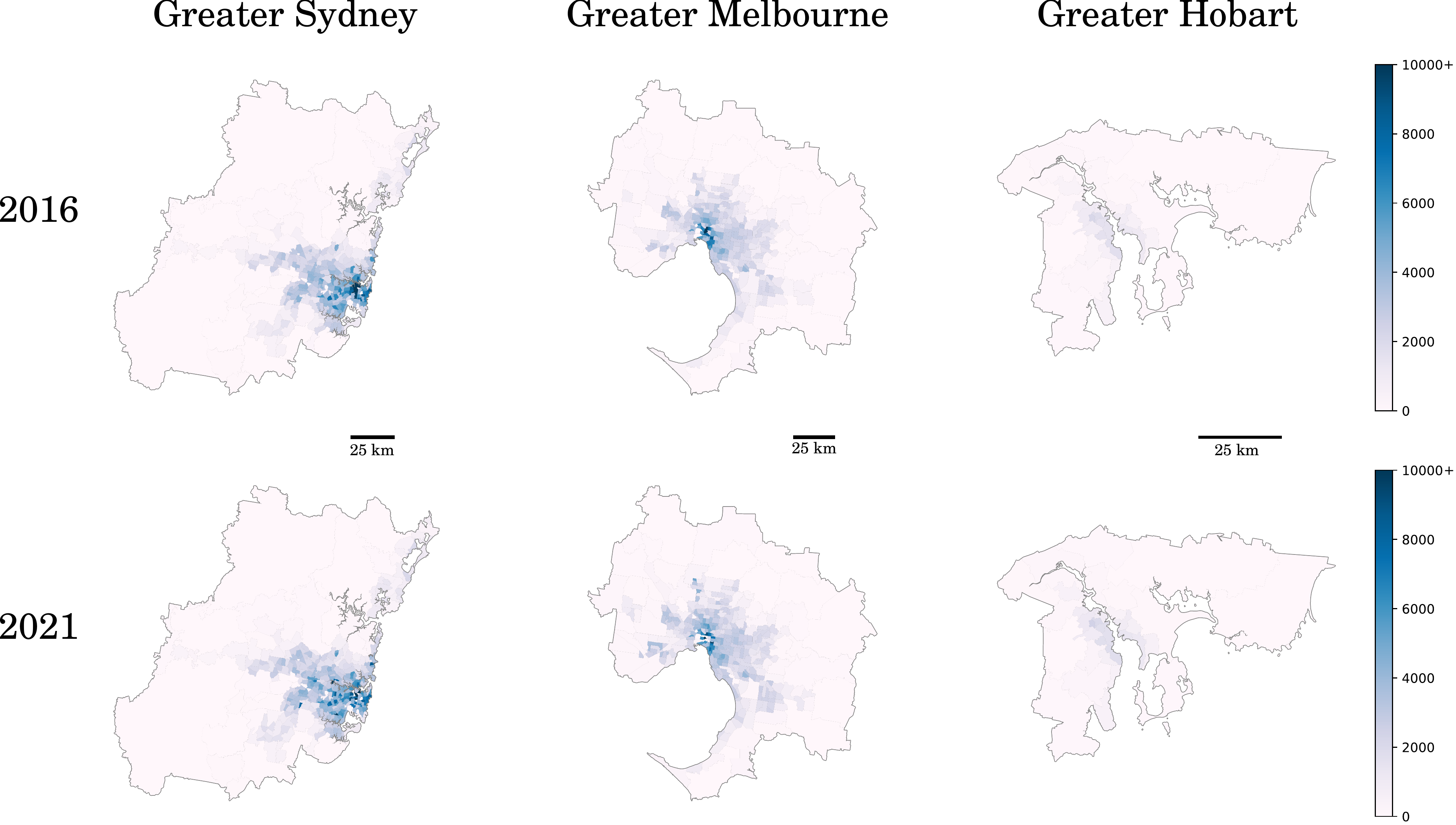}
    \caption{High variation of the population density across three Greater Capital Cities of Australia: Greater Sydney (left), Greater Melbourne (middle), and Greater Hobart (right). Population-dense areas are highlighted in darker colours, with intensity defined by the colour bar on the right. Note that boundaries in these figures are shown at SA2 resolution.}
    \label{fig_sup:pop_density_maps}
\end{figure}

 Firstly, the population densities of GCCs are considerably higher compared to other areas of Australia, allowing for a greater social interaction from both residents and working commuters. This increased likelihood of mixing aggravates the spread of infectious diseases \cite{dye_health_2008, eubank_modelling_2004, yashima_epidemic_2014}. In addition, both residential and working population densities continued to increase between 2016 and 2021 for all GCCs despite the brief disruption of migration and movement during the COVID-19 pandemic, as shown in Figure \ref{fig_sup:urban_density_graphs}. Furthermore, the urbanisation level in GCCs  greatly varies  across the country. Figure \ref{fig_sup:urban_density_graphs} (a) and Figure \ref{fig_sup:pop_density_maps} illustrate a stark contrast between the population-dense GCCs (Sydney, NSW GCC; and Melbourne, VIC GCC) and cities with noticeably lower population density, e.g., Hobart (TAS GCC). This large variability also suggests a high heterogeneity within the GCC population.

\begin{figure}[ht]
    \centering
    \includegraphics[scale=0.095]{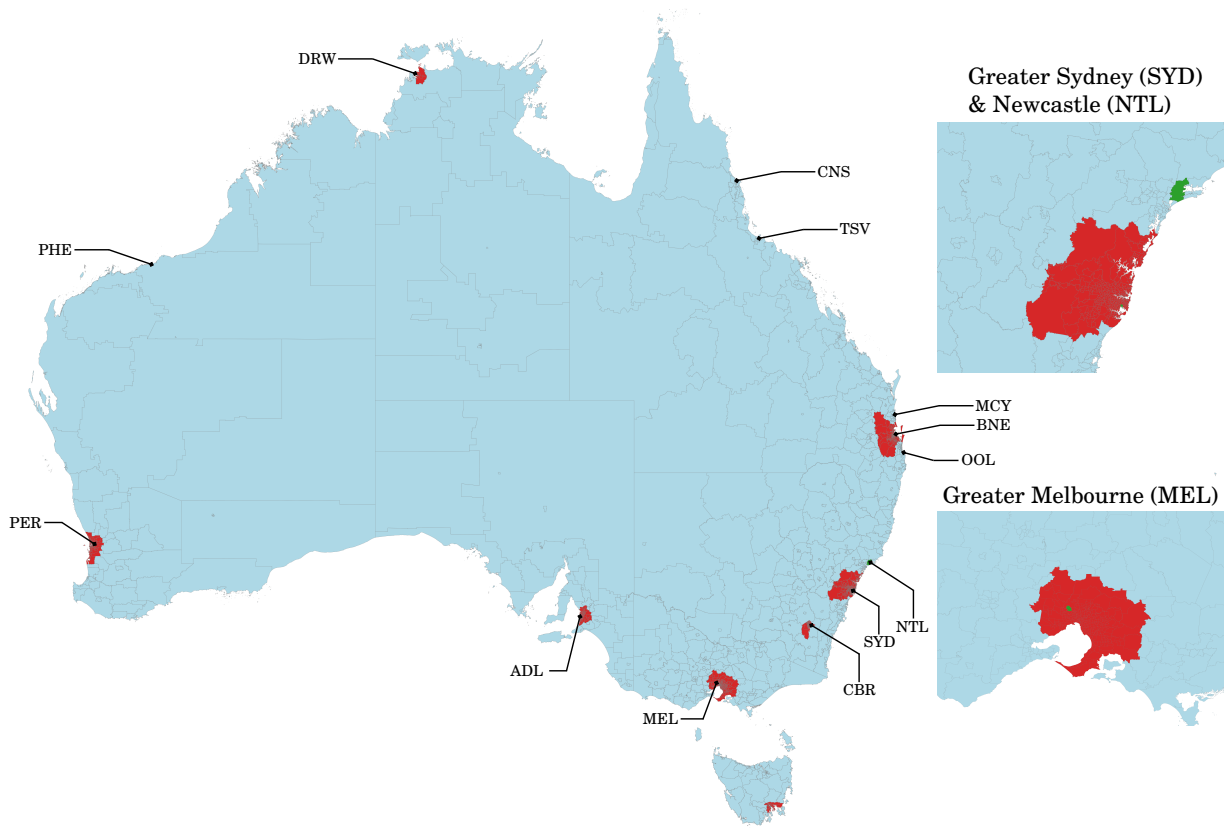}
    \caption{Geographic representation of Greater Capital Cities (in \textit{red}), international airports (having international air traffic in 2019, in \textit{green}), and other regions (in \textit{blue}) across Australia. The boundaries are delineated based on the 2021 SA2-resolution map of Australia. For visual clarity, some Australian islands are not shown. Hobart is not annotated as there were no incoming international passengers to Hobart during the considered period (i.e., in 2019). The 3-letter code names for the annotated international airports are described in Table \ref{tab:int_airport_estimation}.}
    \label{fig_sup:map_with_airports}
\end{figure}

 Secondly, the local and international air traffic is concentrated in GCCs, increasing the potential of disease introduction by long-range travel (i.e., international sources)  \cite{brueckner_airline_2003}, as well as its subsequent spread. Due to Australia's geographic isolation, the international airports located in GCCs take the vast majority of international passenger inflow. We note that airports of several non-GCC cities, such as Newcastle (NTL) and Sunshine Coast (MCY), may also take a relatively small number of international incoming passengers, as shown in Figure \ref{fig_sup:map_with_airports}. The air traffic data is reported by the Australian Bureau of Infrastructure, Transport, and Regional Economics (BITRE) \cite{BITRE_airport_data}, and we use this data to scale the number of initial infections in proportion to the international air traffic inflow, following earlier studies~\cite{cliff_investigating_2018,chang_modelling_2020}. The actual traffic data in 2021 were severely impacted due to travel restrictions imposed during the COVID pandemic. To account for this impact, in this study we use the BITRE air traffic data between 2003 and 2019 (inclusive) in estimating the projected air traffic volume in 2021. See Section \ref{sec_supp:air_travel} below for more details.  

\clearpage
\section{Generation of artificial population}
\label{sec_supp:pop_gene}
This section describes the construction of a surrogate Australian population using high-resolution datasets, including the ABS demographic fields from the Australian census, international air traffic reports from the BITRE, and educational registration records from the Australian Curriculum, and Assessment and Reporting Authority (ACARA). These datasets provide high fidelity details encompassing age, gender, household composition, student enrollment, workforce mobility, and international travel. Using these datasets, we  generate a representative artificial population  capturing  unique demographic characteristics and mobility characterising the Australian population. 

%Each agent in AMTraC-19 is generated with a set of demographic attributes (e.g., age, gender, etc.) and social contexts (e.g., households, household clusters, neighbourhoods, classrooms, workplaces, etc.). This artificially generated surrogate population matches the population characteristics reported in Australian census and the ACARA. 

This section is structured as follows. Section \ref{sec_supp:agent_gene} explains the generation of individual agents and their matching demographic characteristics. Section \ref{sec_supp:TTW} details our approach to re-constructing the commuting network which included travel to school and travel to work depending on agents' age groups. Section \ref{sec_supp:air_travel} elaborates on the estimation of international air traffic in 2021 using traffic records prior to the COVID-19 pandemic. The user guide of our open-source software provides more technical details~\cite{chang_amtract_user_guide_2022_7325675}.

\subsection{Agent-based generation}
\label{sec_supp:agent_gene}
In this section, we describe agent generation procedure. Initially, we generate households and then assign each virtual agent with a household determined by the household composition in their residential SA1 areas. This is followed by assignment of demographic characteristics, including age and gender. The distribution of these agents in each SA1 matches the number reported  in census reports. We then generate agents across all SA1 areas to construct an Australia-wide population. In our population generation process  we excluded all non-geographic regions (e.g., "Migratory/offshore/shipping" and "No usual address") as these regions have no residential population with specific addresses.

\subsubsection{Household composition.}

Each agent is initially assigned to a household. The distribution of household size in each SA1 area matches the census data on household composition (i.e., the count of families in terms of the number of dependent children and/or non-family members, excluding overseas visitors), corresponding to the following five categories:
\begin{itemize}
    \item single adults, 
    \item couples without children (i.e., two adults), 
    \item couples with  one to four children,
    \item single-parent families with one adult and  one to five children, 
    \item non-family groups with two to six adults
\end{itemize}
 A cumulative distribution function (CDF)  of household composition is built for each SA1 area.  

\subsubsection{SA1 population, age, gender, and residency.} 

Within the household, attributes of each agent  (age, gender, and residency) are assigned to match the census records capturing the demographic diversity at the SA1 resolution. Specifically, for each SA1, the ABS summarises the number of individuals residing in this region, and the number of male/female residents categorised by five age groups as follows:

\begin{itemize}
     \item early childhood (0-4 years old),
     \item school age (5-18 years old),
     \item young adult (19-29 years old),
     \item middle age (30-64 years old),
     \item elderly (65 years old and above).
\end{itemize}

CDFs of age groups are produced for male adults (aged 19 and older), female adults (aged 19 and older), male children (aged 18 and younger), and female children (aged 18 and younger). Other CDFs of children genders and adult genders are also created to support the sampling processes requiring specification on agents' gender (see following section).

\subsubsection{Generation of simulated population.}
While each agent is assigned with demographic and household attributes as detailed in earlier sections, the total residential population and household count in each SA1 area are constrained by the aggregate number per SA1 reported by census. We assign the attributes as following:

\begin{itemize}
    \item Step 1: For a given SA1 area, if the number of generated agents is less than the number of residential population reported by census, sample a household based on the CDF of households constructed from the ABS data.
    \item Step 2: Within each of the generated household, assign agents with genders and age groups (male/female adults and/or children) based on the corresponding CDFs. 
    \end{itemize}
    
    For example, if the  household instance sampled from the census data is a couple comprising two adults of opposing gender, with one child, the generated household will consist of three household members:
    \begin{itemize}
        \item an agent representing the father in this household, generated with its age sampled from the male adult CDF of age groups,
        \item an agent representing the mother in this household, generated with its age sampled from the female adult CDF of age groups,
        \item an agent representing the child in this household, generated with its gender sampled from the CDF of genders for children, and its age sampled from the CDF of age groups for children corresponding to the sampled gender.
    \end{itemize}

This process continues until all SA1 areas are considered and the aggregate artificial populations matches the population statistics reported by census. During this process, every three consecutively generated households form a household cluster, representing a residential context with close contacts around the agents' residency.

With the population generated at SA1 level, we follow the partitioning method that the ABS uses to merge SA1 population into SA2 population. We note that no agents are generated during this process as all SA1 areas are included within  SA2 areas. We assume that SA2 areas are spatially distributed and the geographic distance between any pair of SA2 areas is calculated as the distance between the centroids of two SA2 areas defined by the ABS-provided shapefiles representing geographical boundaries. This spatial distribution plays a crucial role in generating school placements for children and working group placements for adults. 

As a result, the generated aggregate artificial population  is a ``digital twin'' of the Australian population, matching key census statistics, including the number of SA2 and SA1 areas, households, household clusters, and total agents. These statistics are summarised in Table \ref{tab:surrogate_pop_stats}. Note that in addition to residential contexts, a number of other mixing contexts is also reported in Table \ref{tab:surrogate_pop_stats}. DZNs and working groups (i.e., work-related environments) are allocated for adults, while schools, grades, and classes (i.e., education-related environments) are allocated for children. The allocation of these contexts is used in generating the commuting network, detailed in Section \ref{sec_supp:TTW}.

\begin{table}[ht]
\centering
% \bgroup
\def \arraystretch{1.2}
\begin{tabular}{|l|r|r|r|}
\hline
\rowcolor[HTML]{C0C0C0} 
\textbf{Number of} & \multicolumn{1}{l|}{\cellcolor[HTML]{C0C0C0}\textbf{Census 2016 based}} & \multicolumn{1}{l|}{\cellcolor[HTML]{C0C0C0}\textbf{Census 2021 based}} & \multicolumn{1}{l|}{\cellcolor[HTML]{C0C0C0}\textbf{Increase (\%)}} \\ \hline
SA2s (SLAs)        & 2,310                                                              & 2,454                                                              & 6.23\%                                                                    \\ 
SA1s (CDs)         & 57,523                                                             & 61,811                                                             & 7.45\%                                                                    \\ 
DZNs               & 9,136                                                              & 9,307                                                              & 1.87\%                                                                    \\ 
Working Groups     & 470,608                                                            & 530,988                                                            & 12.83\%                                                                   \\ 
Schools            & 9,463                                                              & 9,605                                                              & 1.50\%                                                                    \\ 
Grades             & 56,778                                                             & 57,630                                                             & 1.50\%                                                                    \\ 
Classes            & 174,564                                                            & 189,078                                                            & 8.31\%                                                                    \\ 
Households         & 9,656,841                                                           & 10,802,052                                                          & 11.86\%                                                                   \\ 
Household Clusters & 2,435,274                                                           & 2,723,287                                                           & 11.83\%                                                                   \\ 
Population             & 23,406,541                                                          & 25,428,029                                                          & 8.64\%                                                                    \\ \hline
\end{tabular}
\caption{Summary of demographic attributes and mixing contexts of the surrogate population generated from 2016 and 2021 Census. The 4th column shows the relative increase for each  attribute over the five-year period. The population expansion results in an increase across all residential and mixing contexts.}
\label{tab:surrogate_pop_stats}
\end{table}

\subsection{Surrogate domestic commuting network}
\label{sec_supp:TTW}
In addition to the residential mixing contexts (see Section \ref{sec_supp:agent_gene}), our model also considers interactions across spatially distributed agents which occur in workplaces and schools. %We model interactions in these contexts by using the mobility patterns as a proxy to social interactions at work and school. 
The mobility patterns describing commuting to places of work and education are represented by a realistic commuting network constructed using the census commuting data and student registration records from ACARA \cite{ACARA_Data}. In this network, we allocate population flows specifically capturing the student and worker flows between mutually exclusive sets, e.g., between the usual residential areas (SA1 areas) and the places of work (DZNs). In this section, we describe the construction of the student and worker flows in the commuting network in two parts: travel to school, applicable for agents from 5 to 18 years of age; and travel to work, applicable for agents over 18 years of age.

\subsubsection{Travel to school.}
In our surrogate population, nearly 18\%  of the population are school-aged children (from 5 to 18 years old), with the vast majority of them (approximately 95\%) going to school on weekdays during day time and interacting with other school-attending children and teachers. 
These at-school interactions are enabled by integrating ``student flows" into the commuting network. Each flow is set between the student residence (which has been previously assigned during the generation of simulated agents; see Section \ref{sec_supp:agent_gene}) and the destination zones (i.e., schools). 

Using the school enrollment numbers and locations reported by ACARA, the school locations are pseudo-deterministically chosen within destination zones, while accounting for school capacity and geographic proximity to areas with a sufficient number of children and teachers. 

The allocation of schools is based on the following assumptions. The probability of students attending a school is set in proportion to the geographic distance between the school and their residence, within the school's catchment zone. This process emulates the tendency of students to initially prioritise schools in the closest proximity. If schools nearby are unavailable, students are assumed to expand their search to nearby options within the maximum travel distance (150km). This ensures that the vast majority of students are assigned to schools by the end of the generation process. 

Within each school, students are further organised into smaller groups representing grades and classes, where classes represent the primary units of regular interaction for students with the highest frequency of contacts, followed by grades. The contact rates in the school-environment are summarised in Table \ref{tab:daily_transmission_probability}. % , Section \ref{sec_supp:model}.

Upon completion of the student-to-school assignments, we estimate the number of teachers working at a school based on the enrolment numbers of the school, the student-to-staff ratio (2:17) (approximated from historical ABS data), and the estimated number of three teachers per class. We then randomly sample from the ensemble of adults working in the school's DZN to match the estimated number of teachers. The remaining workforce within the DZN is then assigned to non-school-related working groups detailed in the following section. 

\subsubsection{Travel to work.}
\label{sec_supp:TTW_improve}
We then construct a travel-to-work network for the remaining working population, capturing daily frequent contacts at the workplace where each edge represents a worker flow between agents' usual-residential areas (UR, previously assigned as detailed in Section \ref{sec_supp:agent_gene}) and the corresponding places of work (POW). Within the working population, each agent is also assigned to a working group with a maximum of 20 people at work to simulate a realistic working environment where a small group of individuals frequently interact on a daily basis. This assignment process matches the key census attributes reported in travel-to-work (TTW) data  \cite{ABS_census} (i.e., the census table for employment, income, and education representing the number of commuters between UR and POW).

The work-related census reports  partition the population into multiple statistical area (SA) levels and Destination Zones (DZN), in ascending resolution. We note that DZN (i.e., the highest resolution for POW) does not align precisely with SA1 partitions (i.e., the highest resolution for UR). We use DZN and SA1 as two separate partitions simulated in different time cycles, daytime and nighttime respectively (see Section \ref{sec_supp:model} below for more details). Each agent residing in a SA1 is randomly assigned to work in a DZN and the total number of registered commuters between the SA1 (considered as UR) and the DZN (considered as POW) must match the corresponding commuting volume reported by census. This process uses the high-resolution worker flow data between SA1 and DZN to construct a high-fidelity TTW network between all URs and POWs. 
However, the implementation of privacy-protecting policies adopted by the ABS heavily perturbed the data, resulting a sizable discrepancy between high-resolution data (e.g., SA1--DZN) to lower-resolution data (e.g., SA2--SA2). We consider and address this discrepancy in detail, as described below. 

\

\textbf{Inconsistencies in ABS data for travel to work.}
The census data contains inconsistencies between UR and POW data due to the adoption of new privacy protection measures, implemented within the ABS anonymity policy compliance system over different census years~\cite{ABS_data_confidentiality_guide, fair_creating_2019}. Such inconsistencies produce a mismatch in the commuting flows (between UR and POW) reported at different SA partitions. These inconsistencies therefore require a reconstruction of some commuting flows between UR and POW, added to match the aggregate totals. 

\begin{figure}[ht]
    \centering
    \includegraphics[width=\textwidth]{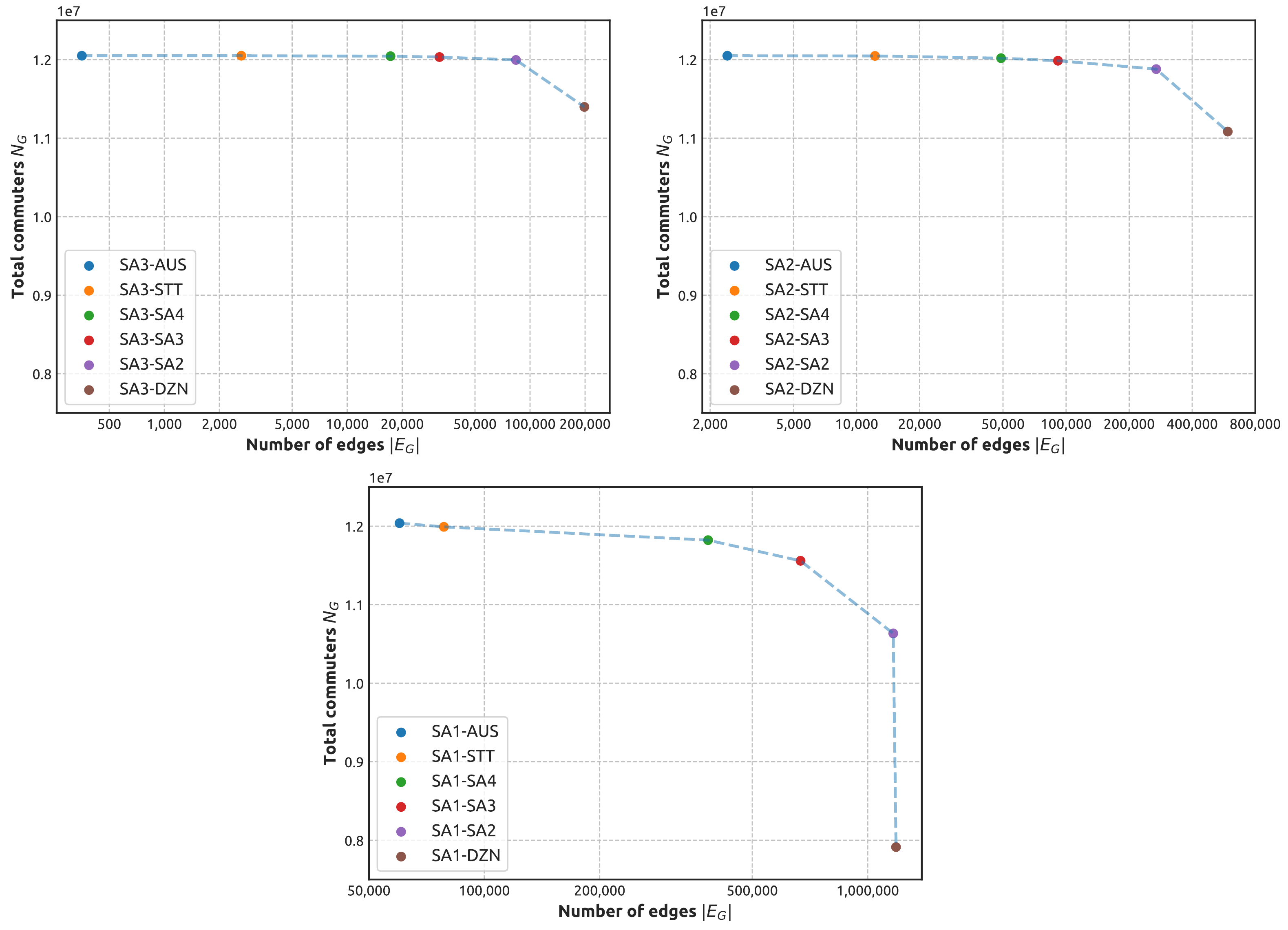}
    \caption{Visualisation of the discrepancies at different resolutions observed in 2021 census commuting data. Each subplot traces combinations between UR and POW for all possible resolutions. Across the  UR $\leftrightarrow$ POW flows, considered at comparable or higher resolutions (i.e., SA3--DZN, SA2--DZN, and SA1--DZN), the edge numbers are noticeably lower towards the right hand side on x-axis, with a reduced number of commuters (y-axis). }
    \label{fig:TTW_discrepancy}
\end{figure}

%Upon close inspection, we find that worker flows in census data have significant discrepancies across different resolutions. 
Specifically, the total number of commuters aggregated at a higher resolution, e.g., between SA1 (UR) and DZN (POW), is usually less than its counterpart obtained from a lower resolution, e.g., between SA2 (UR) and SA2 (POW), as shown in Figure~\ref{fig:TTW_discrepancy}). This results from  various ABS privacy protection measures aimed to avoid re-identification from high-resolution commuting data, particularly when commuting volume is low. As a result, the working mobility in 2016 census shows a 34\%  difference between the aggregated and true commuters \cite{fair_creating_2019}. Here, we report a similar observation for 2021 census, with a difference of a 25.6\% between the SA1--DZN and SA1--SA2 commuting datasets, and a 34.27\% difference between the SA1--DZN and the total number of reported commuters in  Australia (see Table \ref{tab:data_inconsistency_3levels}).

\begin{table}[ht]
\centering
% \bgroup
\def\arraystretch{1.2}
\begin{tabular}{|c|r|r|r|}
\hline
\rowcolor[HTML]{C0C0C0} \textbf{UR-POW} & \multicolumn{1}{r|}{\cellcolor[HTML]{C0C0C0}\textbf{Number of}} & \multicolumn{1}{r|}{\cellcolor[HTML]{C0C0C0}\textbf{Number of}} & \multicolumn{1}{r|}{\cellcolor[HTML]{C0C0C0}\textbf{Reduction in}} \\ 
\rowcolor[HTML]{C0C0C0} \textbf{Resolutions} & \multicolumn{1}{r|}{\cellcolor[HTML]{C0C0C0}\textbf{Edges}} & \multicolumn{1}{r|}{\cellcolor[HTML]{C0C0C0}\textbf{Commuters}} & \multicolumn{1}{r|}{\cellcolor[HTML]{C0C0C0}\textbf{Registered Commuters}} \\
\hline
SA1 -- AUS   &   60,126      &   12,038,052      &       Reference       \\ 
SA1 -- STT   &   78,461      &   11,991,363      &       0.39\%          \\
SA1 -- SA4   &   383,454     &   11,821,831      &       1.80\%          \\
SA1 -- SA3   &   668,132     &   11,558,487      &       3.98\%          \\
SA1 -- SA2   &   1,165,986   &   10,633,702      &       11.67\%         \\
SA1 -- DZN   &   1,185,396   &   7,913,093       &       34.27\%         \\
\hline
SA2 -- AUS   &   2,415       &   12,049,439      &       Reference       \\
SA2 -- STT   &   12,242      &   12,046,956      &       0.02\%          \\
SA2 -- SA4   &   48,929      &   12,018,402      &       0.26\%          \\
SA2 -- SA3   &   91,426      &   11,985,868      &       0.53\%          \\
SA2 -- SA2   &   269,261     &   11,878,866      &       1.42\%          \\
SA2 -- DZN   &   591,375     &   11,083,493      &       8.02\%          \\
\hline
SA3 -- AUS   &   354         &   12,049,397      &       Reference       \\
SA3 -- STT   &   2,637       &   12,049,398      &       0.0\%           \\
SA3 -- SA4   &   17,257      &   12,043,898      &       0.05\%          \\
SA3 -- SA3   &   31,971      &   12,032,775      &       0.14\%          \\
SA3 -- SA2   &   83,778      &   11,995,534      &       0.45\%          \\
SA3 -- DZN   &   198,299     &   11,397,507      &       5.41\%          \\
\hline
\end{tabular}
\caption{Summary of working flows between the usual residential (UR) and the place-of-work (POW), reported at different resolutions of statistical areas defined by census. At each resolution, the number of edges specifies the number of destinations (connections between these regions), while the number of commuters represents the aggregated number of commuters in all edges at this resolution. The last column shows the discrepancy (a reduction measured in percentage) between the specified resolution and the reference resolution. Note that census data are pre-processed, with all non-geographic regions (e.g., "Migratory/offshore/shipping" and "No usual address") excluded.}
\label{tab:data_inconsistency_3levels}
\end{table}

\

\textbf{Prior work resolving inconsistencies in 2016 census.}
In our prior work~\cite{fair_creating_2019} we proposed an algorithm that uses a statistical re-sampling process to generate and add new edges to the SA1-DZN commuting network. This considerably improved the consistency of  commuting flow across different resolutions. This algorithm used the commuting data from the older 2011 census as a baseline (existing prior to the implementation of privacy protection protocols), to supplement the commuting data in different resolutions in 2016 census. The main steps of the algorithm are summarised as follows:

\begin{itemize}
    \item Step 1: Calculate the historical (2011) distribution $P(w|N_{SA1})$ of the number of commuters between UR (SA1) and POW (DZN) (i.e., edge weights of the edges (SA1, DZN) in the TTW network, denoted by $w$), given a working population size in SA1 level (denoted by $N_{SA1}$).
    \item Step 2: Sample additional edges for each UR unit (SA1) based on $P(w|N_{SA1})$. The list of partial candidate edges, $L=\{(SA1_1, w_1), ..., (SA1_n, w_n)\}$, is generated. The total sampled edge weights for an SA1 must be bounded by the inconsistencies between SA2 and SA1 levels (i.e., the difference between the number of commuters residing in an SA2 and the aggregated number of commuters residing in the corresponding decomposed SA1s).
    \item Step 3: Assign workplaces (DZN) for the newly sampled edges, i.e., to complete the list $L$: $L=\{(SA1_1, DZN_1, w_1)$, ..., $(SA1_n, DZN_1, w_n)\}$. These assignments are constrained by the number of remaining unassigned commuters in each DZN and in the SA2-SA2 commuting network when compared with the SA1-DZN network.
    \item Step 4: Eliminate reclaimed edges (i.e., the SA1-DZN edges that the ABS already declared) and concatenate the remaining generated edges with the existing edges in 2016 census.
\end{itemize}

\textbf{An optimised approach to address the inconsistencies in 2021 census.} We optimised the algorithm originally presented in \cite{fair_creating_2019} and applied the optimisation to construct the travel-to-work network using 2021 census data. The optimised algorithm reduces the discrepancy in SA1--DZN travel-to-work data by using the lower resolution commuting data as a reference point, since the privacy protocols mostly affect the higher resolution commuting data. For example, there is only a marginal reduction (1.42\%) in the number of commuters observed in the SA2--SA2 network, compared to a significantly more sizable 34.27\% reduction observed in the higher-resolution SA1--DZN network, as summarised in Table \ref{tab:data_inconsistency_3levels}. 

The optimised re-sampling algorithm reconstructing the surrogate TTW network is detailed in Algorithm~\ref{algo:refined_TTW}, with the changes to the original algorithm~\cite{fair_creating_2019} highlighted in blue. Similar to the original approach, in the optimised algorithm, a number of unassigned working-age agents in an $\text{SA1}_i$ region is selected at random and assigned to a random working group in a $\text{DZN}_i$ region. The assignment of the residential area and the destination zone is constrained by the edge weight $w_i$ of the corresponding edge ($\text{SA1}_i,\text{DZN}_i, w_i)$ in the surrogate network. We introduced several additional constraints to check the eligibility of newly generated edges  (step (3) of  Algorithm~\ref{algo:refined_TTW}), before adding them to the edge-list of the surrogate network. This modification ensures that the sampled artificial edges are not duplicates of existing edges and improves the quality of the final surrogate edge-list. This however had a cost of significantly lengthening the processing time relative to the original algorithm~\cite{fair_creating_2019}. To speed up the computation, we adjusted the sampling process by sampling a uniformly random number of edges ($\text{rand}(N)$)  in the existing edge-list for each DZN at a time. This method significantly reduced the computational cost.

\begin{algorithm}[ht]
\caption{Optimised algorithm generating the 2021 travel-to-work (TTW) surrogate network}
\label{algo:refined_TTW}
\begin{algorithmic}
\Require \\
\begin{tabbing}
    {[A]} \= 2016 -- List of \= surrogate UR-POW (SA1-DZN) TTW edges (reconstructed by \cite{fair_creating_2019}),\\
    {[B]} \> 2016 -- List of working population size in SA1 resolution,\\
    {[C]} \> 2021 -- List of working population size in SA1 resolution,\\
    {[D]} \> 2021 -- List of UR-POW (SA1-DZN) TTW edges (from ABS census data),\\
    {[E]} \> 2021 -- List of POW-UR (DZN-SA2) TTW edges (from ABS census data),\\
    {[F]} \> 2021 -- List of numbers of remaining unassigned commuters (not in D) for each UR-POW (SA2-SA2) TTW edge,\\
    {[G]} \> 2021 -- List of numbers of remaining unassigned working adults (not in D) residing in particular SA1s,\\
    {[I]} \> 2021 -- List of numbers of remaining unassigned working adults (not in D) working in particular DZNs,\\
    {[K]} \> 2021 -- DZN--SA2 correspondence list,\\
    {[L]} \> 2021 -- SA1--SA2 correspondence list,\\
    {[N]} \> Maximum number of edges allowed to be selected in step 3.g.\\ 
\end{tabbing}
% \vspace{-0.5cm}
\Ensure \\
    {[O]} 2021 - List of additional UR-POW (SA1-DZN) edges supplementing D to represent reconstructed TTW network\\\\
% \vspace{0.5cm}
\hspace*{-0.35cm}\textbf{Procedure:}

\hspace*{-0.8cm} \textit{\textbf{(1) Calculate the historical (2016) distribution $\boldsymbol{P(w|N_{\text{SA1}})}$ from {[A,B]}:}}\\
\begin{tabbing}
    \hspace{1cm} $w$: \quad \quad \= An edge weight in the SA1-DZN travel-to-work network, representing the number of commuters\\ \> \quad \quad living in a SA1 and working in a DZN.\\
    \hspace{1cm} $N_{SA1}$: \> Total number of commuters living in a SA1, or working population size of a SA1. $N_{SA1}$ is grouped \\
    \> \quad \quad into bins, used in the construction of $P(w|N_\text{SA1})$.
\end{tabbing}
An illustrative example of the distribution $P(w|N_{SA1})$ is shown below. Each column represents a distribution of edge weight given a bin value of $N_{SA1}$.

\renewcommand{\arraystretch}{1.025}
\vspace*{0.25cm}\hspace*{4.8cm}\begin{tabular}{ |c|c|c|c|  }
 \hline
 & \multicolumn{3}{|c|}{$N_{SA1}$} \\
 \hline
 $w$        & [0,9]     & [10,19]   & ...\\
 \hline
 1          & 0.01      & 0.08      & ...\\
 2          & 0.02      & 0.05      & ...\\
 3          & 0.10      & 0.01      & ...\\
 ...        & ...       & ...       & ...\\
 \hline
\end{tabular}

\vspace*{0.25cm}\hspace*{-0.8cm} \textit{\textbf{(2) Construct lists of additional partial edge candidates (SA1,w) for each SA1 based on $\boldsymbol{P(w|N_{SA1})}$ and {[C,G]}:}}\\
\vspace*{0.2cm}\hspace{1cm} For each $\text{SA1}_i$ listed in G:\\
\hspace{1.5cm} a. From C: determine its bin size $N_{\text{SA1}_i}$.\\
\hspace{1.5cm} b. From G: determine the number of working adults residing in $\text{SA1}_i$, but not assigned a DZN: $R(\text{SA1}_i)$.\\
\hspace{1.5cm} c. Determine $P(w|N_{\text{SA1}_i})$ from step 1.\\
\hspace{1.5cm} d. Construct $L(\text{SA1}_i) = \left\{w_i \;\middle\vert\;
   \begin{array}{@{}l@{}}
   w_i \sim P(w|N_{\text{SA1}_i}) \\ 
   \sum_{w_i \in L(\text{SA1}_i)} w_i \leq R(\text{SA1}_i)
   \end{array}
\right\}$\\

\vspace*{0.25cm}\hspace*{0cm} \textit{\textbf{(3) Construct additional complete edges $\boldsymbol{(\text{SA1}_i, \text{DZN}_i, w_i)}$ based on {[E,I,K,L]} and $\boldsymbol{L(\text{SA1}_i)}$:}}\\
\hspace*{1cm} Repeat until convergence. For each $\text{DZN}_i$ in I:\\
\hspace*{1.5cm} a. From K: search for $\text{SA2}_{\text{POW}}$ that $\text{DZN}_i$ locates.\\
\hspace*{1.5cm} b. From E: construct $L_{\text{SA2-UR}} = \left\{\text{SA2}_i | (\text{DZN}_i, \text{SA2}_i) \in E \right\}$.\\
%of SA2s where commuters working in this DZN may reside: \\
\hspace*{1.5cm} c. From [L,$L_{\text{SA2-UR}}$]: construct $L_{\text{SA1-UR}} = \left\{\text{SA1}_i \;\middle\vert\;
   \begin{array}{@{}l@{}}
   \text{SA1}_i \in \text{SA2}_i, \text{ where SA2}_i \in L_{\text{SA2-UR}}\\ 
   \textcolor{blue}{(\text{SA1}_i,\text{DZN}_i) \notin D \cup O} % No reclaimed edges.
   \end{array}
\right\}$\\

\hspace*{1.5cm} d. From [I,\textcolor{blue}{O}]: determine the remaining unassigned working adults (not in \textcolor{blue}{$D \cup O$}) in $\text{DZN}_i$: $R(\text{DZN}_i)$.\\

\hspace*{1.48cm} e. From [F,\textcolor{blue}{O}]: determine the remaining unassigned commuters (not in \textcolor{blue}{$D \cup O$}) in the UR--POW ($\text{SA2}_{\text{UR},i}-\text{SA2}_\text{POW}$) TTW edge: $R(\text{SA2}_{\text{UR},i}, \text{SA2}_\text{POW})$.\\

\hspace*{1.5cm} f. Construct $L_{f} = \left\{(\text{SA1}_i, \text{DNZ}_i, w_i) \;\middle\vert\;
   \begin{array}{@{}l@{}}
        \text{SA1}_i \in L_{\text{SA1-UR}}\\ 
        w_i \in L(\text{SA1}_i): \text{ sampling without replacement}\\
        \textcolor{blue}{w_i \leq R(\text{DZN}_i)}\\
        \textcolor{blue}{w_i \leq R(\text{SA2}_{\text{UR},i}, \text{SA2}_\text{POW}), \text{ where } \text{SA2}_{\text{UR},i} \in L_{\text{SA2-UR}}}\\
   \end{array}
\right\}$\\

\vspace*{0.1cm}\hspace*{1.5cm} \textcolor{blue}{g. Sample for rand(N) candidate edges without replacement: $(\text{SA1}_i, \text{DNZ}_i, w_i)$ from $L_{f}$.} Add sampled edge(s) to O.\\\\

\hspace*{0cm} \textit{\textbf{(4) Concatenate O and D to reconstruct the SA1--DZN TTW network.}}\\
\end{algorithmic}
\end{algorithm}

\textbf{Comparison of the network reconstruction results.} 
Our optimised algorithm significantly improved the mapping between SA2--SA2 and SA1--DZN networks  from 69.77\% (between the census SA2--SA2 and SA1--DZN networks)  to 92.87\% (between our reconstructed SA2--SA2 and census SA2--SA2 networks), thus accounting for missing edges. It provided further improvement in comparison to the prior algorithm~\cite{fair_creating_2019}, which  reconstructed 90.66\% of the mapping between census SA2-SA2 and  SA2-SA2 networks, as shown in Figure~\ref{fig:TTW_reconstruction_results}.a). The re-sampling process reduces the number of unassigned commuters in the SA2-SA2 TTW network from 3,469,391 to 835,534, as shown in Figure~ \ref{fig:TTW_reconstruction_results}.(b). 

The optimised algorithm reconstructs the SA2--SA2 surrogate network with a smaller mean squared error (MSE) and a higher correlation with the true commuter flow (i.e., census SA2--SA2 network), as shown in Figure~\ref{fig:TTW_reconstruction_results} (c) and \ref{fig:TTW_reconstruction_results} (d).

\begin{figure}
    \centering
    \includegraphics[width=\textwidth]{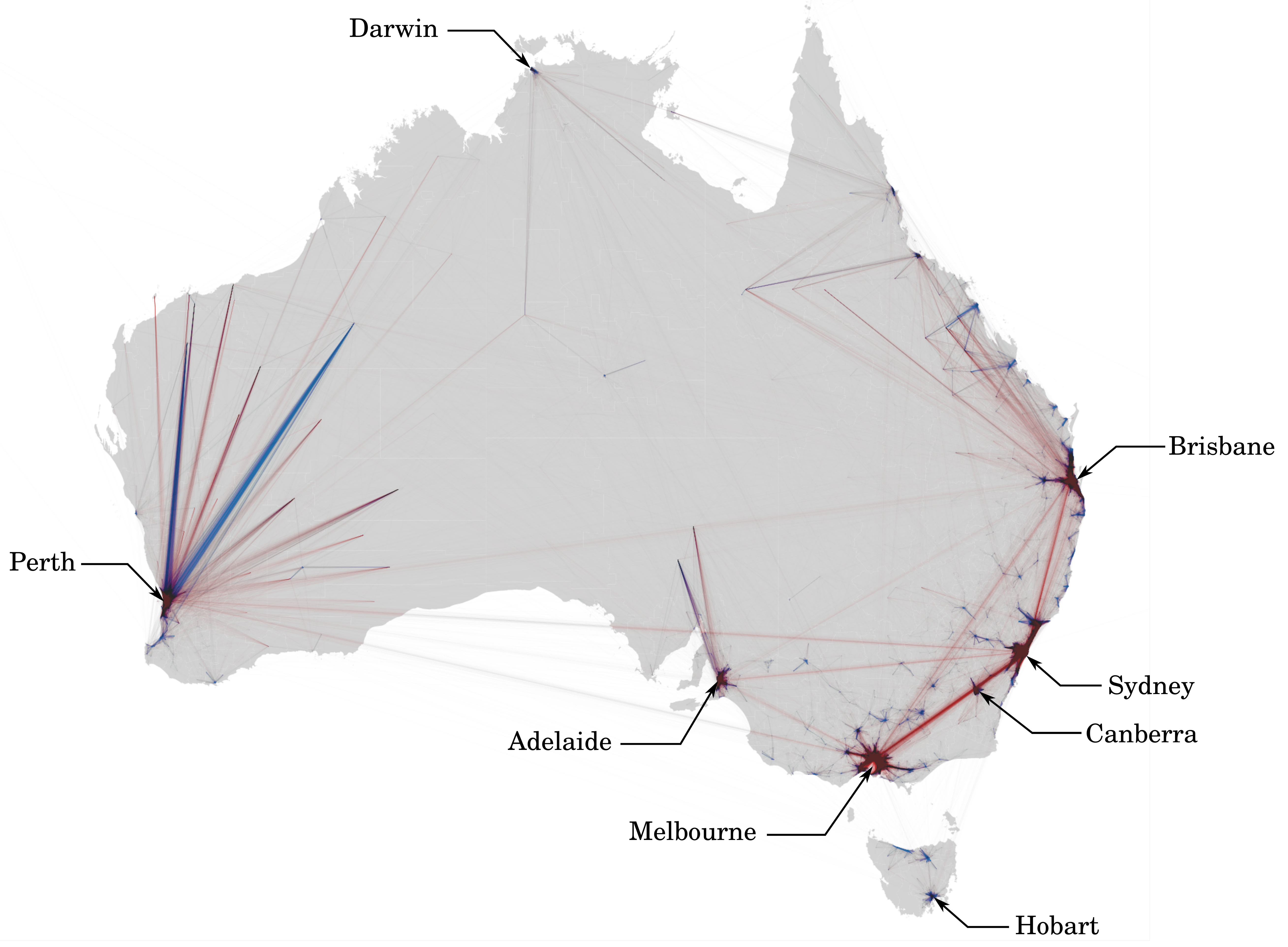}
    \caption{Visualisation of commuting patterns between usual residences (URs) and places of work (POWs), using 2021 census data. Each commuting flow is represented by an edge connecting the centroids of the corresponding UR and POW. Red lines represent the commuting edges between URs (in SA1 level) and POWs (in DZN level) directly taken from 2021 census, while blue lines depict the reconstructed edges created by our algorithm to match the aggregate totals.}
    \label{fig:TTW_map}
\end{figure}

\begin{figure}
    \centering
    \includegraphics[width=\textwidth]{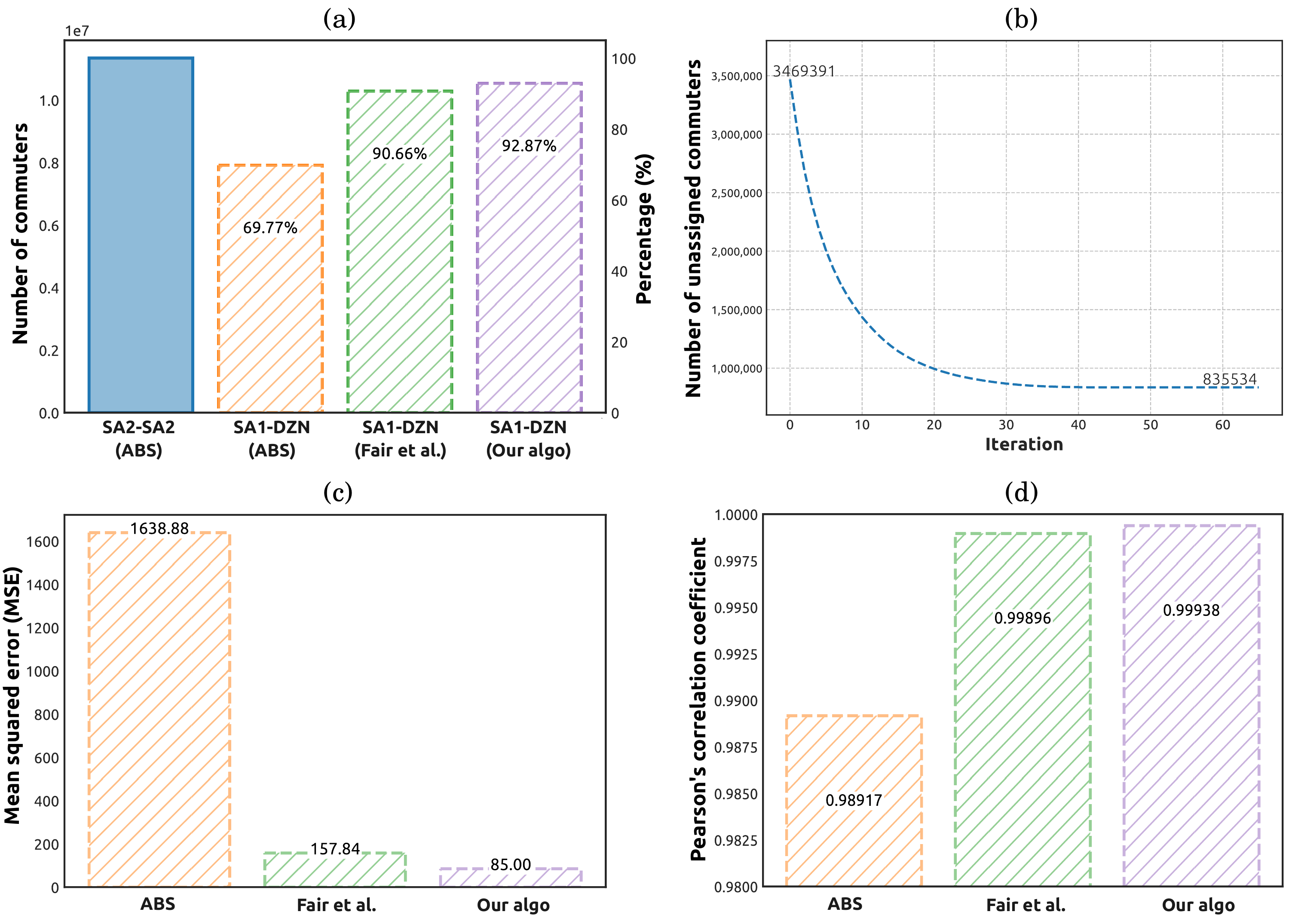}
    \caption{Results of network reconstruction using the 2021 TTW data. (a) Number of commuters between the reference TTW SA2--SA2 network (from ABS Data), the TTW SA1--DZN network (from ABS Data), the TTW SA1--DZN network (refined by the original algorithm of Fair et al.~\cite{fair_creating_2019}, adapted for census 2021), and the TTW SA1--DZN network (refined by the optimised algorithm). (b) Number of unassigned commuters, counted in SA2--SA2 network but not in SA1--DZN network, over refining iterations. (c) Mean squared error between the SA2--SA2 TTW network from ABS and the reconstructed SA2--SA2 TTW network from the refined SA1-DZN networks. (d) Pearson's correlation coefficient between the SA2--SA2 TTW network from ABS and the reconstructed SA2--SA2 TTW network from the refined SA1--DZN networks.}
    \label{fig:TTW_reconstruction_results}
\end{figure}

%In this study, we improved upon a reconstruction technique previously developed in \cite{fair_creating_2019}, aiming to generate a surrogate population matching the latest 2021 census data. In doing so, we achieved a significant improvement in the consistency of commuting flows between UR and POW: the fraction of  commuters consistently represented at both low and high resolution levels increased from 69.77\% (observed in the published ABS 2021 census data) to 92.87\% (refined with our improved algorithm), as described in Section 2 of Supplementary Material. 
The resultant commuting patterns generated from 2021 census are shown in Figure~\ref{fig:TTW_map} with the Greater Capital Cities (GCC) annotated within each state.

We note that although our solution effectively rectified most of the discrepancies between SA1--DZN and SA2--SA2 networks using the re-sampling procedure, it cannot provide a complete mapping between the SA2--SA2 network and the lower resolution networks because the algorithm does not introduce new edges (i.e., new commuting flows) beyond the edges already present in the census SA2--SA2 network (see Table \ref{tab:data_inconsistency_3levels} and Fig. \ref{fig:TTW_discrepancy}.b). However, the proposed re-sampling algorithm is able to produce a surrogate travel-to-work network that adequately represents the census data by supplementing the vast majority of  missing edges, as shown in Figures~\ref{fig:TTW_map} and~\ref{fig:TTW_reconstruction_results}. 

\subsection{International air traffic}
\label{sec_supp:air_travel}
In our model, the initial infection cases are seeded proportionally to the international air traffic inflow. The scenarios for 2016 use the BITRE air traffic data for this year without any modifications. In contrast, to simulate the pandemic scenarios for 2021, we estimate the international air travel for this year based on the BITRE air traffic data between 2003 and 2019. We intentionally do not use the air traffic data between 2020 to 2021 due to the severe disruption of international travel due to the travel restrictions during the COVID-19 pandemic. 

\begin{figure}[ht]
    \centering
    \includegraphics[width=0.85\columnwidth]{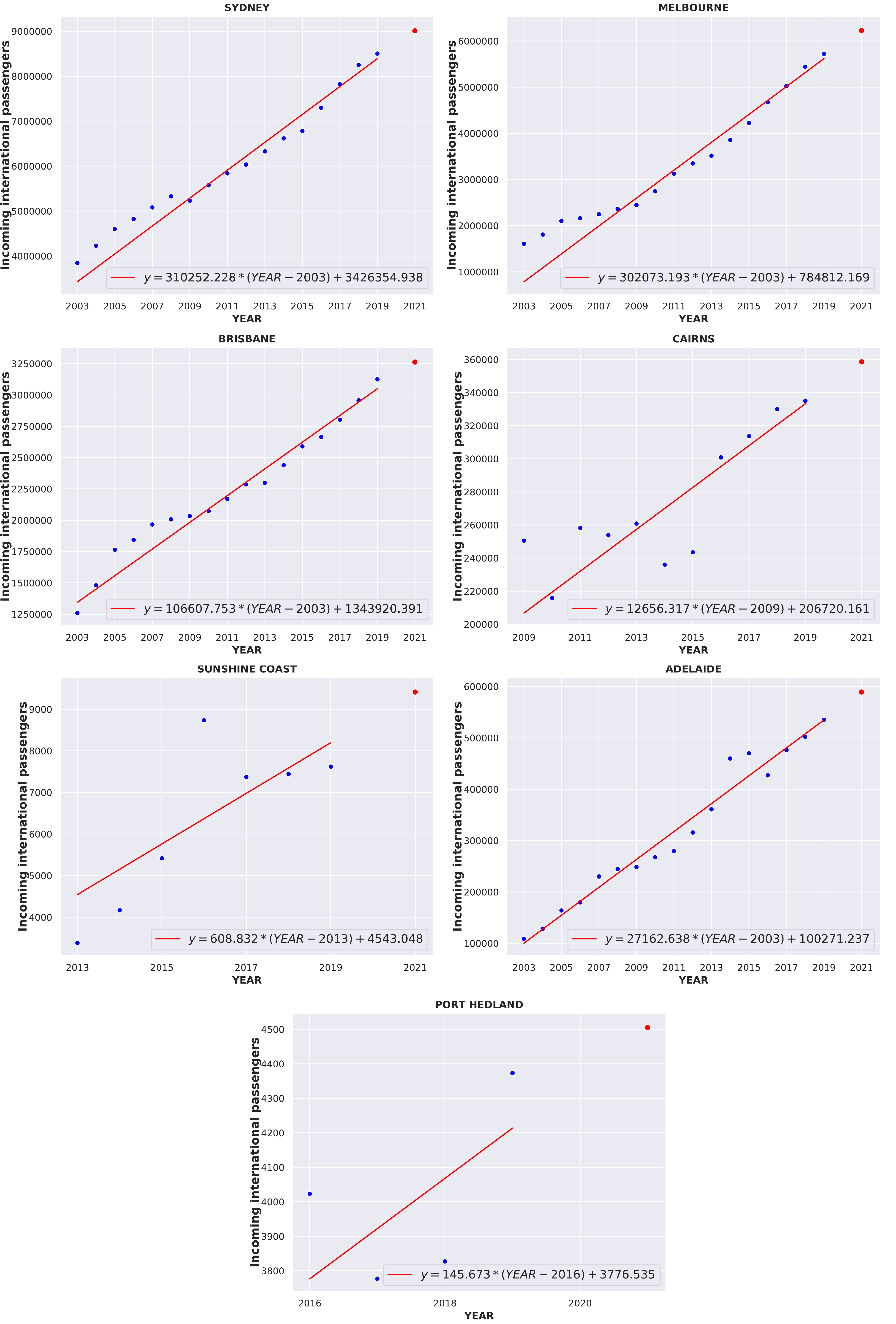}
    \caption{Weighted linear regression models for the considered airports across Australia. The incoming international passengers (IIP) numbers in previous years up to 2019 are shown in blue; the projected IIP numbers in 2021 are shown in red. Solid red line represents the fitted linear regression model.}
    \label{fig:WLR_for_airport}
\end{figure}

We approximated the daily average number of incoming international passengers (IIP) at the international airports across Australia 2021 by utilising the international air traffic data up to 2019~\cite{BITRE_airport_data} which reports the total number of IIP for a given year (reported by June) at various international airports across the country. We then extrapolated the number of IIP in this period by using a weighted multimodal linear regression model. We placed a higher weight ($\gamma$ = 0.8) on more recent data points and constructed the linear regression model which produced the most optimised estimates with small weighted losses. We note that while larger international airports (e.g., Sydney, Melbourne, Brisbane, and Adelaide, shown in Figure \ref{fig:WLR_for_airport}) have a sustained inflow of IIP since 2003, small airports (e.g., Darwin) may not have a continuous IIP until the more recent years. In such cases, we could not produce an accurate weighted linear model due to the lack of data; instead, we estimated the IIP in 2021 by averaging the IIP in available years. A comprehensive summary of the estimation methods employed for each international airport is provided in Table \ref{tab:int_airport_estimation}. The linear regression model for eligible airports is shown in Figure \ref{fig:WLR_for_airport}. 

\begin{table}[ht]
\centering
% \bgroup
\def\arraystretch{1.2}
\begin{tabular}{|l|c|c|c|r|l|}
\hline
\rowcolor[HTML]{C0C0C0} 
\textbf{Airport}   & \textbf{Code}           & \textbf{State}   & \textbf{Location (SA2)}          & \multicolumn{1}{l|}{\cellcolor[HTML]{C0C0C0}\textbf{Estimated IIP}} & \textbf{Estimation Method}                                      \\ \hline
\rowcolor[HTML]{FFFFFF} 
{\color[HTML]{000000} Sydney}    & {\color[HTML]{000000} SYD}     & {\color[HTML]{000000} NSW} & {\color[HTML]{000000} 117011325} & {\color[HTML]{000000} 9,010,895}                                           & {\color[HTML]{000000} Weighted Linear Regression}               \\
\rowcolor[HTML]{FFFFFF} 
{\color[HTML]{000000} Newcastle} & {\color[HTML]{000000} NTL}     & {\color[HTML]{000000} NSW} & {\color[HTML]{000000} 106031125} & {\color[HTML]{000000} 3,493}                                               & {\color[HTML]{000000} Same as 2019}                             \\
\rowcolor[HTML]{FFFFFF} 
{\color[HTML]{000000} Melbourne} & {\color[HTML]{000000} MEL}     & {\color[HTML]{000000} VIC} & {\color[HTML]{000000} 210051248} & {\color[HTML]{000000} 6,222,130}                                           & {\color[HTML]{000000} Weighted Linear Regression}               \\
\rowcolor[HTML]{FFFFFF} 
{\color[HTML]{000000} Brisbane}  & {\color[HTML]{000000} BNE}     & {\color[HTML]{000000} QLD} & {\color[HTML]{000000} 302031036} & {\color[HTML]{000000} 3,262,860}                                           & {\color[HTML]{000000} Weighted Linear Regression}               \\
\rowcolor[HTML]{FFFFFF} 
{\color[HTML]{000000} Cairns}    & {\color[HTML]{000000} CNS}     & {\color[HTML]{000000} QLD} & {\color[HTML]{000000} 306011140} & {\color[HTML]{000000} 358,596}                                             & {\color[HTML]{000000} Weighted Linear Regression}               \\
\rowcolor[HTML]{FFFFFF} 
{\color[HTML]{000000} Gold Coast} & {\color[HTML]{000000} OOL}    & {\color[HTML]{000000} QLD} & {\color[HTML]{000000} 309021231} & {\color[HTML]{000000} 534,733}                                             & {\color[HTML]{000000} Averaging data from 2017, 2018, and 2019} \\
\rowcolor[HTML]{FFFFFF} 
{\color[HTML]{000000} Sunshine Coast} & {\color[HTML]{000000} MCY} & {\color[HTML]{000000} QLD} & {\color[HTML]{000000} 316031426} & {\color[HTML]{000000} 9,414}                                               & {\color[HTML]{000000} Weighted Linear Regression}               \\
\rowcolor[HTML]{FFFFFF} 
{\color[HTML]{000000} Townsville}  & {\color[HTML]{000000} TSV}   & {\color[HTML]{000000} QLD} & {\color[HTML]{000000} 318021475} & {\color[HTML]{000000} 1,031}                                               & {\color[HTML]{000000} Same as 2019}                             \\
\rowcolor[HTML]{FFFFFF} 
{\color[HTML]{000000} Adelaide}   & {\color[HTML]{000000} ADL}    & {\color[HTML]{000000} SA}  & {\color[HTML]{000000} 404031104} & {\color[HTML]{000000} 589,199}                                             & {\color[HTML]{000000} Weighted Linear Regression}               \\
\rowcolor[HTML]{FFFFFF} 
{\color[HTML]{000000} Perth}    & {\color[HTML]{000000} PER}      & {\color[HTML]{000000} WA}  & {\color[HTML]{000000} 506021121} & {\color[HTML]{000000} 2,206,807}                                           & {\color[HTML]{000000} Averaging data from 2017, 2018, and 2019} \\
\rowcolor[HTML]{FFFFFF} 
{\color[HTML]{000000} Port Hedland} & {\color[HTML]{000000} PHE}  & {\color[HTML]{000000} WA}  & {\color[HTML]{000000} 510021269} & {\color[HTML]{000000} 4,505}                                               & {\color[HTML]{000000} Weighted Linear Regression}               \\
\rowcolor[HTML]{FFFFFF} 
{\color[HTML]{000000} Darwin}   & {\color[HTML]{000000} DRW}      & {\color[HTML]{000000} NT}  & {\color[HTML]{000000} 701011001} & {\color[HTML]{000000} 128,585}                                             & {\color[HTML]{000000} Averaging data from 2017, 2018, and 2019} \\
\rowcolor[HTML]{FFFFFF} 
{\color[HTML]{000000} Canberra}  & {\color[HTML]{000000} CBR}     & {\color[HTML]{000000} ACT} & {\color[HTML]{000000} 801031114} & {\color[HTML]{000000} 39,374}                                              & {\color[HTML]{000000} Averaging data from 2017, 2018, and 2019} \\ \hline
\end{tabular}
\caption{Estimated number of incoming international passengers (IIP) in the international airports across in Australia in 2021, projected from the pre-pandemic BITRE data. The estimated IIP is used for the initial infection seeding (see Section 3).}
\label{tab:int_airport_estimation}
\end{table}

\clearpage
\section{Transmission and control model}
\label{sec_supp:model}
\subsection{Transmission model}
The transmission of SARS-CoV-2 in our Agent-based Model of Transmission and Control of the COVID-19 pandemic in Australia (AMTraC-19) is simulated using agent interactions in the generated surrogate population.  
The model operates in discrete time steps, with each step representing a 12-hour cycle (either daytime cycle or nighttime cycle).  In daytime cycle, interactions occur within the working/studying groups, i.e., in the work and education contexts; in nighttime cycle, interactions occur in the residential mixing contexts within communities (SA2 wide), neighborhoods (SA1 wide), household clusters, and households. We also account for differences between weekdays and weekends. A weekday (Monday to Friday) consists of a daytime cycle and a nighttime cycle, while a weekend consists of two nighttime cycles (i.e., no work-related activities). 

The initial infections are seeded as imported infections dependent on the air traffic using international passenger data published by the Bureau of Infrastructure and Transport Research Economics (BITRE). At each time step, until border restrictions are imposed, new infections are introduced  within a 50-km radius of major international airports, in proportion to the international travel influx.

A simulation scenario begins with an initial distribution of infections determined by a binomial distribution with the probability proportional to the average number of incoming passengers at the international airports (see section A.2.3). These infected agents are then assigned to a randomly selected residential area within some distance to an international airport. %For example, in the Greater Sydney area, the daily generated initial infections are distributed within a 50 km radius of Sydney International Airport. 
This seeding process takes place at the start of each simulation day and continues daily until the international border closures are triggered by a defined cumulative incidence threshold.

 % These interactions are simulated in two discrete 12-hour cycles (referred to as ``day" and ``night"), with the work/study-related activities taking place within day cycles and the residential activities occurring during night cycles. There is also a differentiation between weekdays and weekends. On a weekday, the simulator runs a day cycle and a night cycle, whereas on a weekend, two night cycles are simulated instead (i.e., no work- or study-related activities are assumed on weekends).

If an agent is exposed to the disease in one of their mixing contexts, it goes through several health states following the natural history disease model: Susceptible, Latent, Infectious (asymptomatic or symptomatic), and Removed (recovered or deceased). For an agent $i$, let us denote the set $G_i$ consisting of all mixing contexts in which this agent interacts (e.g., workplace or school, grade and class in daytime cycles; and household, household cluster, neighbourhood and community in nighttime cycles). At time cycle $n$, the probability for an susceptible agent $i$ becoming infected across context $g \in G_i$ is determined as follows:
\begin{eqnarray}
    p^g_i(n) = 1 - \prod_{j \in A_g\backslash\{i\}} (1 - p^g_{j \rightarrow i}(n))
    \label{eq:infection_prob_within_a_group}
\end{eqnarray}
where $A_g\backslash\{i\}$ represents the set of agents in the context $g \in G_i$ excluding agent $i$, and  $p^g_{j \rightarrow i}(n)$ denotes the instantaneous probability that an infectious agent $j$, who shares the context $g$ with susceptible agent $i$, transmits the infection to agent $i$:
\begin{eqnarray}
\label{eq:individual_transmission_prob}
    p^g_{j \rightarrow i}(n) = \kappa \ f(n-n_j | j) \ q^g_{j \rightarrow i}    
\end{eqnarray}

A global transmission scalar, denoted by $\kappa$, is used to calibrate the reproductive number $R_0$. The time cycle $n_j$ indicates the moment when agent j becomes infected, while the function $f(n-n_j| j)$ represents the natural history of the disease, reflecting the infectivity of agent $j$ over time. If agent $j$ is not infected, $n < n_j$ and $f(n-n_j | j)=0$. If agent $j$ is infected, $n \geq n_j$ and $f(n-n_j| j) \geq 0$. The infectivity increases exponentially since the latent period ends (the latent period can be set to zero for some variants),  until it reaches its peak at $f(n-n_j | j) = 1.0$. Subsequently, during the recovery period, the infectivity decreases linearly to zero, leading to the agent transitioning to the Removed state. The changes in infectivity of an infected agent are plotted in the main manuscript, see Figure 3. The upper bounds of the daily probabilities of transmission from agent $j$ to agent $i$, denoted by $q^g_{j \rightarrow i}$, which depends on age and mixing contexts. These daily transmission probabilities are summarised in Table~\ref{tab:daily_transmission_probability}, following prior studies \cite{chang_modelling_2020,zachreson_how_2021,chang_simulating_2022,chang_persistence_2023}.

\begin{table}
    \centering
    % \begin{adjustwidth}{-0.5in}{0in}
    \begin{tabular}{l|l|l}
    Mixing context &  Type of interaction & Daily transmission probability ($q^g_{j \rightarrow i}$)\\
        \hline
    \hline
    \rule{0pt}{3ex}Household (size 2)   & Any to child (0 - 18)             &  \hspace{1.7cm} 0.09335\\
                                        & Any to adult (19+)                &  \hspace{1.7cm} 0.02420\\
    \rule{0pt}{3ex}Household (size 3)   & Any to child (0 - 18)             &  \hspace{1.7cm} 0.05847\\
                                       & Any to adult (19+)                &  \hspace{1.7cm} 0.01495\\
    \rule{0pt}{3ex}Household (size 4)   & Any to child (0 - 18)             &  \hspace{1.7cm} 0.04176\\
                                        & Any to adult (19+)                &  \hspace{1.7cm} 0.01061\\
    \rule{0pt}{3ex}Household (size 5)   & Any to child (0 - 18)             &  \hspace{1.7cm} 0.03211\\
                                        & Any to adult (19+)                &  \hspace{1.7cm} 0.00813\\
    \rule{0pt}{3ex}Household (size 6)   & Any to child (0 - 18)             &  \hspace{1.7cm} 0.02588\\
                                        & Any to adult (19+)                &  \hspace{1.7cm} 0.00653\\[0.1cm]
    
    \hline                              
    \rule{0pt}{3ex}Household Cluster    & Child (0 - 18) to child (0 - 18)  &  \hspace{1.7cm} 0.00400\\
                                        & Child (0 - 18) to adult (19+)     &  \hspace{1.7cm} 0.00400\\
                                        & Adult (19+)  to child (0 - 18)    &  \hspace{1.7cm} 0.00400\\
                                        & Adult (19+)  to adult (19+)       &  \hspace{1.7cm} 0.00400\\[0.1cm]
    \hline
    \rule{0pt}{3ex}Working Group        & Adult (19+)  to adult (19+)       &  \hspace{1.7cm} 0.00400\\[0.1cm]
    
    \hline
    \rule{0pt}{3ex}School               & Child (0 - 18) to child (0 - 18)  &  \hspace{1.7cm} 0.00029\\
    Grade                               & Child (0 - 18) to child (0 - 18)  &  \hspace{1.7cm} 0.00158\\
    Class                               & Child (0 - 18) to child (0 - 18)  &  \hspace{1.7cm} 0.00865\\[0.1cm]
                           
   \hline
    \rule{0pt}{3ex}Neighborhood         & Any to child (0 - 4)          &  \hspace{1.7cm} $0.035 \times 10^{-5}$\\
                                        & Any to child (5 - 18)         &  \hspace{1.7cm} $1.044 \times 10^{-5}$\\
                                        & Any to adult (19 - 64)        &  \hspace{1.7cm} $2.784 \times 10^{-5}$\\
                                        & Any to adult (65+)            &  \hspace{1.7cm} $5.568 \times 10^{-5}$\\[0.1cm]
    
    \hline
    \rule{0pt}{3ex}Community            & Any to child (0 - 4)          &  \hspace{1.7cm} $0.872 \times 10^{-6}$\\
                                        & Any to child (5 - 18)         &  \hspace{1.7cm} $2.608 \times 10^{-6}$\\
                                        & Any to adult (19 - 64)        &  \hspace{1.7cm} $6.960 \times 10^{-6}$\\
                                        & Any to adult (65+)            &  \hspace{1.7cm} $13.92 \times 10^{-6}$\\[0.1cm]
    
    \hline
    \end{tabular}
    \caption{Daily transmission probabilities $q^g_{j \rightarrow i}$ from infected agent $j$ to susceptible agent $i$ for different mixing contexts and interaction types. Numbers in brackets show age groups.\vspace*{-0.5cm}}
    \label{tab:daily_transmission_probability}
    % \end{adjustwidth}
\end{table}

The infection probability for agent $i$  across all mixing contexts is calculated as follows:
\begin{eqnarray}
\begin{aligned}
    p_i(n)  &= 1 - \prod_{g \in G_i(n)} \left( 1 - p^g_i(n) \right) \\
            &= 1 - \prod_{g \in G_i(n)} \prod_{j \in A_g\backslash\{i\}} \left( 1 - p^g_{j \rightarrow i}(n) \right)
\end{aligned}
\label{eq:general_infection_prob}
\end{eqnarray}
At the end of each time cycle, a Bernoulli sampling process based on the probability $p_i(n)$ is used to decide whether a susceptible agent $i$ acquires the infection.

An infected agent can be either symptomatic or asymptomatic. The probability of symptomatic illness is determined with respect to the infection probability $p_i(n)$, by incorporating a scaling factor $\sigma$ that represents the fraction of symptomatic cases among the total cases:
\begin{eqnarray}
    p_i^d(n) = \sigma(i) \ p_i(n)
    \label{eq:infection_prob_symptomatic}
\end{eqnarray}
where $\sigma(i)$ is defined based on the age of  agent $i$ using a piece-wise approach: for adults ($\text{age} > 18$), $\sigma_a = 0.67$;  and for children ($\text{age} \leq 18$), $\sigma_c = 0.268$, following previous studies ~\cite{chang_modelling_2020,zachreson_how_2021,chang_simulating_2022,chang_persistence_2023}.
Asymptomatic agents in the model have a lower infectivity compared to their symptomatic counterparts, governed by a scaling factor of $\alpha_{asymp}$ ($0 \leq \alpha_{asymp} \leq 1$).

In reality, not all infection cases are detected, particularly when case detection relies on voluntary self-reporting. To capture this feature, the  model employes two variables adjusting the extent of case detection as the probability of detecting symptomatic cases ($\pi_{symp}$) and the probability of detecting asymptomatic cases ($\pi_{asymp}$).

In this study, we model the transmission of three COVID-19 variants of concern in a wide of range of scenarios in combination with different intervention policies. Some epidemiological parameters are set differently across the three variants, while others are kept constant for all  variants. A summary of parameterisation is provided in Table~\ref{tab_supp:epi_para}.

\renewcommand{\arraystretch}{1.35}
\begin{table}[ht]
    \centering
     \resizebox{\textwidth}{!}{
    \begin{tabular}{l|c|c|c|l}
     Model parameters         & Ancestral & Delta & Omicron & Note \\
         \hline \hline
         $\kappa$                       & 2.75      & 5.3   & 23  & Global transmission scalar \\
         \hline
         $T_{inc}$, mean                & 5 ($\mu$=1.609 $\sigma$=0.00001)         & 4.4 ($\mu$=1.396 $\sigma$=0.413)  & 3 ($\mu$=1.013 $\sigma$=0.413) & Incubation period (log-normal) \\
         \hline
         $T_{rec}$, mean and range      & 12 [12,12] & 10.5 [7, 14] & 9 [7,11] &Recovery period, mean and range (uniform)\\
         \hline
         $T_{lat}$, fixed                  & 2         & 0     & 0 & Latent period \\
         \hline
         $\alpha_{asymp}$               & \multicolumn{3}{c|}{0.3}   & Asymptomatic transmission scalar \\
         \hline
         $\sigma_a$                     & \multicolumn{3}{c|}{0.67}  & Probability of symptoms (age > 18) \\ 
         \hline
         $\sigma_c$                     & \multicolumn{3}{c|}{0.268} & Probability of symptoms (age $\leq$ 18)  \\ 
         \hline
         $\pi_{symp}$                   & \multicolumn{3}{c|}{0.1}   & Daily case detection probability (symptomatic)  \\ 
         \hline
         $\pi_{asymp}$                  & \multicolumn{3}{c|}{0.01}  & Daily case detection probability (asymptomatic)  \\ 
         \hline
         \hline
         $R_0$, mean and 95\% CI        & 2.77 [2.73,2.83]  & 5.97 [5.93, 6.00] & 19.56 [19.12, 19.65]  &Basic reproductive number\\
         \hline
         $T_{gen}$, mean and and 95\% CI& 7.62 [7.53, 7.70] & 6.88 [6.81, 6.94] & 5.42 [5.38, 5.44]  & Generation/serial interval\\
         \hline
    \end{tabular}}
    \caption{Model parameterisation for the three considered variants of concern (i.e., ancestral, Delta, and Omicron) adopted by AMTraC-19. The last two rows show the corresponding basic reproductive number ($R_0$) and generation/serial interval ($T_{gen}$), calibrated in our previous studies~\cite{chang_modelling_2020, chang_simulating_2022, zachreson_how_2021,chang_persistence_2023}.\vspace*{-0.5cm}}
    \label{tab_supp:epi_para}
\end{table}

\subsection{Non-pharmaceutical interventions}
The model incorporates various non-pharmaceutical interventions (NPIs) aimed to mitigate SARS-CoV-2 transmission, including case isolation (CI), home quarantine (HQ), school closures (SC), and social distancing (SD). Each NPI is characterised by: (i) macro-distancing level, setting a fraction of the population that adopts, or complies with, the intervention, and (ii) micro-distancing parameters which quantify the adjusted (usually reduced) interaction strengths between the agents who have adopted the NPI and other agents in the shared mixing groups. The infection probability $p_i(n)$ for individuals who have adopted an NPI is adjusted as follows:
\begin{eqnarray}
    p_i(n) =  1 - \prod_{g \in G_i(n)} \left [1 - F_g(i) \left ( 1 - \prod_{j \in A_g\backslash\{i\}} (1 - F_g(j) \ p^g_{j \rightarrow i}(n)) \right ) \right ]
    \label{eq:infection_prob_compliant_agent}
\end{eqnarray}
where $F_g(j)$ is the strength of the interaction between agent $j$ and other agents in the mixing context $g$. For NPI-adopting agents $j$, the interaction strength is modified: $F_g(j) \neq 1$. For non-adopting agents $j$, the interaction strength is unchanged: $F_g(j) = 1$. 

While CI and HQ are activated from the beginning of the simulation and last throughout the entire simulation period, SD and SC are triggered only if the cumulative incidence exceeds certain threshold. Thus, in general, an NPI is activated once the cumulative incidence exceeds a defined, NPI-dependent, threshold. Agents are assigned to adopt NPIs according to a Bernoulli process. Agents can adopt more than one NPIs when multiple NPIs are active, but their interaction strengths are set to the value of $F_g(j)$ following a descending priority assignment order: CI, HQ, SD, and SC. In this study, we assume static adoption/compliance-with NPI for CI, HQ, SC, and SD (i.e., adoption/compliance level remains constant throughout the simulation). The NPI parameterisation, encompassing both macro-distancing and micro-distancing levels, is presented in Table~\ref{tab:NPI}.

\renewcommand{\arraystretch}{1.3}
\begin{table}[ht]
    \centering
    \resizebox{\textwidth}{!}{
    \begin{tabular}{c|c|c|c|c|c|c|c}
         & \multicolumn{3}{c|}{Macro-distancing (population fractions)} & \multicolumn{4}{c}{Micro-distancing (interaction strengths)}\\
        \hline 
    Intervention & Compliance level & Duration $T$ & Threshold &    Household    & Community & Workplace\textbackslash{School} & Duration t  \\ \hline \hline
    CI      & 0.7 & 196 & 0 & 1.0 & 0.25 & 0.25 & $D(i)$ \\ \hline
    HQ      & 0.5 & 196 & 0 & 2.0 & 0.25 & 0.25 & 7 \\ \hline
    $SC^{s/t}$  & 1.0 & 110 & 100 & 1.0 &0.5 &0 & 110 \\ \hline
    $SC^p$  & 0.25& 110 & 100 & 1.0 &0.5 &0 & 110 \\ \hline
    SD      & 0.7 & 196 & 400 & 1.0 &0.25 &0.1 & 196 \\ \hline
    \end{tabular}}
    \caption{The macro-distancing parameters (population fractions) and micro-distancing (interaction strengths) for the considered NPIs. The micro-duration of CI is limited by the disease progression in the affected agent $i$,  $D(i)$. Compliance levels for SC for students/teachers and parents can be different, as indicated by distinct settings for $SC^{s/t}$ and $SC^p$ respectively. }
   \label{tab:NPI}
\end{table}

\vspace*{-0.5cm}
\subsection{Vaccination}
\label{sec_supp:vac}
In this study, we consider several scenarios with a partial vaccination preemptively rolled out before a pandemic wave (Policies 3, 4, and 5 as shown in Figure \ref{fig:policies_diamond_flow_chart} in the main manuscript). In these scenarios, the vaccination coverage is set to 50\% of the population prior to the  simulation start, corresponding to 11.7 million agents using 2016 census, or 12.712 million agents using 2021 census. The vaccines are distributed to different age groups as follows: 8.33\% of vaccinated agents are aged 18 years or younger ($age \leq 18$), 83.34\% are aged between 18 and 65 years ($18 < age <65$), and 8.33\% are aged 65 years or older ($age \geq 65$). In our model, the vaccination rollout contains two types of vaccines: the ``priority'' vaccine with a higher clinical efficacy (Priority, $VE^c=0.7$), representing BNT162b2 (Pfizer/BioNTech) and Moderna;  and the ``general'' vaccine with a medium efficacy (General, $VE^c=0.5$), representing ChAdOx1 nCoV-19 (Oxford/AstraZeneca). These values align with clinical observations following booster vaccination against the Omicron variant~\cite{Andrews2022}. For simplicity, we use the clinical vaccine efficacy against Omicron as a reference point for all considered variants, aiming to investigate the impact of imperfect vaccines in a comparative way. 

We further decompose $VE^c$ into two efficacy components: the susceptibility-reducing efficacy ($VE^s$) and the disease-preventing efficacy ($VE^d$), following \cite{zachreson_how_2021, chang_simulating_2022, chang_persistence_2023}:
\begin{eqnarray}
    VE^c = VE^d + VE^s - VE^s \times VE^d
    \label{eq:VEc}
\end{eqnarray}
with the efficacy components set as specified in Table~\ref{tab:vac_parameters}.

We also consider the transmission-limiting efficacy ($VE^t$) set at $VE^t = 0.4$, observed for the considered vaccines against the Omicron variant~\cite{jamanetworkopen.2022}. We carried out the sensitivity analysis in terms of vaccine efficacy components in prior studies~\cite{zachreson_how_2021,chang_simulating_2022}, testing a range of  $VE^t$ and $VE^c$ values and showing  robustness of the model to changes in these efficacy components. The model parameterisation with respect to vaccination is provided in Table \ref{tab:vac_parameters}.

The transmission probability of infecting a susceptible agent $i$, accounting for both vaccination and NPIs, is derived as follows:
\begin{eqnarray}
    p_i(n) =  1 - \prod_{g \in G_i(n)} \left [1 - (1 - VE^s_i) F_g(i) \left( 1 - \prod_{j \in A_g\setminus \{i\}} (1 - (1 - VE^t_j) F_g(j) \ p^g_{j \rightarrow i}(n))  \right) \right ]
		\label{eq-ve}
\end{eqnarray}
where for all vaccinated agents $j$: $VE^t_j = VE^t$, $VE^s_j = VE^s$ and $VE^d_j = VE^d$, and for all unvaccinated agents $j$:  $VE^t_j = VE^s_j = VE^d_j = 0$. The values of $VE^s$, $VE^d$, and $VE^t$ are selected according to the vaccine type (priority or general) allocated to the agent.

The probability of an agent experiencing illness (symptomatic) is further influenced by the disease-preventing efficacy, ($VE^d_i$), as follows: $p^d_i(n) = (1 - VE^d_i) \ \sigma_{a|c} \ p_i(n)$, given the fractions of symptomatic adults and children, denoted by $\sigma_a$ and $\sigma_c$ respectively.

\renewcommand{\arraystretch}{1.35}
\begin{table}
    \centering
    \resizebox{\textwidth}{!}{\begin{tabular}{c|c|c|c}
        Parameter                                  & Value (census 2016) & Value (census 2021) & Reference \\\hline \hline
        Priority vaccine coverage                  & 5.850M                   & 6.356M                 & 25\% of total population\\  \hline
        General vaccine coverage                   & 5.850M                   & 6.356M                 & 25\% of total population \\ \hline
        Vaccine allocation $[age \leq 18]$             & 0.975M                   & 1.059M                 & 8.33\% of vaccinated population \\ \hline
        Vaccine allocation $[18 < age < 65]$ & 9.750M                   & 10.594M                & 83.34\% of vaccinated population \\  \hline
        Vaccine allocation $[age \geq 65]$         & 0.975M                   & 1.059M                 & 8.33\% of vaccinated population \\ \hline
        Priority, $VE^c$                           & \multicolumn{2}{c|}{0.7}                          & \cite{Andrews2022} \\ \hline
        Priority, $VE^s$                           & \multicolumn{2}{c|}{0.452}                        & derived \\ \hline
        Priority, $VE^d$                           & \multicolumn{2}{c|}{0.452}                        & derived \\ \hline
        Priority, $VE^i$                           & \multicolumn{2}{c|}{0.4}                          & \cite{jamanetworkopen.2022} \\ \hline
        General, $VE^c$                            & \multicolumn{2}{c|}{0.5}                          & \cite{Andrews2022}  \\ \hline
        General, $VE^s$                            & \multicolumn{2}{c|}{0.293}                        & derived \\ \hline
        General, $VE^d$                            & \multicolumn{2}{c|}{0.293}                        & derived \\ \hline
        General, $VE^i$                            & \multicolumn{2}{c|}{0.4}                          & \cite{jamanetworkopen.2022} \\ \hline
    \end{tabular}}
    \caption{Simulation parameters for the preemptive vaccine rollout simulated within the ABM.\vspace*{-0.5cm}}
    \label{tab:vac_parameters}
\end{table}

\subsection{Sensitivity analysis}
We have carried out an extensive sensitivity analysis for a wide range of model parameters and multiple variants of concern, including  different profiles of the disease natural history~\cite{chang_modelling_2020,chang_simulating_2022,nguyen_general_2022}, the global transmission scalar~\cite{chang_modelling_2020,chang_persistence_2023,nguyen_general_2022}, the asymptomatic infectivity~\cite{chang_modelling_2020,nguyen_general_2022,chang_persistence_2023}, the fractions of symptomatic adults and children~\cite{chang_modelling_2020,chang_simulating_2022,nguyen_general_2022,chang_persistence_2023},  the NPI compliance levels (e.g., SD compliance~\cite{chang_modelling_2020,chang_simulating_2022,chang_persistence_2023}), the SD intervention threshold~\cite{nguyen_general_2022},  the interaction strength in different mixing contexts (at home, work, or community~\cite{chang_modelling_2020}), and the vaccination coverage and vaccine efficacy components~\cite{zachreson_how_2021,chang_modelling_2020}.
These studies demonstrated robustness of the  model to parameter changes within acceptable ranges supported by epidemiological evidence.

\clearpage
\section{Lorenz curves}
\label{sec_supp:lorenz}
% Should we specify what is this Lorenz curve for? e.g. "attach rate" or "measuring population heterogeneity on pandemic severity" (like the main text)?
The pandemic Lorenz curves measure unequal contributions of SA2 areas towards the aggregate nationwide pandemic severity (introduced in Section \ref{sec:Lorenz} of the main manuscript) and are constructed as follows:
\begin{enumerate}
    \item Record the national cumulative incidence at the end of  simulation (196 days), $CI_{AUS,T=196}$.
    \item Compute local attack rate, $AR_{SA2}$, as the ratio between the cumulative incidence and the total population, both taken at SA2 level,  as $AR_{SA2}=\frac{CI_{SA2,T=196}}{UR_{SA2}}$, where $CI_{SA2, T=196}$ is the local cumulative incidence at the end of the simulation (196 days) and $UR_{SA2}$ is the usual residential population.
    \item Rank SA2 areas by their local attack rate in ascending order. Using this rank, add one SA2 at a time to accumulate the population fraction reaching the national total, from 0\% to 100\%, until all SA2 areas are considered. This forms the x-axis. 
    \item Following the rank of SA2 areas in terms of their local attack rate, determined in the previous step, add one SA2 at a time to compute the fraction of the national cumulative incidence for the considered SA2 areas, from 0\% to 100\%, until all SA2 areas are considered. This forms the y-axis.
    \item Connect all data points to form a pandemic Lorenz curve.
\end{enumerate}

Let us consider a simplified example of the national population, comprising the populations of three SA2 areas, namely $SA_x$, $SA_y$, and $SA_z$, with demographic and epidemic characteristics specified in Table~\ref{tab:lorenz_dummy}.

\renewcommand{\arraystretch}{1.1}
\begin{table}[hb]
\centering
    \begin{tabular}{c|c|c|c|c}
    \hline
    SA2 & $CI$ & UR & $AR$ & $AR_{SA2}$ rank\\
    \hline \hline 
         $SA_x$ & 12 & 40 & 12/40 = 0.3 & 1 \\
         $SA_y$ & 15 & 24 & 15/24  =0.625 & 3\\
         $SA_z$ & 18 & 36 & 18/36  =0.5 & 2\\
         National total  & 45 & 100 & 45/100 = 0.45 & NA \\
         \hline
    \end{tabular}
     \caption{Lorenz curves setup: demographic and epidemic characteristics of a simplified example.}
     \label{tab:lorenz_dummy}
\end{table}

Using Table~\ref{tab:lorenz_dummy}, we can compute the corresponding pandemic Lorenz curve, shown in Figure \ref{fig_sup:SM_lorenz}:

\begin{figure}[ht]
    \centering 
    \includegraphics[scale=0.5]{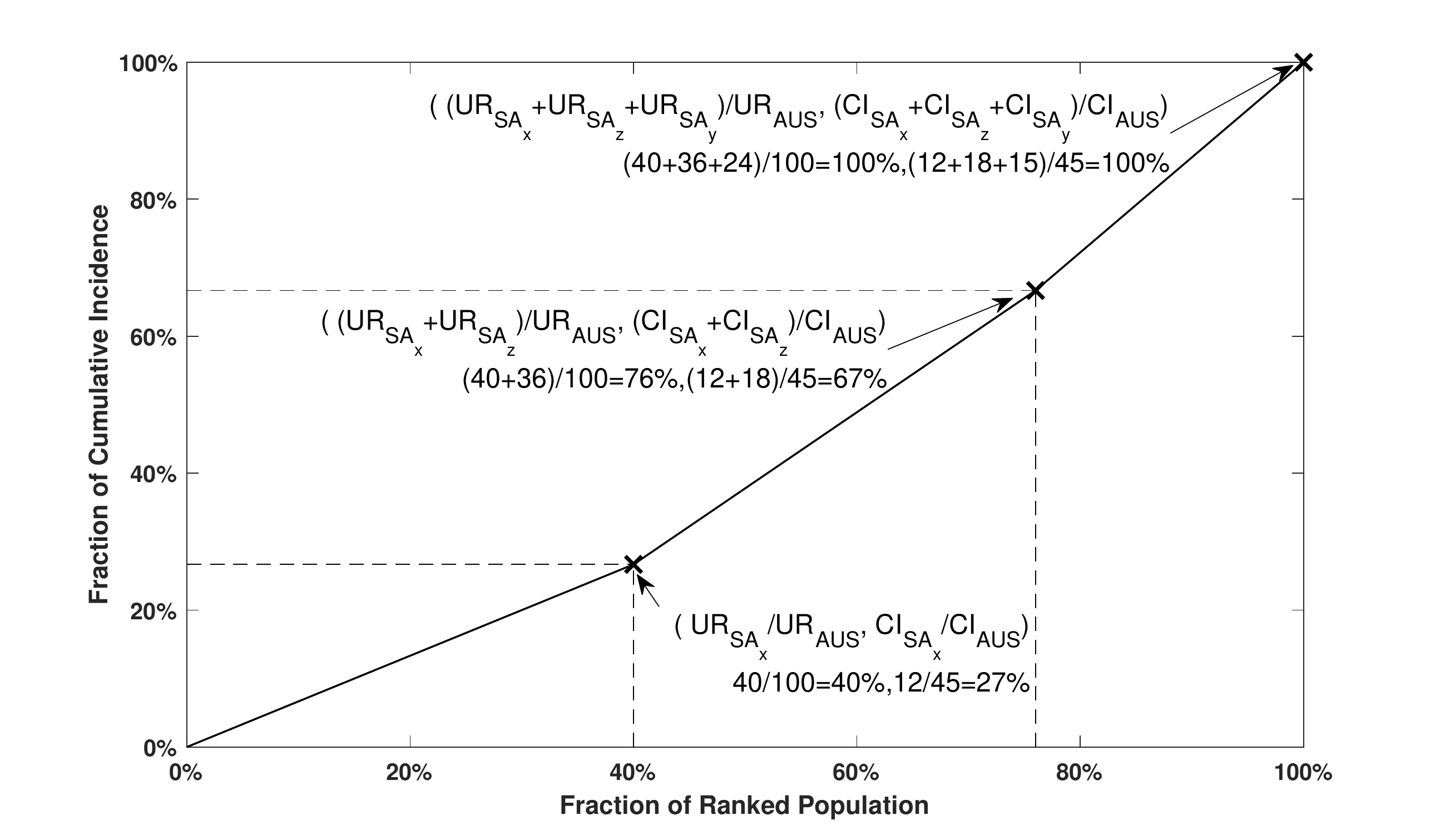}
    \caption{Constructing pandemic Lorenz curve for the simplified example, using the data presented in Table~\ref{tab:lorenz_dummy}.}
    \label{fig_sup:SM_lorenz}
\end{figure}

In this example, the three considered SA2 areas  contribute unequally towards the global attack rate, with $SA_x$ having the smallest gradient and thus deviating furthest from the line of equality (i.e., diagonal line). When using the census data, we follow this process for all SA2 areas across the country: 2,310 SA2 areas in 2016 and 2,454 SA2 areas in 2021.

\clearpage
\section{Supporting results}
\label{sm:supporting_results}
In this study, we systematically compared various COVID-19 pandemic scenarios across different census years, intervention strategies, and variants of concern. This section complements results presented in Section \ref{sec:results} in the main manuscript and follows the same structure, providing supporting results in terms of the effects of population heterogeneity on pandemic severity (Section \ref{sec_supp:pop_hetero}), the impact of urbanisation on the spread of the virus (Section \ref{sec_supp:GCC}), and the effects of school closures across variants of concern (Section \ref{sec_supp:sc}).   

\subsection{Effects of population heterogeneity on pandemic severity}
\label{sec_supp:pop_hetero}
\begin{figure}[ht]
    \centering
    \includegraphics[width=\textwidth]{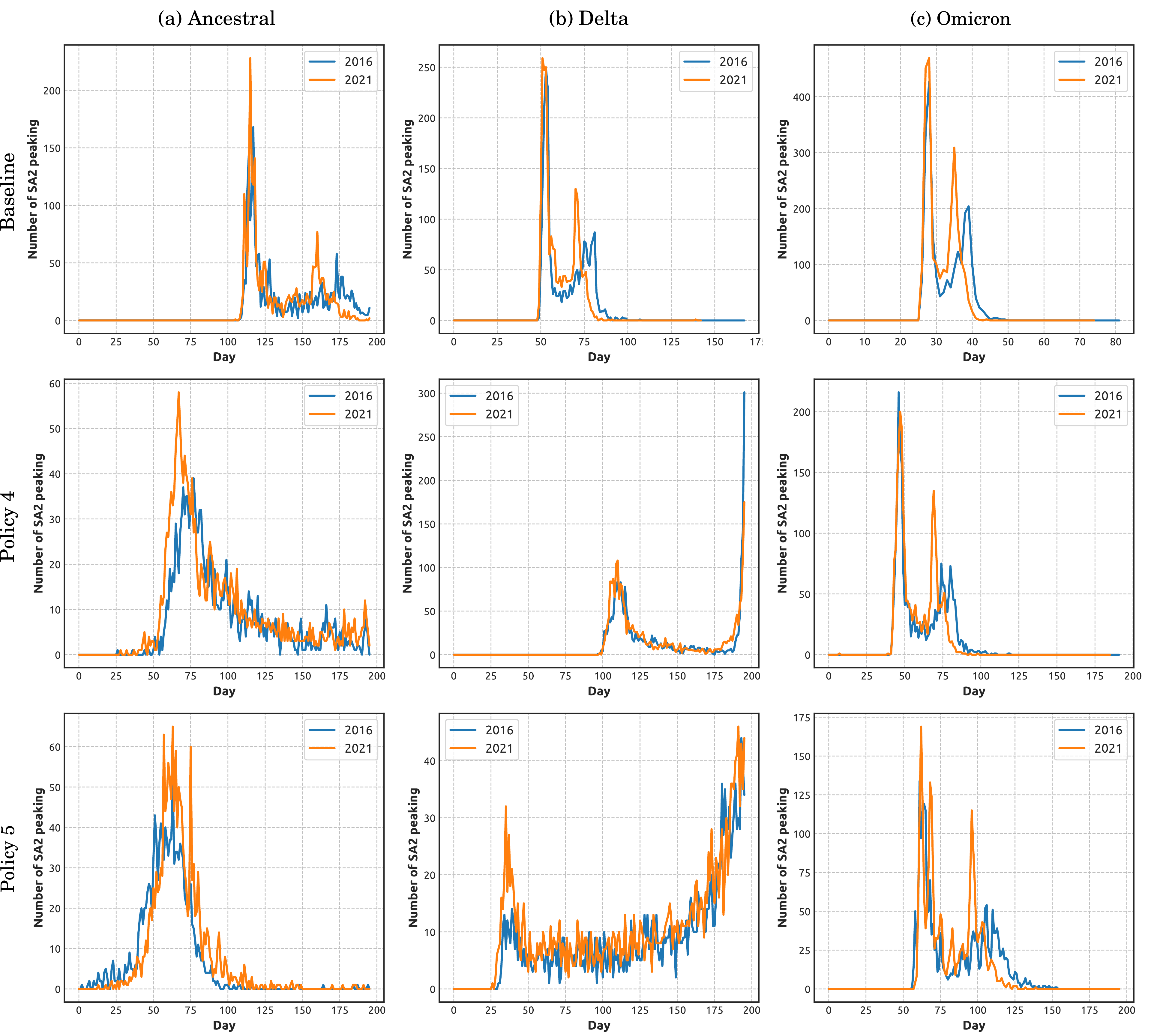}
    \caption{Number of SA2 areas exhibiting an incidence peak in the simulated time period, across three considered policies Policy 1, Policy 4, and Policy 5) and three variants: (a) ancestral; (b) Delta; (c) Omicron. Each plot is averaged over 100 runs.}
    \label{fig_sup:SA2_peak}
\end{figure}

\begin{figure}[ht]
    \centering
    \includegraphics[width=0.8\textwidth]{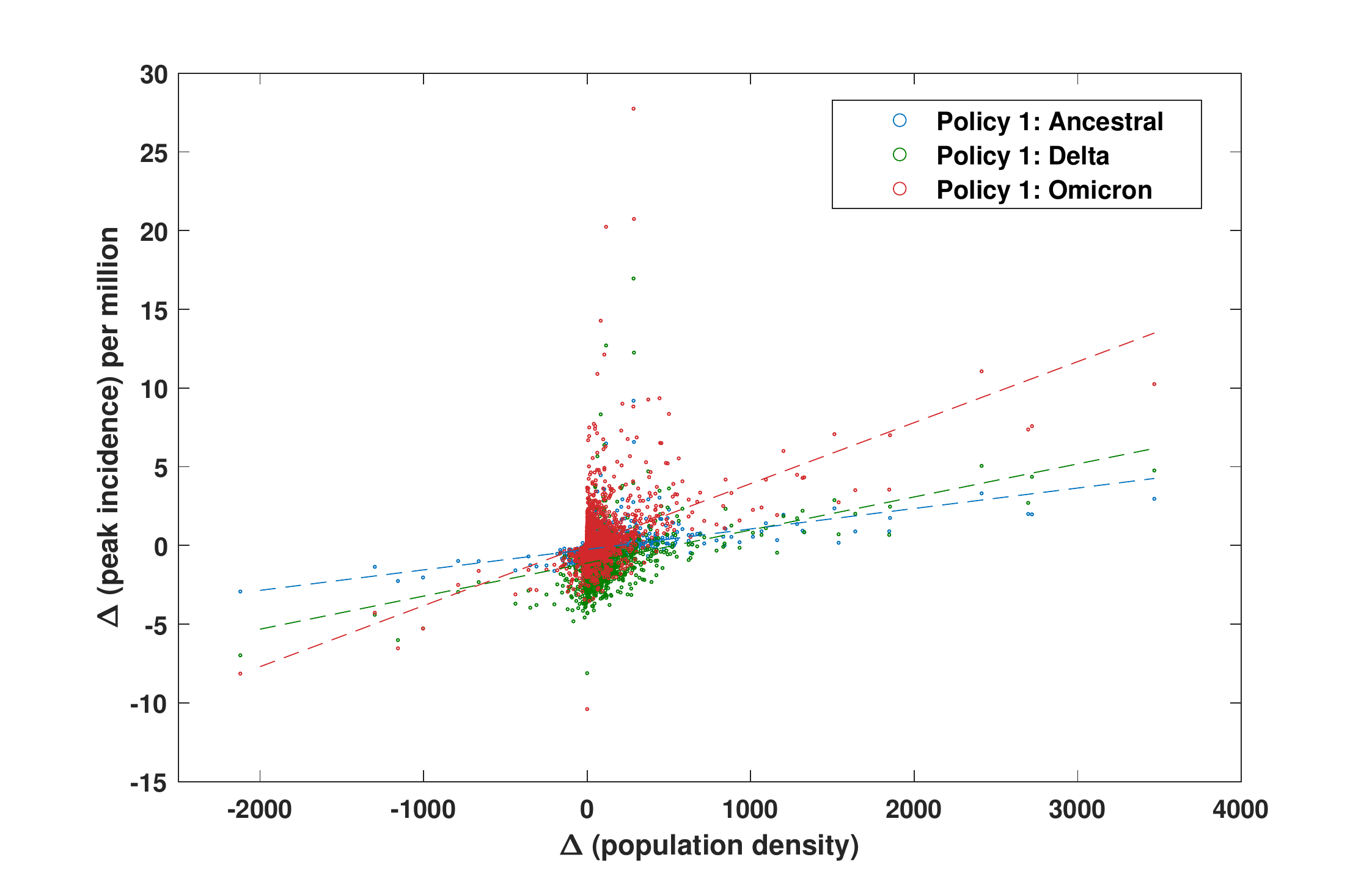}
    \caption{Correlation between the usual residential population density difference and the peak-incidence  difference, computed between
2016 and 2021 at the SA2 resolution for the three considered variants: ancestral (blue), Delta (green), and Omicron (red).
Data points corresponding to each SA2 area are derived as the averages over 100 runs. Dashed lines represent linear fit for each of the profiles. Total number of overlapping SA2 areas between 2016 and 2021 census years: 2,147. Pearson correlation coefficients:
$r_{ancestral} = 0.4150$, $r_{Delta} = 0.3426$, $r_{Omicron} = 0.4509$. These are lower correlation coefficients than the ones obtained for the population size difference, as reported in main manuscript, Figure \ref{fig:inc_pop_delta}.}
    \label{fig_sup:pop_density}
\end{figure}

\paragraph{Population growth amplifies pandemic peaks.}
Figure \ref{fig_sup:SA2_peak} shows that in comparison to 2016 results (solid blue line), there are more SA2 areas in 2021 that peak around similar times across all variants and policies (solid orange line). In cases where two peaks are observed (e.g., Figure~\ref{fig_sup:SA2_peak} (c)), in 2021, the time difference between the two peaks is shortened, explaining the weakened bimodal dynamics (Figures \ref{fig:sim2} and \ref{fig:urbanisation} in main manuscript) as the two incidence peaks tend to merge into a single but wider incidence wave.    
\paragraph{Changes in population size amplify incidence peak more than changes in density.}
Figure \ref{fig_sup:pop_density} shows a weaker yet significant correlation between the usual residential population density difference and the peak-incidence difference, computed between 2016 and 2021 at the SA2 resolution for three considered variants. The statistical analysis of the linear fits, shown in Figure~\ref{fig_sup:pop_density} and Figure \ref{fig:inc_pop_delta}, is summarised in Table~\ref{tab_sup:t10} below.

\renewcommand{\arraystretch}{1.2}
\begin{table}[hb]
\centering
    \begin{tabular}{l|c|c|c|c}
    \hline
        Variant of concern & \multicolumn{2}{c|}{Usual residential population} & \multicolumn{2}{c}{Population density}   \\
        \hline \hline 
                           & $R^2$ & RMSE &$R^2$ & RMSE\\
                           \hline
         Ancestral & 0.607 & 0.296  & 0.206 & 0.393\\
         Delta & 0.432 & 0.694  &  0.15 & 0.788\\
         Omicron & 0.81 & 0.561 & 0.25 & 1.02 \\
         \hline
    \end{tabular}
    \caption{Statistical analysis of the linear fits shown in Figure \ref{fig:school_closure} in main manuscript (effects of population size difference) and Appendix Figure \ref{fig_sup:pop_density} (effects of population density difference), in terms of the square of the correlation ($R^2$) and the  root mean squared error (RMSE).}
    \label{tab_sup:t10}
\end{table}

\begin{figure}[ht]
    \centering 
    \includegraphics[height=0.8025\textheight]{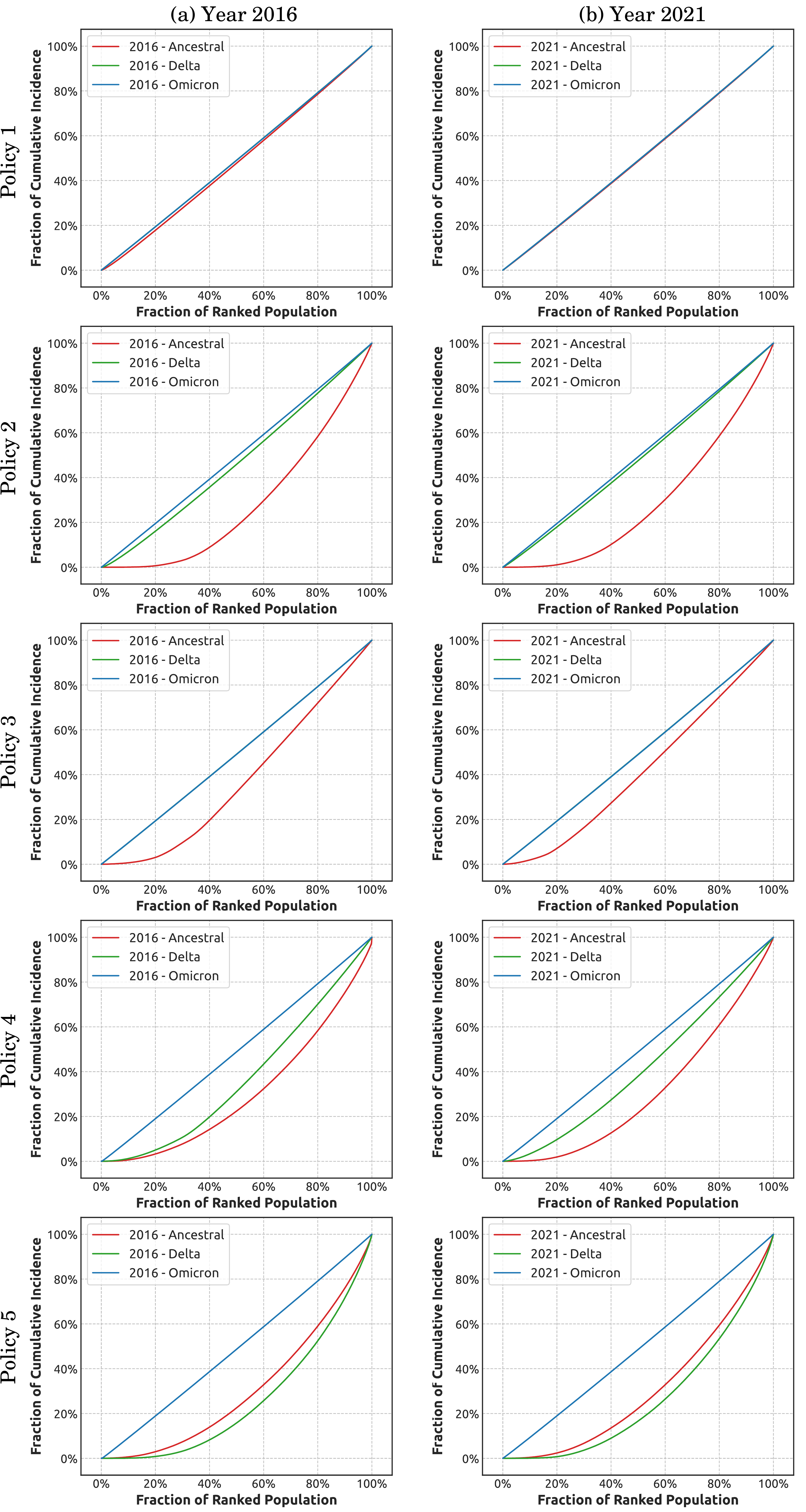}
    \caption{Pandemic Lorenz curves  for the considered policies across the three variants and two census years. Each row compares the impact of a policy using 2016 census (left) and 2021 census (right) for three variants. Refer to Figure 2 in the main manuscript for a detailed description of the considered intervention policies. Each profile corresponds to one intervention policy and is computed as the average over 100 runs.}
    \label{fig_sup:Lorenz_policy}
\end{figure}

\paragraph{Pandemic severity is distributed unequally across local (SA2) areas.}
In this section, we present pandemic Lorenz curves directly comparing variants of concern for different policies and census years (Figure \ref{fig_sup:Lorenz_policy}).

\clearpage
\subsection{Effects of urbanisation on pandemic spread} 
\label{sec_supp:GCC}
\begin{figure}[ht]
    \centering
    \includegraphics[width=\textwidth]{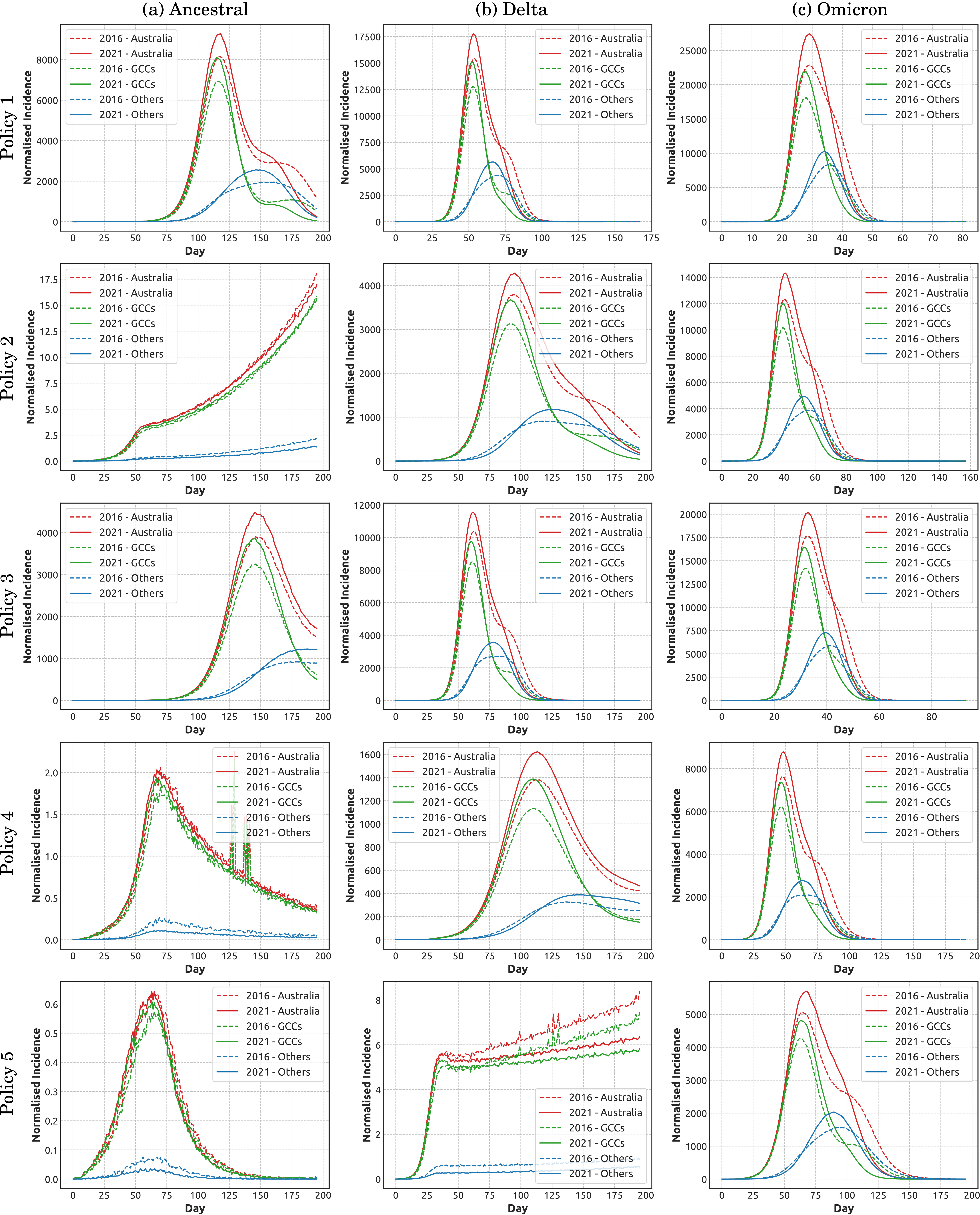}
    \caption{GCCs versus other non-urban areas. Effects of urbanisation on pandemic dynamics for the considered policies, across three variants, using 2016 and 2021 census years. We partition SA2 areas in Australia as Greater Capital Cities (GCCs), and other non-urban areas. Each column compares the impact of five intervention policies for a variant of concern: (a) ancestral; (b) Delta; (c) Omicron. Each plot is averaged over 100 runs.}
    \label{fig_sup:GCC_inc}
\end{figure}

\begin{figure}[ht]
    \centering
    \includegraphics[width=\columnwidth]{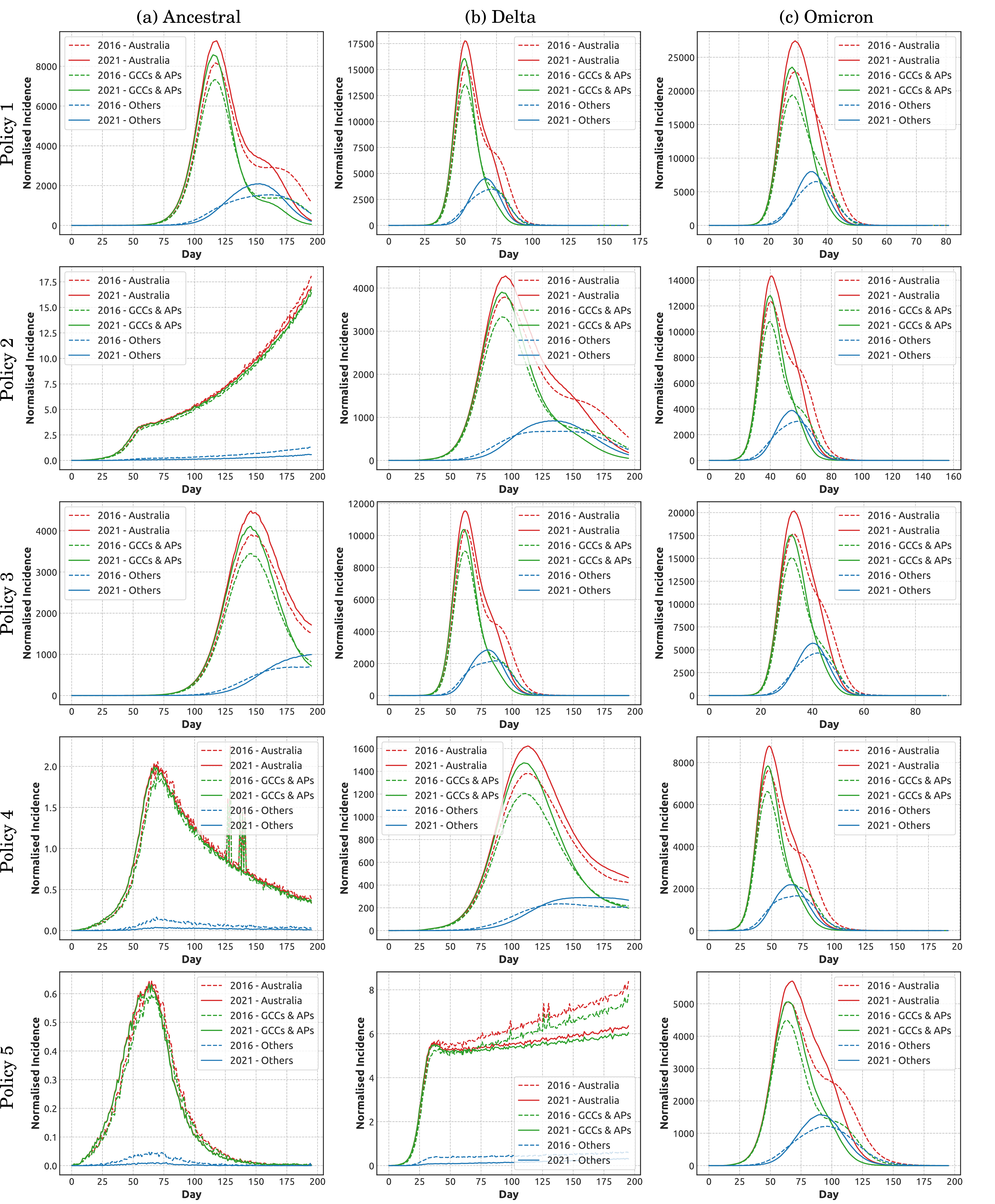}
    \caption{GCCs and APs versus other non-urban areas. Effects of urbanisation on pandemic dynamics for considered policies, across three variants using 2016 and 2021 census years. We partition SA2 areas in Australia as Greater Capital Cities (GCCs) and those close to international airports (APs), and other non-urban areas. Each column compares the impact of five intervention policies for a variant of concern: (a) ancestral; (b) Delta; (c) Omicron. Each plot is averaged over 100 runs.}
    \label{fig_sup:GCC_AP_inc}
\end{figure}

\begin{figure} [ht]
    \centering
    \includegraphics[width=0.9\columnwidth]{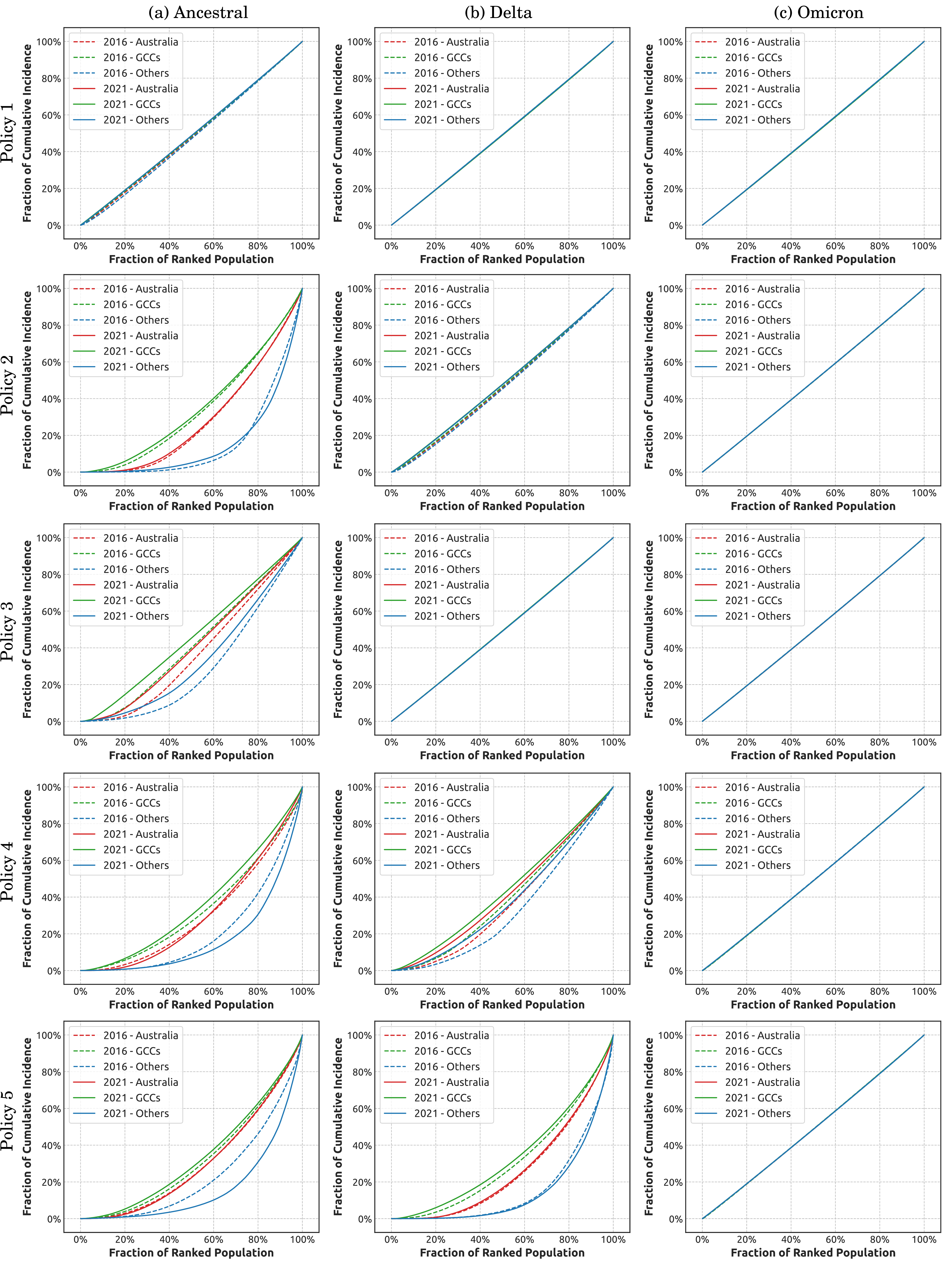}
    \caption{Pandemic Lorenz curves  for considered variants across policies and census years. Each column compares the impact of five intervention policies for a variant of concern for GCCs and non-GCCs: (a) ancestral; (b) Delta; (c) Omicron. Each profile corresponds to one intervention policy for Australia (red), GCCs (green), or other non-urban areas (blue), and is computed as the average over 100 runs.}
    \label{fig_sup:Lorenz_urban}
\end{figure}

Over the last five years, the Australian population has increased in both urban and non-urban (i.e., regional and rural) areas, leading to redefinition of SA2 boundaries (see Table \ref{tab_sup:GCC_others}). There are several non-GCC cities in Australia with international airports (see Figure \ref{fig_sup:map_with_airports} in Supplementary Material, shown in green). The inclusion of the SA2 areas surrounding these cities into analysis of the urban areas does not affect the observations reported in Section~\ref{sec:urbanisation} of main manuscript.  In this section, we compare the effects of 
growing urbanisation on pandemic dynamics for Greater Capital Cities (GCC) considered without and with cities having international airports (AP), as shown by Figures~\ref{fig_sup:GCC_inc} and~\ref{fig_sup:GCC_AP_inc} respectively.  
We also examine pandemic Lorenz curves in context of this urbanisation (Figure \ref{fig_sup:Lorenz_urban}).

\renewcommand{\arraystretch}{1.2}
\begin{table}[ht]
\centering
    \begin{tabular}{c|c|c|c}
         & GCCs / non-urban areas  & GCCs \& APs  / non-urban areas & Total   \\
        \hline \hline 
        \multicolumn{4}{c}{2016 census} \\
        \hline
     Population & 15,612,130 / 7,794,205 & 17,359,223 / 6,047,112 & 23,406,335 \\
     SA2 areas       & 1,319 / 991 & 1,498 / 812  & 2,310 \\
     \hline
     \multicolumn{4}{c}{2021 census} \\
        \hline
     Population & 17,059,063 / 8,368,966 & 18,973,019 / 6,455,010  &  25,428,029\\
     SA2 areas       & 1,460 / 994 & 1,650 / 804 & 2,454 \\
     \hline
    \end{tabular}
    \caption{Population distribution of 2016 and 2021 census years, following the urban and non-urban partition explained in Section 3.2 in main manuscript, for Greater Capital Cities (GCCs) and cities with international airports (APs).}
    \label{tab_sup:GCC_others}
\end{table}

\clearpage
\subsubsection{Effects of school closures across variants of concern} 
\label{sec_supp:sc}
\begin{figure}[ht]
    \centering
    \includegraphics[width=\columnwidth]{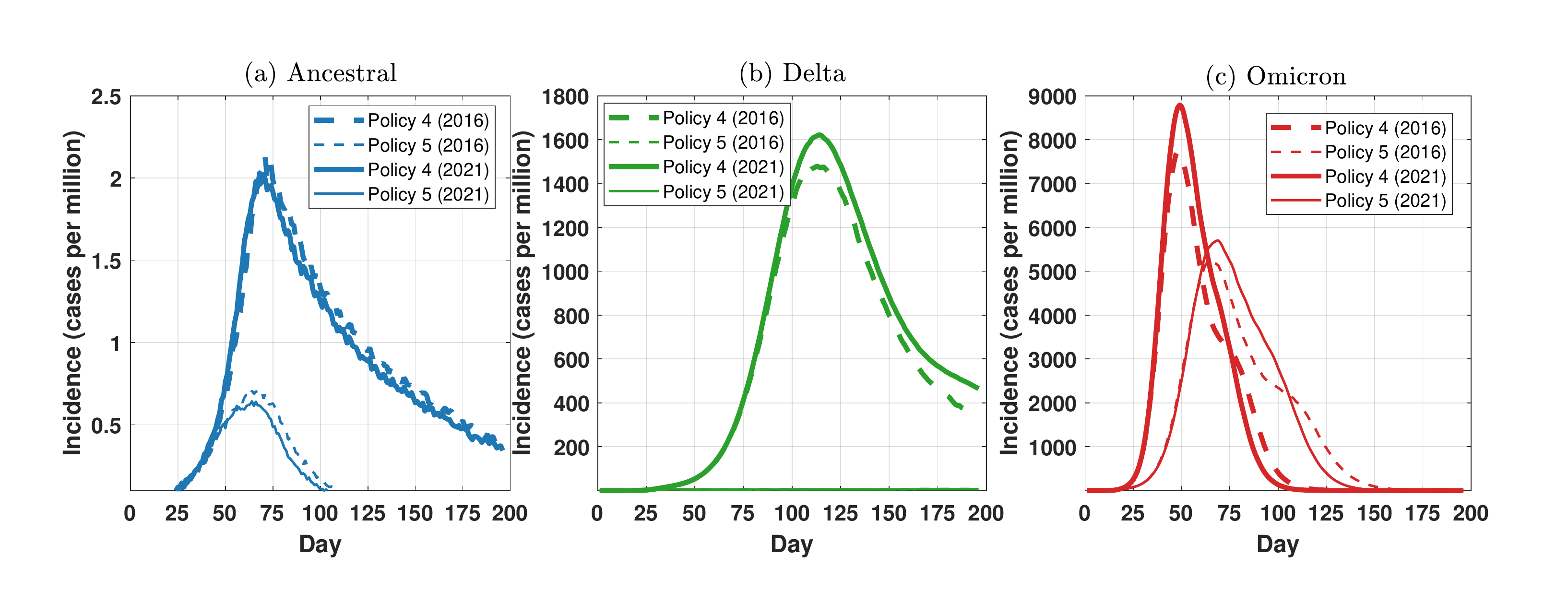}
    \caption{Effects of school closures combined with NPIs for three considered variants (on linear scale). (a) ancestral; (b) Delta; (c) Omicron. Figure \ref{fig:school_closure} in main manuscript shows these plots on log scale. Each profile corresponds to one intervention policy and is computed as the average over 100 runs.}
    \label{fig_sup:school_linear}
\end{figure}

 Figure \ref{fig_sup:school_linear} shows the effects of school closures combined with NPIs for three considered variants on linear scale.

\end{appendices}

\clearpage
\bibliographystyle{unsrt}  
\bibliography{references}

\end{document}